\documentclass[12pt]{article}

\usepackage{amsmath,amssymb,multirow,jheppub,graphicx}
\usepackage{verbatim}
\usepackage[all]{xy}
\usepackage{bm}  
\usepackage{slashed}

\allowdisplaybreaks[1]

\def\Z{\mathbb{Z}} 
\def\R{\mathbb{R}} 
\def\C{\mathbb{C}} 
\def\B{\mathbb{B}} 
\def\P{\mathbb{P}} 
\def\gf{\mathfrak{g}}
\def\tf{\mathfrak{t}}
\def\Im{\text{Im}}
\def\Re{\text{Re}}
\def\tr{\text{tr}}
\def\ad{\text{ad}}
\def\Gad{{G_\ad}}
\def\v{{\vee}}
\def\vev#1{{\langle{#1}\rangle}} 
\def\tsum{{\textstyle{\sum}}}
\def\til{\widetilde}
\def\hat{\widehat}
\def\bar{\overline}
\def\del{{\partial}}
\def\cA{{\mathcal A}}
\def\cB{{\mathcal B}}
\def\cC{{\mathcal C}}
\def\cG{{\mathcal G}}

\def\cL{{\mathcal L}}
\def\cM{{\mathcal M}}
\def\cN{{\mathcal N}}
\def\cO{{\mathcal O}}
\def\cP{{\mathcal P}}
\def\cV{{\mathcal V}}

\def\hT{{\hat T}}
\def\hW{{\hat W}}
\def\tcG{{\til\cG}}

\def\tG{{\til G}}

\def\tV{{\til V}}
\def\a{{\alpha}}
\def\ba{{\bar\a}}
\def\b{{\beta}}
\def\g{{\gamma}}
\def\G{{\Gamma}}
\def\d{{\delta}}
\def\D{{\Delta}}
\def\e{{\epsilon}}
\def\z{{\zeta}}

\def\th{{\theta}}

\def\l{{\lambda}}
\def\L{{\Lambda}}
\def\m{{\mu}}
\def\n{{\nu}}
\def\x{{\xi}}
\def\r{{\rho}}
\def\s{{\sigma}}
\def\t{{\tau}}
\def\f{{\varphi}}
\def\o{{\omega}}
\def\O{{\Omega}}

\title{
The semi-classical expansion and resurgence in gauge theories:   
new perturbative, instanton, bion, and renormalon effects}
\author[1]{Philip C. Argyres}
\author[2]{and Mithat \"Unsal}
\affiliation[1]{Physics Dept., Univ.\ of Cincinnati, Cincinnati OH 45221-0011}
\affiliation[2]{Department of Physics and Astronomy, SFSU, San Francisco, CA 94132}
\affiliation[2]{SLAC and Department of Physics, Stanford University, CA 94025}
\emailAdd{philip.argyres@gmail.com}
\emailAdd{unsal.mithat@gmail.com}
\abstract{
We study the dynamics of four dimensional gauge theories with adjoint fermions for all gauge groups, both in perturbation theory and non-perturbatively, by using circle compactification with periodic boundary conditions for the fermions.  There are new gauge phenomena.  We show that, to all orders in perturbation theory, many gauge groups are Higgsed by the gauge holonomy around the circle to a product of both abelian and nonabelian gauge group factors.  Non-perturbatively there are monopole-instantons with fermion zero modes and two types of monopole--anti-monopole molecules, called bions. One type are \emph{magnetic bions} which carry net magnetic charge and induce a mass gap for gauge fluctuations.  Another type are \emph{neutral bions} which are magnetically neutral, and their understanding requires a generalization of multi-instanton techniques in quantum mechanics --- which we refer to as the Bogomolny--Zinn-Justin (BZJ) prescription --- to  compactified field theory.  The BZJ  prescription applied to bion--anti-bion topological molecules predicts a singularity on the positive real axis of the Borel plane (i.e., a divergence from summing large orders in peturbation theory) which is of order $N$ times closer to the origin than the leading 4-d BPST instanton--anti-instanton singularity, where $N$ is the rank of the gauge group.  The position of the bion--anti-bion singularity is thus qualitatively similar to that of the 4-d IR renormalon singularity, and we conjecture that they are continuously related as the compactification radius is changed.  By making use of transseries and \'Ecalle's resurgence theory we argue that a non-perturbative continuum definition of a class of field theories which admit semi-classical expansions may be possible. 
}
\arxivnumber{...}

\begin{document}
\maketitle

\section{Introduction and results}

Circle compactification with periodic fermions, as opposed to thermal compactification, provides an effective framework to study the non-perturbative dynamics of four dimensional gauge theories.  In particular, it has been recently realized \cite{Kovtun:2007py} that $SU(N)$ gauge theory with $n_f$ light adjoint representation fermions---commonly called QCD(adj)---compactified on $\R^3 \times S^1$ does not undergo a center-symmetry changing phase transition provided the fermions are endowed with periodic boundary conditions.  Furthermore, at sufficiently small circle size (with respect to the strong coupling scale of the 4-d theory) this theory is weakly coupled and the gauge group abelianizes (is Higgsed down to $U(1)$ gauge factors).  In this situation difficult properties such as confinement and the mass gap can be studied analytically through semi-classical methods \cite{Unsal:2007vu, Unsal:2007jx}.  At large $N$ this theory on a small circle is in the same universality class as the theory on $\R^4$, and provides a controlled approximation for studying its gauge dynamics. 

The Euclidean partition function with periodic fermions on a circle of circumference $L$ corresponds to a twisted (non-thermal) partition function, $\til Z (L) = \tr [e^{-L H } (-1)^F]$, where $H$ is the gauge theory Hamiltonian and $F$ is fermion number.  For supersymmetric theories, like QCD(adj) with $n_f=1$, this is the Witten index \cite{Witten:1982df} which is famously independent of $L$.  Recent work has shown that there are also non-supersymmetric gauge theories, like $SU(N)$ QCD(adj) with $n_f>1$ and with large-enough $N$, which do not undergo any phase transition as the radius of the circle is varied \cite{Bringoltz:2009kb, Azeyanagi:2010ne, Hietanen:2009ex}.  This is due to large-$N$ volume independence:  at large $N$ an $SU(N)$ gauge theory on $\R^4$ is non-perturbatively equivalent to its compactified version on $T^d \times \R^{4-d}$, where $T^d$ is a $d$-dimensional torus, provided center and translation symmetries are unbroken \cite{Eguchi:1982nm, Kovtun:2007py}.  This implies, for example, a large-$N$ equivalence among a matrix quantum mechanics for small $T^3 \times \R$, compactified field theory on $\R^3 \times S^1$,  and quantum field theory on $\R^4$.  Furthermore, there is a large-$N$ orientifold equivalence between $SU(N)$ QCD(adj) and $SU(N)$ gauge theory with two-index antisymmetric representation fermions, QCD(AS) \cite{Armoni:2003gp}, provided charge conjugation symmetry is unbroken \cite{Unsal:2006pj}.  QCD(AS) is of special interest as it provides a different large-$N$ limit of $SU(3)$ QCD with fundamental (or, equivalently, antisymmetric) Dirac fermions.   

It is a natural hope that the interconnected ideas of center-stabilizing abelianizing compactifications and large-$N$ volume independence will provide effective alternative ways to think about 4-d gauge dynamics in general, for example by using equivalent matrix models.   In this work we take a small step towards evaluating this idea by systematically studying QCD(adj) for general simple gauge group $G$ on $\R^3 \times S^1$.  We uncover new gauge phenomena compared to the $G=SU(N)$ case.  In particular, we find that although perturbative effects lead to center-stabilizing potentials for the gauge holonomy, their minima  do not always abelianize the gauge dynamics.  We also argue that a topological molecule that we refer to as a \emph{neutral bion} with the same quantum numbers as the perturbative vacuum gives important and calculable contributions to the holonomy effective potential.  This effect is also present in supersymmetric theories, the $n_f=1$ case, as explained in \cite{Poppitz:2011wy}, and can also be deduced from the bosonic potential which arises from the superpotential for $n_f=1$ \cite{Davies:2000nw,Seiberg:1996nz, Katz:1996th}.   Finally, we argue that bion--anti-bion contributions to the semiclassical expansion of vacuum quantities are associated to poles in the Borel plane responsible for the leading divergence of perturbation theory---the so-called \emph{IR renormalon divergence}.  We show how an extension of methods used to control the semiclassical expansion in double-well quantum mechanics can also be used to give unambiguous results for the dilute 3-d monopole-instanton gas that appears in the semiclassical expansion

In the rest of this introduction, we review the perturbative and non-perturbative behavior of $G=SU(N)$ QCD(adj) on a small circle, and summarize and contrast our results for other choices of gauge group $G$.

\subsection{Perturbation theory}

In a 4-d gauge theory with gauge group $G$, $n_f$ massless  adjoint fermions, and compactified on a periodic circle of circumference $L$, denote the gauge holonomy (the open Wilson line) around the circle by $\O:=\exp\{2\pi i \f\}$ where $\f$ is an element of the Lie algebra $\gf$ associated to $G$.  Gauge transformations change $\f$ by conjugation in $\gf$, so the gauge-invariant information in the holonomy is the conjugacy class, $[\f]$, of $\f$.  One way of characterizing this conjugacy class is by giving the set of eigenvalues, $\{\f_i\}$, of $\f$ in a given representation of $\gf$.  (In later sections, though, we will use a more invariant description of $[\f]$ that does not depend on a choice of representation.)  For $SU(N)$, choosing the fundamental representation, the $\f_i$ are the $N$ eigenvalues of $\f$ which are defined only up to integer shifts and obey $\sum_{i=1}^N\f_i\in\Z$; equivalently, $\exp\{2\pi i\f_i\}$ are $N$ eigenphases of $\O$ which are constrained to multiply to one.

For pure Yang-Mills theory in the small-$S^1$, weak coupling regime, the bosonic gauge fluctuations induce an attraction between eigenvalues causing them to clump at $\f_i=0$ \cite{Gross:1980br}.  When periodic adjoint fermions are added to the $G=SU(N)$ theory, they generate an eigenvalue interaction of the form $\sum_{1\le i<j\le N} g(\f_i - \f_j)$ which is repulsive between any pair of eigenvalues.  The minimum of this potential is a uniform distribution of the eigenphases over the unit circle, and is the unique configuration which is invariant under the $\Z_N$ center symmetry.  Since $\O$ behaves as an adjoint Higgs field, this configuration leads to the abelianization of long-distance gauge dynamics, Higgsing $SU(N) \to U(1)^{N-1}$.

For general gauge group, adjoint fermions still induce an effect which negates that of the bosonic fluctuations and favors $\f$ which preserve the center symmetry.  However for groups other than $SU(N)$ the fermion-induced eigenvalue repulsion is no longer uniform between all pairs of eigenvalues, but has more structure, and, except for $Sp(N)$, has the effect of forcing some pairs of eigenvalues to coincide.  When there are coincident eigenvalues, there are nonabelian factors in the un-Higgsed gauge group.  In particular, we find through a combination of analytical and numerical techniques the gauge symmetry-breaking patterns shown in table \ref{tab-1loop}, valid at all orders in perturbation theory.   The eigenvalue distributions which minimize the perturbative potential for the rank-9 classical groups are plotted as examples in figure \ref{fig-phases} in section \ref{sec2.1}.

\begin{table}[ht]
\begin{center}
\begin{tabular}{|rcccl|} \hline
 & $G$ &$\to$& $H$ & \\
\hline\hline
$\text{\it SU}(N{+}1)\ \simeq$ & $A_N$ &$\to$& 
$U(1)^N$ & for $N\ge1$ \\ \hline
$\text{\it SO}(2N{+}1)\ \simeq$ & $B_N$ &$\to$& 
$U(1)^{N-1}\times\text{\it SO}(3)$ & for $N=2,3$ \\
 & &$\to$&
$\text{\it SO}(4)\times U(1)^{N-3}\times\text{\it SO}(3)$ & for $N\ge4$ \\ \hline
$\text{\it Sp}(2N)\ \simeq$ & $C_N$ &$\to$& 
$U(1)^N$ & for $N\ge3$ \\ \hline
$\text{\it SO}(2N)\ \simeq$ & $D_N$ &$\to$& 
$\text{\it SO}(4)\times U(1)^{N-4}\times\text{\it SO}(4)$ & for $N\ge4$ \\ \hline
 & $E_6$	&$\to$& 
$\text{\it SU}(3)\times\text{\it SU}(3)\times\text{\it SU}(3)$
&\\ 
 & $E_7$	&$\to$& 
$\text{\it SU}(2)\times\text{\it SU}(4)\times\text{\it SU}(4)$&
\\ 
 & $E_8$	&$\to$& 
$\text{\it SU}(2)\times\text{\it SU}(3)\times\text{\it SU}(6)$&
\\ 
 & $F_4$ 	&$\to$& 
$\text{\it SU}(3)\times\text{\it SU}(2)\times U(1)$&
\\ 
 & $G_2$ 	&$\to$& 
$\text{\it SU}(2)\times U(1)$&\\ 
\hline
\end{tabular}
\caption{Perturbative patterns of Higgsing of the gauge group $G$ to an unbroken group $H$ for $n_f>1$ adjoint fermions with periodic boundary conditions on $\R^3 \times S^1$.\label{tab-1loop}} 
\end{center}
\end{table}

Note that the rank of the nonabelian factors does not grow with increasing $N$, and is at most four for the $SO(N)$ groups.  Also, the unbroken nonabelian factors are all $SU(n)$ factors (since $SO(4)\simeq SU(2)\times SU(2)$ and $SO(3)\simeq SU(2)$).  This may seem surprising, since the $SU(n)$ theories abelianize, but there is no contradiction since the unbroken $SU(n)$ factors are in the low-energy effectively 3-d theory which already has integrated out the Kaluza-Klein states that were responsible for generating the gauge holonomy potential in the first place.  

Importantly, a qualitative difference between the $SU(N)$ groups and the other groups is that the $SU(N)$ $\Z_N$ center symmetry group has order comparable to the rank of $SU(N)$ and uniquely determines the center-symmetric gauge holonomy, while all other groups have small center symmetries ($\Z_2$, $\Z_3$, $\Z_4$, or $\Z_2\times\Z_2$) which do not grow with rank and for which there are whole manifolds of center-symmetric holonomies.  Despite the small order of the center symmetry groups, the eigenvalues of the Wilson lines for the large-rank Lie algebras are almost uniformly distributed, with $O(1/N)$ spacing between the eigenphases.  The uniformity of eigenphases implies that at $N=\infty$ the center symmetry for the infinite Lie algebras may accidentally enhance to $\Z_{\infty} \equiv U(1)$, much like in $SU(N)$ QCD(AS) which has an exact $\Z_2$ (for even $N$) or $\Z_1$ (for odd $N$) but an emergent $\Z_{\infty}$ center symmetry at large $N$ \cite{Armoni:2007kd, Kovtun:2007py}.  Both are a consequence of a large-$N$ orientifold equivalence \cite{Armoni:2003gp, Unsal:2006pj}.  

On the other hand, the smallness of the centers of the $SO(N)$ and $Sp(2N)$ groups implies that it is possible to engineer sequences of gauge theories (by choosing appropriate fermion content or by adding Wilson line potentials) such that the eigenphase distribution does not approach a uniform limit as $N\to\infty$ even though the center symmetry remains unbroken.  This implies that for groups other than $SU(N)$, unbroken center symmetry is not a sufficient condition by itself for large-$N$ volume independence.

Gauge symmetry breaking by Wilson lines has appeared previously in models of gauge-Higgs unification in extra-dimensional model building  \cite{Hosotani:1988bm}, and examples of  Higgsing  patterns with non-abelian gauge factors appeared in examination of phases with partial center-symmetry breaking \cite{Ogilvie:2007tj}.

\paragraph{Fate of the non-abelianized theories.}

For QCD(adj) with gauge groups different from SU(N) and Sp(2N), the 3-d couplings of the non-abelian factors quickly run to strong coupling, rendering 3-d semiclassical methods ineffective.  It seems likely that these 3-d versions of QCD(adj) themselves confine; see, for example \cite{Armoni:2011dw} for a discussion of the evidence from small spatial circle compactification and large-N volume independence arguments (and of the problems with continuing from small to large circle radius).  This does suggest, however, that compactification of QCD(adj) on small 2-tori will result in a 2-d effective theory amenable to a semi-classical treatment for all gauge groups $G$.

Note that abelianizing Wilson line dynamics can be arranged for gauge theories with groups other than $SU(N)$ and $Sp(2N)$ by appropriately changing the fermion content or by modifying the theory with single-trace Wilson line deformations.  For instance, for $SO(N)$ gauge groups if one puts in $n_\text{s}=n_\ad-1$ massless Majorana fermions in the symmetric-traceless representation, where $n_\ad$ is the number in the adjoint representation, then a uniform distribution of Wilson line eigenphases results.

\subsection{Topological molecules} 

The long-distance dynamics of theories which abelianize in the small-$L$ domain is analytically tractable.  In this regime, a semi-classical treatment of elementary and molecular monopole-instanton events reveals the existence of a mass gap for gauge fluctuations and confinement of electric charges.  The semi-classical expansion is an expansion in the diluteness (or fugacity) of these defects.  The leading topological defects which play non-trivial roles in the dynamics are \\
\indent (i) monopole-instantons (or 3-d instantons and the twisted instanton) $\cM_i$, \\
\indent (ii) magnetic bions  $\cB_{ij} = [\cM_i\bar\cM_j]$, \\
\indent (iii) neutral bions   $\cB_{ii} = [\cM_i\bar\cM_i]$, and \\ 
\indent(iv) multi-bion molecular events  $[\cB_{ij} \cB_{ji}],  [\cB_{ii} \cB_{ij} \cB_{ji}]$  etc.  \\
The index $i, j$ is explained below.  We describe the physics associated with the proliferation of each type of topological defect briefly.  The third type gives a new instanton effect in compactified gauge theories, and the fourth type gives a semi-classical realization of IR renormalons that we describe below. 


\paragraph{Scales in the low-energy effective theory.}

\begin{figure}[t]
\begin{center}
\includegraphics[angle=0, width=4in]{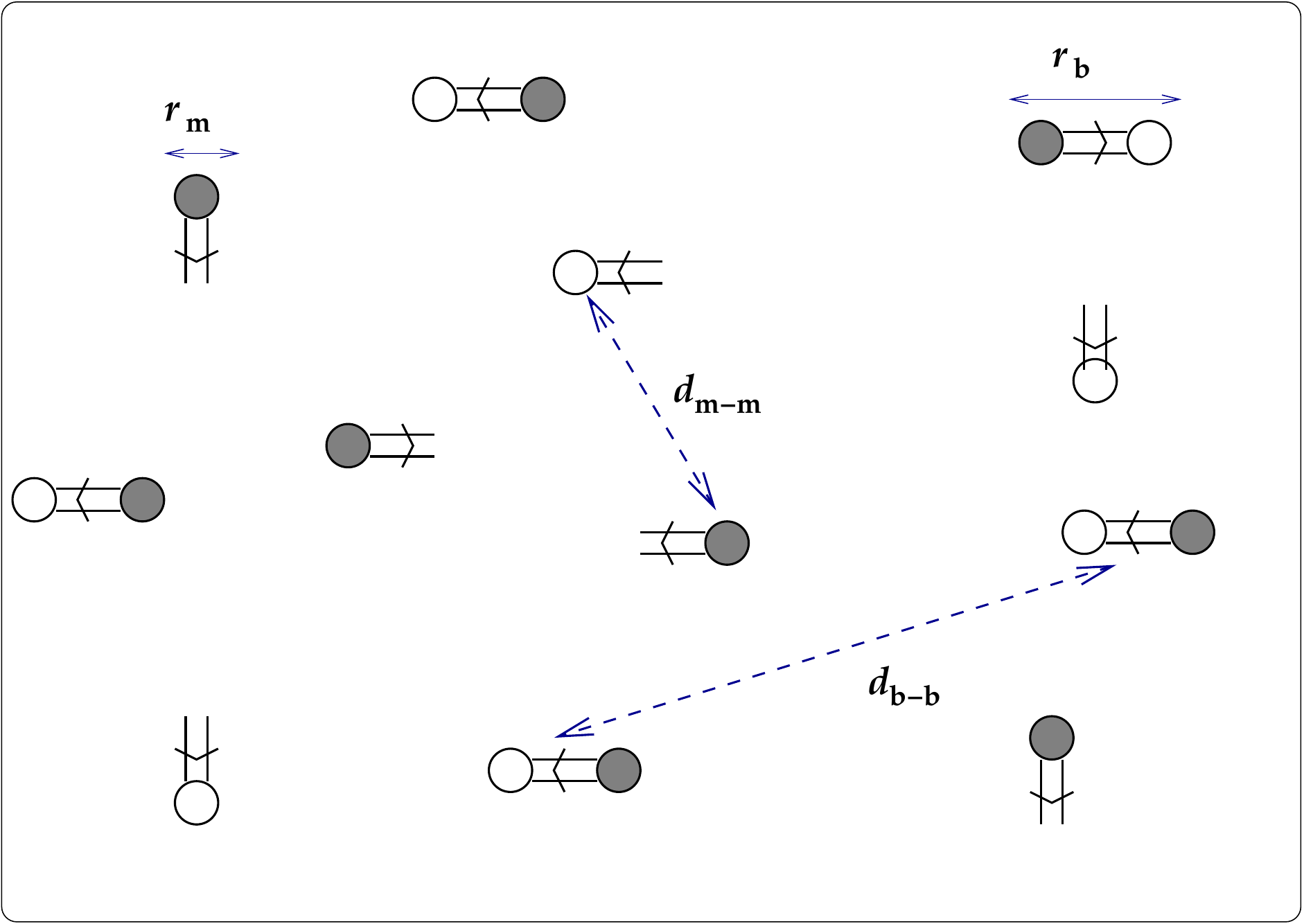}
\caption{A cartoon of the leading topological defects and molecules in the small-$S^1$ domain.  Gray and white circles represent monopole-instantons and anti--monopole-instantons.  Unpaired arrows represent fermion zero modes, paired monopole-instanton events are magnetic and neutral bions. See text for explanations.}
\label {fig:plasma}
\end{center}
\end{figure}

First, though, we explain the separation of scales,
\begin{align}  
r_\text{m} \ll r_\text{b} \ll d_\text{m-m} \ll  d_\text{b-b},
\end{align}
which makes the dilute gas of monopole-instantons and topological molecules and the effective long-distance theory derived from them reliable.  Here $r_\text{m}$ is the maximum size of a monopole-instanton, $r_\text{b}$ is the size of a bion, $d_\text{m-m}$ is the inter-monopole-instanton separation, and $d_\text{b-b}$ is the inter-bion separation.  The resulting picture of the Euclidean vacuum structure of the abelianizing QCD(adj) theories is shown in figure \ref{fig:plasma}.

This hierarchy arises as follows.  The maximum size of a monopole-instanton is fixed by the vev of the gauge holonomy.  In an abelianizing theory this gives $r_\text{m}\sim L$, where $L$ is the size of the $S^1$.  This is unlike 4-d QCD-like theories where 4-d instantons come in all sizes at no action cost, and there is no clear meaning to the long-distance description of a Euclidean instanton gas.  Because of this, 4-d instantons are unable to describe many aspects of 4-d physics, for example the mass gap or the $\th$ angle dependence of the vacuum energy.  On $\R^3 \times S^1$, however, the gauge symmetry breaking provides an IR cutoff to the size of 4-d instanton events, rendering the semi-classical analysis reliable.  

The Euclidean instanton gas is dilute when the monopole-instanton action is large, $S_0\sim (g^{2}N)^{-1} \gg1$, which is valid for small $S^1$ in asymptotically free theories since then the effective 4-d gauge coupling at the scale of the $S^1$, $g^2 := g_4^2(L)$, is small.  The density of monopole-instantons is proportional to $e^{-S_0}$, so the typical separation between monopole-instantons is $d_\text{m-m} \sim L e^{S_0/3}$ and they are rare in the limit of small fugacities $e^{-S_0}\ll 1$ (or, small $S^1$).

The size of the magnetic bion is calculated in \cite{Unsal:2007jx, Anber:2011de} and found to be $r_\text{b} \sim L g^{-2}$.  The size of the neutral bion is calculated here through the BZJ-prescription, described below, and is the same as the magnetic bion size.  The typical bion action is twice the monopole-instanton action, so the separation between bions is $d_\text{b-b} \sim L e^{2S_0/3}$, and they are even rarer than monopole-instantons.

\paragraph{Monopoles and bions.}

There are $\text{rank}(\gf)+1=N+1$ types of self-dual monopole-instantons which can be associated with the simple roots $\a_j$, $j=1, \ldots, N$ and the affine (or lowest) root $\a_0$ of the gauge algebra.  The first $N$ are sometimes referred to as 3-d instantons and the last one as the twisted instanton.  The twisted instanton  owes its existence to the locally 4-d nature of the theory, and it would not exist in a microscopically 3-d theory.  These defects carry a certain number of fermionic zero modes dictated by the Nye-Singer index theorem \cite{Nye:2000eg, Poppitz:2008hr}.\footnote{See also \cite{Bruckmann:2003ag,GarciaPerez:2009mg} on the boundary condition dependence of zero modes.} Consequently, in theories with adjoint fermions, the self-dual defects do not induce a mass gap or confinement \cite{Unsal:2007jx}. 
 
At second order in the semi-classical expansion, there are correlated instanton--anti-instanton events of various types.  In  Euclidean space, where 3-d instantons are viewed as  particles forming a dilute classical plasma, the correlated instanton--anti-instanton events should be viewed as molecular structures.  We refer to these topological molecules as bions, as they are composites of two 3-d instantons.  They fall into two classes both according to their physical effects and according to their Lie-algebraic properties.  In particular, we will see that bions are in  one-to-one correspondence with the the non-vanishing entries of the extended (or untwisted affine) Cartan matrix, $\hat A_{ij} \neq 0$, data that one can easily read off from the extended Dynkin diagram.  

For each non-vanishing off-diagonal element, $\hat A_{ij} < 0$, there exists a magnetic bion.  These carry a net magnetic charge and no fermionic zero modes.  The monopole-instanton constituents of magnetic bions have both repulsive and attractive interactions which counter-balance each other at a characteristic size much larger than that of the constituents themselves, and thus lead to a picture of the magnetic bion as a loosely bound topological molecule \cite{Unsal:2007jx, Anber:2011de}.  The plasma of magnetic bions induces a mass gap for gauge fluctuations (which is strictly forbidden to all orders in perturbation theory) and hence confinement of electric charge, similar to the way that instantons in the 3-d Polyakov model induce these phenomena \cite{Polyakov:1976fu}. 

We refer to the bions associated with the diagonal elements of the extended Cartan matrix, $\hat A_{ii}>0$, as neutral bions.  They are quite elusive in the sense that they carry neither magnetic nor topological charge, just like the perturbative vacuum.  Yet they induce a net repulsion between pairs of gauge holonomy eigenvalues, and a center-stabilizing potential, whose global minimum is at a point which leads to abelianization of the gauge group.  This is familiar from the supersymmetric ($n_f=1$) QCD(adj) theories where this non-perturbative effect is the only contribution to the superpotential and  effective bosonic potential  \cite{Seiberg:1996nz, Katz:1996th, Davies:2000nw, Poppitz:2011wy}.  Since, for $n_f>1$, perturbative effects also induce a potential for the gauge holonomy, the two effects mix.  Note that this is unlike the potential for the dual photons non-perturbatively induced by magnetic bions, which gets no contribution at any order in perturbation theory.

It may at first seem hard to make sense out of neutral bions due to their mixing with perturbation theory.  Moreover, and as it turns out relatedly, the interaction between the constituents of neutral bions are all attractive, seemingly making any notion of a topological instanton--anti-instanton molecule  meaningless.  This does not turn out to be the case.  We give a detailed description of how neutral bions arise through a generalization of multi-instanton techniques in quantum mechanics \cite{Bogomolny:1980ur, ZinnJustin:1981dx}---that we refer to as the {\it Bogomolny-Zinn-Justin (BZJ) prescription}---to compactified field theory on $\R^3 \times S^1$.  This prescription tells us how to make sense of neutral bions through an analytic continuation in coupling constant space.  The result of the BZJ-prescription agrees with the WKB-approximation in bosonic quantum mechanics \cite{Bogomolny:1980ur, ZinnJustin:1981dx}, and with exact results in supersymmetric quantum mechanics \cite{Balitsky:1985in, Balitsky:1986qn} and supersymmetric field theory on $\R^4$ \cite{Yung:1987zp, Poppitz:2011wy}.  

\subsection{Resurgence and Borel-\'Ecalle summability} 

We believe that the BZJ-prescription can be systematically extended to all orders of the semi-classical expansion.  At fourth order and beyond in the semi-classical expansion of QCD(adj) a new gauge phenomenon appears.  We find an ambiguity in the non-perturbative bion--anti-bion $[\cB_{ij} \cB_{ji}]$ contribution to the instanton expansion.   According to Lipatov \cite{Lipatov:1976ny}, this predicts a divergence in the leading zeroth order part of the semi-classical expansion, which is the purely perturbative part of the expansion around the perturbative vacuum.  This divergence corresponds to a singularity in the Borel plane which is of order $N = \text{rank}(G)$ closer to the origin than the one associated to the 4-d BPST instanton.  This is similar to the location of the ``IR renormalon" singularity in 4-d gauge theories \cite{'tHooft:1977am}, and we conjecture, as already reported in \cite{Argyres:2012vv}, that this singularity is in fact continuously connected to the 4-d IR renormalon singularity.

If, furthermore, this is the \emph{leading} singularity in the Borel plane of abelianizing QCD(adj) on a small circle, then the extension of the BZJ prescription to all orders in the instanton expansion together with the technique of \emph{resurgence} and Borel-\'Ecalle summation of semi-classical transseries \cite{Ecalle:1981,  Sternin:1996, Costin:2009, s07} offers the promise of a finite definition of this class of field theories from their semi-classical expansions.  
Resurgence theory provides detailed information on Borel transforms and sums, their inter-connection to Stokes phenomena and a set of general summation rules for asymptotic perturbative expansions  which are otherwise known to be non-Borel summable.  

Recently, in a class of matrix models which do not involve infrared renormalons, Mari\~no showed that the BZJ prescription, used to cancel ambiguities, can indeed be systematically extended to all orders via resurgence \cite{Marino:2008ya}.  Schiappa et.\ al.\ provides a generalization of this to any one-parameter transseries  \cite{Aniceto:2011nu}.  In theories with renormalons, the present work on gauge theory on $\R^3 \times S^1$, and its companion paper on the ${\C\P}^{N-1}$ model on $\R^1 \times S^1$ \cite{Dunne}, are the first attempts to combine the perturbative and semi-classical expansions into a well-defined transseries expansion. However, it is  currently not clear to us whether a one-parameter transseries will suffice for the extension of the BZJ prescription to all orders, or a multi-parameter transseries is needed. 

The main physical idea underlying resurgence can be explained for ordinary integrals with multiple saddle points.  The most intuitive and physical explanation that we have found is due to Berry and Howls \cite{bh91}.  A key point---and a surprising one---is that our interpretation and analysis of the path integral of quantum field theory, in particular QCD(adj), fits well with that of \cite{bh91}, despite the fact that a path integral is infinitely many coupled ordinary integrals!  

Consider an ordinary integral with a certain number of saddle points, and let $\cC_n$ denote a contour passing through the $n^{\rm th}$ saddle point.  The $n=0$ saddle point may be considered as a zero-dimensional analog of the perturbative vacuum, and we may set the its action to zero to emulate the field theory construction.  Each integral associated with contour $\cC_n$ can be treated by refining the method of steepest descent.  The result for a small expansion parameter $\l$ (or large parameter $1/\l$) is of the form $\exp(-nA/\l) P_n(\l) \sim \exp(-nA/\l) \sum_{q=0}^{\infty} a_{n, q} \l^q$ where $A$ is a positive constant.  $P_0(\l)$ is thus the perturbative expansion, and all the series $P_n(\l)$ are asymptotic.  Ref.\ \cite{bh91} shows that the divergence of the asymptotic series $P_n(\l)$ is a consequence of the existence of other saddle points 
$n' \neq n$, through which the contour $\cC_n$ \emph{does not} pass.  In particular, the non-perturbative data from non-trivial ($n\neq0$) saddle points (``instantons") are encoded into the \emph{universal} late terms of the divergent series $P_0(\l)$.  In other words, the late terms of the perturbative expansion ``knows" of the existence of all the other saddle points.  For general $n$,  
again, the late terms are dictated by the existence of the other saddle points, meaning that the late terms of all series $P_n(\l)$ are interconnected by the requirement mutual consistency.  There is a 
universality associated with late terms, regardless of what the value of $n$ is, encoded in the positions of all saddle points (or, the instanton actions).  Thus, there is a sense in which all perturbative fluctuations around all non-perturbative sectors are interconnected.  The perpetual \emph{reappearance} of the universal form in the late terms of the asymptotic expansions around non-trivial saddle points is called the \emph{principle of resurgence}.

It seems to us that the concept of resurgence and resurgent functions  is the natural language of semi-classical expansions in quantum field theory.  We anticipate that it will play a crucial role in making sense of general continuum field theories, especially if the theory admits a semi-classical expansion. 

\subsection{Outline of the rest of the paper}

The organization of the paper is given in the table of contents.  We have included some review material to help make the paper more self-contained.  In particular, sections 2, 4, and appendix A are mostly review of standard results in effective gauge theories, BPS instantons on $\R^3\times S^1$, and in Lie algebras, respectively.  However the argument in section 2.4 showing that the Higgsing pattern determined at 1-loop is not modified at any higher order in perturbation theory is new, and the discussion of section 4 generalizes earlier discussions for the supersymmetric case to non-supersymmetric theories.

Section 3 contains a combination of analytic arguments and numerical calculations to determine the location of the minima of the one-loop potential for the Wilson line, and the resulting mass spectra and patterns of gauge symmetry breaking.  These calculations rely on explicit descriptions of the gauge cells (affine Weyl chambers) of the simple Lie algebras worked out in appendix B.

Section 5 reviews the description of magnetic bions, then explains and applies the BZJ prescription to the calculation of the neutral bion and bion--anti-bion contributions to the instanton expansion.  

Section 6 briefly explains why there is no consistent regime in which the potential induced by the neutral bion contribution, though giving rise to a strong Wilson line eigenvalue replusion, can overcome the perturbative contributions which force some pairs of eigenvalues to coincide.

Section 7 discusses the implications of the semi-classical analysis for abelianizing QCD(adj) theories on predictions for the mass gap, string tension, and chiral symmetry realization in the 3-d effective theory.  These results are qualitatively similar to previous results obtained for $SU(N)$ QCD(adj).

Finally, section 8 contains a preliminary discussion of the some of the systematics of how the BZJ prescription and the machinery of Borel-\'Ecalle resummation may be applied to higher orders in the semi-classical expansion.

\section{Gauge theory effective actions on $\R^3 \times S^1$\label{sec1}}

\subsection{4-d theory}

Consider an asymptotically free (AF) euclidean gauge theory with gauge group $G$ with Lie algebra $\gf$.  The 4-d microscopic action is
\begin{align}\label{LUV}
\cL_\text{UV} = \frac{1}{2g^2} (F_{\m\n},F_{\m\n}) 
+ \frac{2i}{g^2} (\bar\Psi_f, \bar\s_\m D_\m \Psi_f) 
+ \frac{i\th}{16\pi^2}(F_{\m\n},\til F_{\m\n}) ,
\end{align}
where $f=1,\ldots,n_f$ is an index that runs over Weyl fermions in irreps $R_f$, $\til F_{\m\n} := \tfrac12 \e_{\m\n\r \s} F_{\r \s}$, and $(\cdot,\cdot)$ stands for the Killing form (invariant inner product) on $\gf$.  For simplicity we take $\gf$ to be simple and do not include fermion masses or scalar fields.  Since the fermions are massless, we can use a chiral rotation to set the theta angle to zero, $\th=0$.  For most calculations in later sections we will focus on the QCD(adj) theory with $n_f$ fermions all in the adjoint representation, but will keep the fermion representation content general for now. 

With the theta angle set to zero, there is no need to fix the normalization of the Killing form since it can always be absorbed in the definition of the coupling $g$.  In the next few sections, where we focus on the perturbative properties of the theory, we will refrain from fixing the normalization of the Killing form, and, in particular, will not identify weight spaces with co-weight spaces.  This helps to make the interesting GNO duality structure of the effective theories on $\R^3\times S^1$ clearer.  In later sections where we focus on the semi-classical nonperturbative properties of the theory, however, it is convenient to choose the normalization of the Killing form in which the smallest instanton number is 1, or, equivalently, in which $\th$ is periodic with period $2\pi$ in (\ref{LUV}).  This normalization is discussed in appendix \ref{secA3}.

The coupling $g(\m)$ is a function of energy scale $\m$ given at one loop in perturbation theory by $(\L/\m)^{\b_0} = \exp\{ - 8\pi^2 g^{-2}(\m)\}$, where $\L$ is the strong coupling scale and $\b_0$ is the coefficient of the 1-loop beta function, given by  
\begin{align}\label{b0}
\b_0 = \frac{1}{6} \left[ 11\, T(\ad) - 2 \tsum_f T(R_f) - \tsum_b T(R_b) \right] .
\end{align}
Here $T(R)$ is the Dynkin index of the representation $R$ (see appendix \ref{secA3} for definition and normalization), ``ad" stands for the adjoint irrep, and the sums run over the irreps $R_f$ of Weyl fermions and $R_b$ of complex scalars.  For QCD(adj), where there are only $n_f$ fermions in the adjoint irrep, the beta function becomes in the Killing form normalization mentioned above
\begin{align}\label{b0adj}
\b_0 = h^\v \, \frac{11-2 n_f}{3} ,
\end{align}
where $h^\v$ is the dual Coxeter number of the gauge algebra, defined in appendix \ref{secA3}.  In particular, AF requires $n_f \le 5$.

We are interested in putting the theory on $\R^3 \times S^1$ with the $S^1$ of size $L$ in the $x^4$ direction so that $x^4\simeq x^4+L$, and we impose periodic boundary conditions on the fermions.  Furthermore, we assume that $L^{-1}\gg\L$ so that our AF theory is weakly coupled at the scale of the compactification, $g(L^{-1})\ll 1$.  Most of the rest of this paper will analyze the dynamics of the effective 3-d theory with a cut-off scale $\m$ such that $\L \ll g/L \ll \m \ll 1/L$, where, from now on, 
\begin{align}\label{}
g := g(L^{-1})
\end{align}
denotes the 4-d coupling at the compactification scale.

\subsection{Classical 3-d effective action \label{sec2.4}}

Integrate out the Kaluza-Klein (KK) modes on the circle to get an effective 3-d action at energy scale $\m$.  Since the KK modes are all weakly coupled and massive (with masses of order $2\pi n/L$ for positive integers $n$), they are integrated out simply by setting them to zero.\footnote{This is not quite true; see the discussion around (\ref{Wmass2}) below.}  Only the zero modes of the fields (i.e., those constant on $S^1$) are light, the classical 3-d effective action is the 4-d action with all fields, $\phi$, replaced by their 0-modes, $\phi(x^m) := L^{-1} \int_0^L dx^4 \phi(x^m,x^4)$, giving
\begin{align}\label{3dcea1}
\cL_\text{3d-class.} = \frac{L}{g^2} \left[ 
\tfrac12 F_{mn}^2 
+ |D_m A_4|^2 
+ 2i \bar\Psi_f \slashed{D} \Psi_f 
- 2\bar\Psi_f \bar\s_4 A_4 \Psi_f \right] .
\end{align}

Infinitesimal gauge transformations of the $A_4$ 0-mode are $\d A_4 = L^{-1} \int_0^L dx^4 D_4 h = [A_4,h_0]$ for $h(x)\in\gf$ and periodic around the $S^1$, where $h_0:=L^{-1}\int_0^L dx^4 h$.  These can be used to rotate $A_4$ to a given Cartan subalgebra (CSA) $\tf\subset\gf$, but do not shift $A_4$ within the CSA.   So define the 3-d fields 
\begin{align}\label{3d0modes}
A_4(x) &:= \frac{2\pi}L \f(x), 
\qquad\qquad\quad\  \f\in\tf, 
\nonumber\\
A_m(x) &:= a_m(x) + W_m(x),
\qquad a_m\in\tf, \quad W_m\in\tf^\perp .
\end{align}
$\f$ is a $\gf$-valued scalar field with gauge invariance $\d \f = i [h,\f]$, i.e., $\f$ transforms in the adjoint representation of the gauge group, while the ``$W$-boson" fields can be decomposed as $W_m=\sum_\a e_\a W^\a_m$ where $\{e_\a\}$ is a basis of generators of $\gf$ not in $\tf$ which are in 1-to-1 correspondence with the roots, $\a\in\Phi$, of $\gf$.  Then the 3-d classical action is, keeping only quadratic terms,
\begin{align}\label{3dcea}
\cL_\text{3d-class.} 
&=
\tfrac{L}{2g^2} \left(f_{mn} + d_{[m}W_{n]}
\right)^2 
+ \tfrac{4\pi^2}{g^2L} \left( \del_m \f 
+ \tsum_{\a\in\Phi} \a(\f) W^\a_m e_\a \right)^2 
\nonumber\\
&\qquad\text{}
+ \tfrac{2L}{g^2} \tsum_f \tsum_{\l\in R_f}\bar\Psi_\l
[i\slashed{d}
- \tfrac{2\pi}{L} \bar\s_4 \l(\f)] \Psi_\l  + \cdots .
\end{align}
Here we have defined a CSA-valued gauge field strength, $f_{mn}:=\del_{[m}a_{n]}$, and covariant derivative $d_m := \del_m + i a_m$.   The roots $\a$ and weights $\l$ can be thought of as vectors of charges of the $W_m$ and $\Psi_f$ fields with respect to the CSA gauge fields.

We will use a natural notation where, instead of denoting the weights as vectors, we treat them as elements of the dual CSA, $\tf^*$.  That is, they act as real linear maps on $\tf$: $\l:(\f\in\tf)\mapsto(\l(\f)\in\R)$.  For example, we will write $d_m W_n^\a = [\del_m+i\a(a_m)]W_n^\a$ and $d_m \Psi_\l = [\del_m+i\l(a_m)]\Psi_\l$.  When necessary, we can work with (dual) vector components by going to a basis.  So if $\{e_i\}$ is a basis of $\tf^*$ and $\{e^i\}$ is the dual basis of $\tf$ (so that $e_i(e^j)=\d^j_i$), then for arbitrary elements $\l=\l^i e_i\in\tf^*$ and $\f=\f_je^j\in\tf$ (summations understood, $\l^i,\f_j\in\R$), then $\l(\f)=\l^i \f_i$.   Also, the squares in the first line of (\ref{3dcea}) include not only Lorentz index contractions but also the Killing inner product on the Lie algebra.  (Appendix \ref{secA} reviews needed Lie algebra definitions and concepts.)

Since there is no potential for $\f$, the space of classical vacua are parameterized by $\vev\f\in\tf$.  This moduli space is actually compact, since points on $\tf$ are further identified by a remaining discrete group of gauge transformations, $\hW=W\ltimes\G^\v_r$, so that
\begin{align}\label{phispace}
\f \in \tf/(W\ltimes \G^\v_r)\ :\simeq\  \hT.
\end{align}
Here $W$ is the discrete Weyl group of $\gf$ and $\G^\v_r$ is the co-root lattice (or magnetic root lattice; the definitions of these lattices are reviewed in appendix \ref{secA1}.).  These lattice identifications on $\f$, $\f \simeq \f + \m$ with $\m\in\G^\v_r$, arise from 4-d gauge transformations connected to the identity $A_4 \to g^{-1} A_4 g - i g^{-1} \del_4 g$ with $g(x^4) = \exp\{2\pi i h(x^4)\}$ where $\tf\ni h(x^4+L) = h(x^4) + \m$.  (These lattice identifications are independent of the choice of global from of the gauge group, but do depend on choosing the the group of gauge transformations to include only those continuously connected to the identity; see appendix \ref{secA2}.)  Note that $\tf/\G^\v_r$ is the same as the maximal torus of $G$, $T_G \simeq \tf/\G^*_G$ only for $G=\til G$ the simply connected form of the group; otherwise it is a cover of $T_G$.  The additional Weyl group identifications in (\ref{phispace}) are described in appendix \ref{secA4}.  

We call a fundamental domain in $\tf$ of $\hW$ a ``gauge cell", and denote a canonical choice of gauge cell by $\hT$.  As we discuss in appendix \ref{secA4}, the gauge cell is also known as an affine Weyl chamber, and has a simple description as the region of $\tf$
\begin{align}\label{cellwalls}
\hT := \{\ \f \ | \ \a_i(\f)\ge0, \  i=1,\ldots,r,
\quad\text{and}\quad 
-\a_0(\f)\le 1 \ \},
\end{align}
where the $\a_i$ are a basis of simple roots, and $\a_0$ is the lowest root with respect to this basis.  Here 
\begin{align}\label{}
r := \text{rank}(\gf) .
\end{align} 
$\hT$ is a convex $r$-dimensional region bounded by the $r+1$ hyperplanes $\a_i(\f)=0$ and $\a_0(\f)=-1$, an $r$-dimensional generalization of a tetrahedron.  In particular, there are $r+1$ vertices, each of which is opposite to one of the hyperplanes and is where the remaining $r$ hyperplanes intersect.  Some examples of gauge cells are given in figure \ref{fig-cells} in section \ref{sec2.1}. The gauge cells of all simple Lie algebras are explicitly described in appendix \ref{secB}.

So we take $\f\in \hT$ to parameterize the inequivalent vacua.  $\f$ can also be considered as a gauge-invariant order parameter in the following sense.  The gauge holonomy in the 4-d theory around the $S^1$ (the open Wilson line) is $\O(x) := \exp\{i\int_x^{x+L}A_4\}\in G$.  Under a periodic gauge transformation $g(x)\in G$, $\O(x) \to g^{-1}(x) \O(x) g(x)$, so the conjugacy class of $\O(x)$ is a gauge-invariant order parameter distinguishing the different vacua.  But conjugation in $G$ can take any element to a given maximal torus of $G$, so we can write a representative in the conjugacy class of any holonomy as $[\O(x)] = \exp\{2\pi i \f\}$ with $\f\in \hT$.  Thus we will treat $\f$ as our gauge-invariant order parameter, even though it actually depends on the gauge-dependent choice of CSA $\tf\subset\gf$ and of a fundamental domain $\hT\subset \tf$ of the remaining discrete gauge identifications.

At interior points of $\hT$ there are no roots for which $\a(\f)=0$ so the gauge group is Higgsed to abelian factors,
\begin{align}\label{}
\f: G \to U(1)^r
\qquad\text{for}\quad \f\in\text{interior}(\hT).
\end{align}
From (\ref{3dcea}) it follows that the W-bosons and fermions have masses 
\begin{align}\label{Wmass}
m_{W^\a} = \frac{2\pi}{L}|\a(\f)|,
\qquad
m_{\Psi_\l} = \frac{2\pi}{L} |\l(\f)| .
\end{align}

We restrict ourselves to QCD(adj) --- the theories with only adjoint fermions --- for the rest of the paper.  In this case the fermions are in the adjoint representation, there will be $r$ massless components of $\Psi$ in the CSA---which we will denote by $\psi$---and the remaining $\Psi_\a$ components will have the same masses as the $W^\a_m$.

At boundary points of $\hT$ saturating one or more of the inequalities (\ref{cellwalls}), the unbroken gauge symmetry is enhanced to contain nonabelian factors, and some of the $W^\a$-bosons and $\Psi_\a$ fermions become massless.  One slightly subtle point is that even at the lowest root boundary, where $\a_0(\f)=-1$, $W^{\a_0}$-bosons and $\Psi_{\a_0}$ fermions will also become massless.  It is actually the first Kaluza-Klein mode of these fields which becomes massless there.  The proper formula for the mass \emph{gap} in $\hT$, replacing (\ref{Wmass}), is
\begin{align}\label{Wmass2}
m_{W^\a} = m_{\Psi_\a} = 
\frac{2\pi}{L} \cdot \text{min}
\Bigl\{\, |\a(\f)|\ ,\, 1{-}|\a(\f)| \,\Bigr\} .
\end{align}

Away from the boundaries of $\hT$, the 3-d classical effective action for the massless modes of QCD(adj) is then simply
\begin{align}\label{3dcea-int}
\cL_\text{3d-class.}^\text{int.} =
\tfrac{L}{2g^2} \left(f_{mn}\,,f_{mn}\right) 
+ \tfrac{4\pi^2}{g^2L} \left( \del_m\f \,,\del_m\f \right) 
+ i \tfrac{2L}{g^2} \left( \bar\psi_f , \slashed{\del} \psi_f \right),
\end{align}
where $(\cdot,\cdot)$ is the Killing form restricted to the CSA.   This is a 3-d $U(1)^r$ gauge theory with $r$ real, massless, neutral scalars and Weyl fermions.  Note, however, that at the boundaries of $\hT$ the associated massless charged $W^\a$'s and $\Psi_\a$'s must be included as well in a consistent effective action, giving rise to a nonabelian gauge theory. 

\subsubsection*{Charge lattices}

We now describe the spectrum of charged operators and probes in QCD(adj) on $\R^3 \times S^1$.

The 4-d UV theory has fields charged in representations of the gauge group $G$ and, when $G$ is Higgsed to $U(1)$ factors---as when $\f$ is in the interior of $\hT$---the theory also admits magnetic monopoles.   These fields create states whose possible electric and magnetic $U(1)$ charges lie in lattices (i.e., are quantized).  An external (massive) electrically or magnetically charged source corresponds to the insertion of a Wilson or 't Hooft line operator, respectively, in the path integral.   Upon compactification on a spatial circle, these line operators will give rise to point and line operators in the effective 3-d $U(1)^r$ theory that also carry quantized $U(1)$ electric and magnetic charges.

We define electric ($\l\in\tf^*$) and magnetic ($\m\in\tf$) charges in a 4-d $U(1)^r$ theory by
\begin{align}\label{}
\l := \int_{S^2_\infty} *F \ ,
\qquad\qquad
\m := \frac{1}{2\pi} \int_{S^2_\infty} F \ ,
\end{align}
where $F := \frac12 F_{\m\n}dx^\m\wedge dx^\n\in\tf$ is the $U(1)^r$ field strength, and the dual field strength, $*F := \frac12 \til F^*_{\m\n} dx^\m \wedge dx^\n\in\tf^*$, is both Hodge-dualized,
\begin{align}\label{}
\til F_{\m\n} := \frac12 \e_{\m\n\r\s} F_{\r\s},
\end{align}
and dualized with respect to the Killing form 
\begin{align}\label{normKF}
\frac{1}{g^2}
(\,\cdot\,,\,\cdot\,)
\end{align}
which appears in the microscopic Lagrangian (\ref{LUV}).  Thus,
\begin{align}\label{}
F^*(\cdot) := \frac{1}{g^2} (F,\cdot) .
\end{align}
Thus a particle with worldline $C$ and electric and magnetic charges $\l$, $\m$, has
\begin{align}\label{}
F = \frac{\l^*}{4\pi} \frac{1}{r^2}dr\wedge dz 
+ \frac{\m}{2} \sin\th d\th\wedge d\phi,
\qquad
*F = \frac{\l}{4\pi} \sin\th d\th\wedge d\phi 
+ \frac{\m^*}{2} \frac{1}{r^2}dr\wedge dz,
\end{align}
where $z$ is a coordinate along $C$, $r$ the coordinate perpendicular to $C$, and $\th$ and $\phi$ are the polar and azimuthal angles on the $S^2$ linking $C$.

With the gauge field normalization of (\ref{LUV}), electric charges defined in this way are the same as the weights, $\l$, of representations that enter into the covariant derivative as $D_\m = \del_\m + i \l(A_\m)$.  Note that a more conventional definition of electric charge would be $\l^*$, not $\l$.  Also, both the electric and magnetic charges are commonly divided by $g$ to be charges for canonically normalized gauge fields (i.e., without the $g^{-2}$ factor multiplying the action).

\subsubsection*{Electric operators and center symmetry}

By the definition of the gauge group $G$, all fields and probes transform in representations of $G$, and so have electric charges, $\l$, under a $U(1)^r \subset G$ maximal torus which span the gauge lattice $\G_G\subset \tf^*$,
\begin{align}\label{}
\l \in \G_G \quad\text{for all electric charges.}
\end{align}

For QCD(adj) where all dynamical fields are in the adjoint representation, the electric charges of the fields are thus in the root lattice, $\G_r=\G_\Gad$ (see appendix \ref{secA1} for the definitions of and relations among the various possible charge lattices),
\begin{align}\label{}
\l \in \G_r \quad\text{for electric charges of dynamical fields in QCD(adj).}
\end{align}
When the gauge group $G$ is taken to be larger than the adjoint group, $\Gad$, then the group lattice is larger (finer) than the root lattice, $\G_G \supset \G_r$.  In this case electric probe operators, like $E[\l,P]$ and $W[\l,C]$ defined below, are allowed in representations with weights other than those of the adjoint representation (or, more generally, weights not in the root lattice). 

We saw in (\ref{phispace}) that in a gauge theory with gauge group $G$ on $\R^3 \times S^1$, the 0-mode of the $A_4$ gauge field, $\f\in\tf$, is defined only up to gauge transformations which act as translations in the co-root lattice, $\f \simeq \f + \m$, $\m\in\G_r^\v$.  A Wilson loop wrapping the $S^1$ at a point $P\in\R^3$ (a.k.a.\ the gauge holonomy or Polyakov loop) descends in the 3-d effective theory to the electric point operator 
\begin{align}\label{elecop}
E[\l,P] := \exp 2\pi i\l(\f)(P)
\end{align}
for some $\l\in\G_G$.\footnote{We have ignored above, for simplicity, the discrete Weyl group of gauge equivalences.  In fact, the Wilson loop in the 4-d theory will be in some irrep $R$ of $G$, $\tr_R{\cP}\!\exp i\int_{S^1} A$, which gives in the 3-d effective theory $\sum_{\l\in R}\exp 2\pi i \l(\f)$.  The weights $\l\in R$ fill out Weyl orbits, and the sum then enforces the invariance of the electric operator under the Weyl group identifications on $\f$.}
Likewise, an external (massive) electrically charged source with worldline $C\subset\R^3$ (at a point on the $S^1$ in the 4-d theory) is accompanied by the insertion of the Wilson line operator,
\begin{align}\label{wilsonline}
W[\l,C]=\exp{i \int_C \l(a)},
\end{align}
in the 3-d effective $U(1)^r$ theory, where again $\l\in\G_G$, and $a:= a_m dx_m\in\tf$ is the one-form $U(1)^r$ gauge potential.

As described in appendix \ref{secA2}, the center symmetry acts by large gauge maps $g_c=g_\m$ given by (\ref{gmu}), which are in the disconnected component $c$ of the group of gauge transformations according to $c\simeq [\m]\in\G_w^\v/\G_r^\v$.  Repeating the argument after (\ref{phispace}) with $g(x)=g_c(x)$  shows that the action of the center symmetry on $\f$ is to shift 
\begin{align}\label{largegaugeshifts}
\f \to \f^{g_c} = \f + \m \quad\text{with}\quad 
c \simeq [\m]\in\G_w^\v/\G_r^\v,
\end{align}
which in turn multiplies the electric point operators by a phase,
\begin{align}\label{centerphase}
g_c : E[\l,P] \to e^{2\pi i \l(\m)} E[\l,P],
\quad \l\in\G_G ,\ c\simeq[\m]\in\G_w^\v/\G_r^\v .
\end{align}
The electric operators $E[\l,P]$ can thus be taken as order parameters for the center symmetry.  For example, for $G=SU(N)$ and $\l$ a weight of the fundamental representation, say $\l=e_i - \frac1N\sum_j e_j$ in the basis of appendix \ref{secB1}, then for $\m$ a weight of the fundamental representation of $G^\v$, say $\m=e^k - \frac1N\sum_j e^j$, the phase in (\ref{centerphase}) is $\exp -2\pi i/N$.  The center symmetry acts trivially on the Wilson loop operators $W[\l,C]$ simply because they come from 4-d operators which do not wrap the $S^1$.

\subsubsection*{Magnetic operators and charges}

A classical magnetic charge in the 4-d theory with worldline $C$ is represented by the insertion of a line operator along $C$.  This operator is described by boundary conditions for the gauge field along $C$ corresponding to inserting a GNO monopole \cite{Goddard:1976qe, Kapustin:2005py} (a Dirac monopole embedded in the gauge group $G$).  Explicitly, if $\th$ and $\phi$ are the usual polar coordinates on a small $S^2$ linking $C$, then the boundary condition is that, up to a gauge transformation, the 4-d gauge potential has the singularity
\begin{align}\label{GNObcs}
\lim_{r\to0}A_\pm = 
-\frac{\m}{2} \left( \cos\th \mp 1\right) d\phi, \qquad \m \in \tf.
\end{align}
The $\pm$ indices denote the $1\ge \pm\cos\th \ge0$ coordinate patches (the northern and southern hemispheres of the $S^2$) respectively.  Along the equatorial $S^1$ overlap of the two patches at $\th=\frac{\pi}{2}$, $A_+-A_- = d(\m \,\phi)$ which is a continuous gauge transformation only if $e^{2\pi i \m}=1$ in $G$, which is true when the magnetic charge is in the dual of the group lattice,
\begin{align}\label{m-charges}
\m\in\G^*_G .
\end{align}
This is the Dirac quantization condition \cite{Dirac:1931,  Goddard:1976qe}.\footnote{The Weyl group of additional discrete gauge identifications on $\tf$ implies that allowed $\m$ are actually classified by their Weyl orbits  which can be put into one-to-one correspondence with highest weights of irreducible representations of the GNO dual group $G^\v$ \cite{Goddard:1976qe}.}

For QCD(adj) if we take $G=\Gad$, so that the group lattice is the root lattice, $\G_G=\G_r$, then allowed magnetic charges are in $\G^*_G=\G^*_r=\G^\v_w$, the co-weight lattice.  On the other hand, if we choose $G=\tG$, so that massive sources are allowed to be charged in the larger weight lattice, $\G_w$, then the allowed magnetic charges can only be in $\G^*_w=\G^\v_r$, the co-root lattice.  But arbitrarily massive probes decouple from the low energy dynamics, so their presence or absence cannot affect the spectrum of light magnetic states in the theory.  Therefore the magnetic fields can only be charged in the co-root lattice, $\G^\v_r$, which is smaller (coarser) than the co-weight lattice, so in fact
\begin{align}\label{}
\m \in \G^\v_r \quad\text{for dynamical fields}.
\end{align}
Thus not all magnetic charges allowed by the Dirac quantization condition are necessarily realized in the spectrum of light states: a dynamical field carrying a magnetic charge in the finer $\G^\v_w$ lattice would imply a violation of decoupling of massive charged states.

In the theory on $\R^3\times S^1$, the 't Hooft line operator will descend to a point or line operator in the 3-d effective theory depending on whether it wraps the $S^1$ or not.  If $C$ wraps the $S^1$ at a point $P\in\R^3$, this becomes a monopole point operator at $P$ in the 3-d $U(1)^r$ theory, 
\begin{align}\label{monopoint}
M[\m,P]\quad \text{creates a gauge field singularity at $P$ such that}\quad 
\int_S f = 2\pi \m
\end{align} 
for any closed surface $S$ which encloses $P$ once, and where $f := \tfrac12 f_{mn}dx_m\wedge dx_n = da$ is the $U(1)^r$ field strength.  Note that if both a Wilson line operator $W[\l,C]$ and a monopole operator $M[\m,P]$ are present, since $\int_C \l(a) = \int_S \l(f)$ for any surface $S$ with $\del S = C$, and since the Wilson line insertion (\ref{wilsonline}) should be independent of the choice of $S$, $\exp{2\pi i \l(\m)} = 1$, and the Dirac condition (\ref{m-charges}) follows.

A 4-d 't Hooft loop operator of charge $\m$ along a curve $C \subset \R^3$ and at a point on the $S^1$ will descend to a 't Hooft operator $T[\m,C]$ in the 3-d $U(1)^r$ theory.  The 4-d operator is characterized by having $\int_S f=2\pi \m$ for any surface $S$ linking $C$ once in $\R^3\times S^1$.  Since $C$ is at a point on the $S^1$, we can take $S$ to be a 2-torus with one cycle wrapping the $S^1$ and the other a curve $C'$ linking $C$ in $\R^3$.  Then $\int_S f = 2\pi \int_{C'} d\f$, so 
\begin{align}\label{thooftop}
T[\m,C] \quad\text{creates a monodromy $\f\to\f+\m$ around $C$.}
\end{align}

The center symmetry acts trivially on the magnetic operators since a large gauge map $g_c=g_{\hat\m}$ given by (\ref{gmu}) does not change the singular part of the boundary conditions (\ref{GNObcs}).  Inserting this magnetic probe operator in the path integral means that we should integrate over all gauge fields with the boundary condition (\ref{GNObcs}), so shifting the non-singular part of the gauge field is just a shift in the integration variable.  

\subsection{3-d dual photon and dual center symmetry}

The $U(1)^r$ CSA photon fields $a_m(x)\in\tf$ can be dualized in 3-d as $r$ derivatively coupled scalars $\s(x)\in\tf^*$ \cite{Polyakov:1976fu, Affleck:1982as, Deligne:1999qp}.  This follows from considering a theory with, in addition to the 3-d $U(1)^r$ gauge field $a_m\in\tf$ with field strength $f_{mn}$, a vector field $b_m\in\tf^*$ and a scalar $\s\in\tf^*/\G_r$ and partition function
\begin{align}\label{predual}
Z &= \int [da_m][db_m][d\s]\ e^{-\int d^3x\, \cL}
\quad\text{with}\quad
\cL := \tfrac{g^2}{4L} (\del_m\s+b_m)^2 
+ \tfrac{i}{2} \e_{mnp}b_m(f_{np})  ,
\end{align}
where in the first term, both a space-time contraction and one on $\tf^*$ using the inverse Killing form is understood.  In addition to the usual $U(1)^r$ gauge invariance for $a_m$ this theory has an additional gauge invariance
\begin{align}\label{dualginv}
\s &\to \s + \s' ,\qquad
b_m \to b_m - \del_m\s'.
\end{align}

Fixing this latter invariance by setting $\s=0$ and then integrating out $b_m$ gives
\begin{align}\label{a-dual}
Z &= \int [da_m]\ \exp \left\{-\frac{L}{2g^2} \int\!\! d^3x\, (f_{mn} , f_{mn} ) \right\},
\end{align}
which is the original $U(1)^r$ gauge theory (\ref{3dcea-int}) that we want to dualize.  Note that the chosen periodicity of $\s$, i.e.\ $\s\in\tf^*/\G_r$, implies that holonomies of $b_m$ are also in $\tf^*/\G_r$.  Then, upon integrating out $b_m$, the periods of $f_{mn}$ can only take values in $2\pi\G^\v_w$, and so allows the largest (finest) lattice of magnetic charges $\m\in\G^\v_w$.  By (\ref{m-charges}) physical (field or probe) magnetic charges only appear in the $\G^*_G$ lattice which may be smaller than $\G^\v_w$.\footnote{Since the $a_m\in\tf$ gauge fields are identified by the discrete Weyl group of gauge equivalences, $\s$ will be too, under the dual action of the Weyl group on $\tf^*$, so, in fact, $\s\in\tf^*/(W\ltimes\G_r)$.}

The choice of $\G_r$ as the periodicity of $\s$ implies that there is a global discrete symmetry
\begin{align}\label{dualcs1}
\G_w/\G_r \simeq Z(\til{G^\v}) \simeq Z(\tG)
\end{align}
which acts on the low energy dual photon by
\begin{align}\label{dualcs2}
\s \to \s^c := \s+\l
\quad\text{with}\quad 
c \simeq [\l]\in\G_w/\G_r,
\end{align}
similar to the action of center symmetry (\ref{largegaugeshifts}) on $\f$.  (The outstanding difference from center symmetry is that there is no microscopic description in terms of a non-abelian $G^\v$ magnetic gauge theory, and so no microscopic derivation of this symmetry as coming from large magnetic gauge transformations.  It has nevertheless been argued \cite{'tHooft:1977hy, KorthalsAltes:2000gs} to be an exact symmetry of 3-d and 4-d gauge theories with adjoint matter, and not just a low-energy accidental symmetry in 3-d abelianizing vacua.)   We will call this symmetry the \emph{dual center symmetry} in what follows.\footnote{It does not seem to have a standard name.  For $\tG=SU(N)$ it is called ``topological global $\Z_N$ symmetry"  in \cite{'tHooft:1977hy} and ``magnetic $\Z_N$ symmetry" in \cite{KorthalsAltes:2000gs}.} 

Integrating out $a_m$ instead sets $db=0$, and then the gauge invariance (\ref{dualginv}) can be used to set $b_m=0$, giving the dual formulation of the theory,
\begin{align}\label{sigma-dual}
Z &= \int [d\s]\ \exp \left\{-\frac{g^2}{4L} \int\!\! d^3x\, 
(\del_m \s , \del_m \s ) \right\}.
\end{align}
Including the fermion and $\f$ fields of (\ref{3dcea-int}) then gives the dual effective 3-d Lagrangian in the interior of the gauge cell, $\hT$, for the theory with $n_f$ adjoint fermions
\begin{align}\label{3dmea}
\cL_\text{3d-mag.}^\text{int.} =
\tfrac{g^2}{4L} (\del_m\s,\del_m\s) 
+ \tfrac{4\pi^2}{g^2L} \left( \del_m\f \,,\del_m\f \right) 
+ i \tfrac{2L}{g^2} \left( \bar\psi_f , \slashed{\del} \psi_f \right).
\end{align}
Note that with this normalization, $\s$ and $\f$ are dimensionless, while $\psi_f$ has dimension $3/2$.

Under this duality, operators map as follows.  The operator $\del_m \s$ is dual to $-\frac{iL}{g^2} \e_{mnp}f^*_{np}$, where $f^*\in\tf^*$ is the dual of $f$ with respect to the Killing form.  This follows from inserting $\del_m \s + b_m$ in the path integral (\ref{predual}) and integrating out as in (\ref{a-dual}) and (\ref{sigma-dual}).

The point monopole operator (\ref{monopoint}) becomes the local operator 
\begin{align}\label{disorderop}
M[\m,P] := \exp 2\pi i \s(\m)(P) 
\end{align}
in the dual variables.  This follows from inserting into (\ref{predual}) the gauge-invariant operator $e^{2\pi i\s(\m)}(P) \cdot \exp\{2\pi i \int_C b(\m)\}$ with the Dirac string $C$ ending at $P$, and doing the duality integrations.  Integrating out $a_m$ sets $b=0$, giving (\ref{disorderop}), while gauge fixing $\s=0$ and integrating out $b_m$ gives (\ref{a-dual}) as before but with the restriction that $f$ satisfies (\ref{monopoint}).  The dual center symmetry (\ref{dualcs2}) acts on the point monopole operators by multiplication by phases
\begin{align}\label{dualcenterphase}
Z(\til{G^\v})\ni c : M[\m,P] \to e^{2\pi i \l(\m)} M[\m,P],
\quad c\simeq[\l]\in\G_w/\G_r ,
\end{align}
analogous to the action of the center symmetry on electric point operators (\ref{centerphase}).

A Wilson line operator (\ref{wilsonline}) is dualized to the operator 
\begin{align}\label{dualWL}
W[\l,C]\quad\text{creates a monodromy $\s\to\s+\l$ around $C$.}
\end{align}
This follows since integrating $a_m$ out of (\ref{predual}) with an insertion of (\ref{wilsonline}) sets $db$ to have delta-function support on $C$ such that $\int_{C'} b = \l$ for any curve $C'$ linking $C$ once.  Equivalently, using the gauge invariance (\ref{dualginv}) we can set $b=0$ at the expense of requiring $\s$ to have the monodromy (\ref{dualWL}).

The electric point operator (\ref{elecop}) and the 't Hooft loop operator (\ref{thooftop}) are unchanged, since they do not involve the $a_m$ fields.

\subsubsection*{Summary}

We can summarize all this for QCD(adj) with gauge group $G$, gauge transformations continuously connected to the identity, and vacuum in the interior of the gauge cell as follows.  The charges and basic operators in the dual 3-d effective theory are:
\begin{list}{$\bullet$}{\itemsep=0pt \parsep=0pt \topsep=2pt}
\item Electric charges $\l \in \G_G$ are allowed, but only $\l\in\G_r$ occur for dynamical fields.
\item Magnetic charges $\m \in \G_G^*$ are allowed, but only $\m \in \G_r^\v$ occur for dynamical fields.
\item The holonomy field $\f \in \tf / \G_r^\v$, \footnote{which descends from the  4-d $A_4$ KK 0-mode.} in addition to local operators made from its derivatives, $\del_m \f$, etc., can be used to construct
\begin{list}{$\circ$}{\itemsep=0pt \parsep=0pt \topsep=0pt}
\item electric operators $E[\l,P]$ which insert $\exp 2\pi i \l(\f)$ at $P$, \footnote{which descends from a 4-d Wilson line wrapping the $S^1$.} and
\item 't Hooft lines $T[\m,C]$ which create $\f \to \f+\m$ monodromy around $C$. \footnote{which descends from a 4-d 't Hooft loop at a point on the $S^1$.}
\end{list}
\item The dual photon field $\s \in \tf^*/ \G_r$, \footnote{which descends from and is dual to the 4-d $A_i$ KK 0-modes.} in addition to local operators made from its derivatives, $\del_m \s$, etc., can be used to construct
\begin{list}{$\circ$}{\itemsep=0pt \parsep=0pt \topsep=0pt}
\item monopole operators $M[\m, P]$ which insert $\exp 2\pi i\m(\s)$ at $P$, \footnote{which descends from a 4-d 't Hooft loop wrapping the $S^1$.} and
\item Wilson lines $W[\l, C]$ which create $\s \to \s + \l$ monodromy around $C$. \footnote{which descends from a 4-d Wilson line at a point on the $S^1$.}
\end{list}
\end{list}
The electric and monopole point operators are order parameters for the center and dual center symmetries, respectively:
\begin{align}\label{}
Z(\tG) \ni c &: E[\l,P] \to e^{2\pi i \l(\m)} E[\l,P]
&&\text{with}\quad c \simeq [\m] \in \G^\v_w/\G^\v_r
\nonumber\\
Z(\til{G^\v}) \ni c^\v &: M[\m,P] \to e^{2\pi i \l(\m)} M[\m,P]
&&\text{with}\quad c^\v \simeq [\l] \in \G_w/\G_r.
\end{align}

This presentation of the low energy dynamics in the interior of the gauge cell in terms of $\f$ and the dual photon $\s$ makes the GNO-duality between the electric and magnetic degrees of freedom manifest.   This does not mean that the dynamics treats these two sets of variables symmetrically.  Indeed, the GNO-duality of the low energy descriptions is a property of any theory with an adjoint Higgs phase, but only in special theories, like $N=4$ SYM where the dynamics is realized in a conformal phase, is GNO-duality realized symmetrically.  

For QCD(adj), as we will see in detail in later sections, the dynamics is not realized in a GNO-symmetric way.  In particular, neither perturbative nor semi-classical non-perturbative effects spontaneously break center symmetry in QCD(adj); while non-perturbatively the dual center symmetry is spontaneously broken in the effective theory, leading to stable domain wall solitons interpolating between the different vacua related by the broken symmetry.  These correspond to the electric flux tubes expected in a confining phase.

The rest of this paper is devoted to computing the effective potential for the $\f$ and $\s$ fields by computing semi-classical contributions from the electric and monopole point operators.

\subsection{Structure of perturbative corrections\label{sec2.6}}

The effective action of QCD(adj) in the interior of the gauge cell is given in (\ref{3dmea}).  This low energy theory has a large IR global symmetry group.  It includes a $U(1)_\s^r$ symmetry under shifts of $\s$,
\begin{align}
U(1)_\s^r : \quad \s \to \s + \e, \quad \e\in\tf^*,
\end{align}
a similar $U(1)_\f^r$ symmetry under shifts of $\f$, and a $U(r\, n_f)$ flavor symmetry of the fermions.

These symmetries are mostly accidental IR symmetries of the classical (tree-level) effective action, and as such will generically be broken by quantum corrections.  For instance, perturbative effects break the flavor symmetry of the adjoint fermion theory to the $U(n_f)=U(1)_A\times SU(n_f)$ chiral symmetry which is present in the microscopic 4-d theory.  The $U(1)_A$ factor is anomalous in the 4-d theory, broken to $\Z_{2h^\v n_f}$ by instantons, where $h^\v$ is the dual Coxeter number of $\gf$.  Thus the $U(1)_A\to \Z_{2h^\v n_f}$ breaking will not occur at any order in perturbation theory, but will be seen in the 3-d effective theory only once non-perturbative effects involving monopole-instantons are included.

Similarly, the $\s$ shift symmetry of the dual photons is broken by coupling to magnetic monopoles via the disorder operators (\ref{disorderop}).  But since there are no magnetically charged states in the microscopic theory, such terms will not arise at any order in perturbation theory, and $\s$ will remain derivatively coupled.  We can thus classify states by an associated conserved magnetic charge (pseudo) quantum number.  But, once non-perturbative effects are included, magnetic-charge non-conserving operators will enter the effective action, and magnetic charge will not be a good quantum number.

On the other hand, the $U(1)_\f^r$ shift symmetry of the $\f$ bosons is broken due to coupling of electrically charged matter, so, in particular, perturbative effects in $g$ can generate an effective potential for $\f$.   

In the special case where $n_f=1$, there is a supersymmetry relating $\f$ and $\s$ as the real and imaginary parts of a complex scalar component of a supermultiplet, corresponding to the enhancement of the $U(1)^r$ low energy gauge group to the complexified gauge group acting on offshell superfields.  This prohibits any perturbative potential from arising, and so, in this case, there is also a perturbatively-conserved pseudo quantum number associated to the $\f$ shift symmetry (sometimes called ``dilaton charge" \cite{Poppitz:2011wy, Harvey:1996ur}).

For $n_f\neq 1$, the effective potential for $\f$ correcting the classical action (\ref{3dcea-int}) or its magnetic dual (\ref{3dmea}) in perturbation theory has the structure
\begin{align}\label{pertpotl}
V_\text{pert}(\f) = L^{-3} \left( v_0(\f) + g^2 v_2(\f) + g^3 v_3(\f) + \cdots \right)
\end{align}
where $v_n$ are dimensionless functions of $\f$.  This effective 3-d potential comes from integrating in loops the massive KK modes as well as the massive charged 0-modes in (\ref{3dcea1}).  To consistently compute $V_\text{pert}(\f)$ in an effective action at scales $\m \lesssim L^{-1}$, we should only integrate out modes with masses greater than $\m$.  In particular, some of the charged 0-modes become massless at the boundaries of $\hT$, as shown by the formula (\ref{Wmass2}) for the charged modes' mass gap derived above.  So, close to these boundaries these modes should not be integrated in loops.  

With no light or massless states being integrated in loops, the $v_n(\f)$ will locally be analytic functions of $\f$, even at the boundaries of $\hT$.  ``Locally" here means locally in $\hT$.  There will be no global analytic expression for the $v_n(\f)$ valid on the whole of $\hT$, since massive modes which should be integrated out in some parts of $\hT$ may be too light to be integrated out in other parts.  We will see this explicitly in the 1-loop calculation in section \ref{sec2}.  

We emphasize that this local analytic behavior is a property of the potential in an effective theory with a finite (nonvanishing) cutoff $\m$.  By contrast, a 1PI effective potential---corresponding to formally taking the cutoff $\m\to0$ in the effective theory---can have nonanalyticities at the boundaries of $\hT$.  But this is not our situation: we are working in the effective theory with cutoff $\m\sim L^{-1} \gg \L$, and cannot take $\m\to0$ without running into strong coupling.

An analytic 1-loop contribution to the effective potential, $v_0$, will have an expansion around its minimum of the form
\begin{align}
v_0 \sim (\f-\f_0)^\v\cdot v_{0,2}\cdot (\f-\f_0) +\cO(\f-\f_n)^3
\end{align}
where $\f_0$ is the position of the minimum, and $v_{0,2}$ is some positive-definite matrix of coefficients.  Then higher order terms can only shift the 0-th order minimum point, $\f_0$, by amounts vanishing as a positive power of $g$.  If some of the eigenvalues of the coefficient matrix $v_{0,2}$ happened to vanish at one loop, then higher order terms could shift the 0-th order minimum point by amounts of order 1; however, we show in section \ref{sec2} that $v_{0,2}$ is, in fact, positive-definite at the unique global minimum for all simple Lie algebras.

We will also see in the next section that for many gauge groups the minimum, $\f_0$, of the one-loop effective potential, $v_0(\f)$, is at a boundary of the gauge cell $\hT$ where the low energy gauge group is not completely abelianized.  These boundaries are fixed hyperplanes of the group of affine Weyl gauge identifications, under which the effective potential is symmetric.  So if $\x$ is a coordinate in $\tf$ measuring the perpendicular distance from one such hyperplane at $\x=0$, we must have $V(-\x)=V(\x)$.  In particular, all analytic contributions to the potential will be even in $\x$, so 
\begin{align}\label{analytic}
v_n \sim \x^2 + \cO(\x^4)
\end{align}
for all $n$.  Thus if the $v_0$ minimum is at $\x=0$, it cannot be shifted away from this point by any contributions at higher orders in perturbation theory.

Note that a similar argument also implies that if center symmetry is not spontaneously broken at 1-loop, it cannot be broken at any higher order in perturbation theory.

Since the 1-loop effective potential in (\ref{pertpotl}) has no $g$-dependence and depends on $L$ only through an overall factor of $L^{-3}$, and since the kinetic term for $\f$ in (\ref{3dcea-int}) has a factor of $(g^2L)^{-1}$, the masses of the $r$ components of $\f$ at its minimum will all be of order
\begin{align}\label{phimass}
m_\f \sim \frac{g}{L} .
\end{align}
(For large $r = \text{rank}(G)$ we will see from explicit calculation in section \ref{sec2} that the $r$ $\f$ masses are distributed in the range $g/(L\sqrt r) \sim (g\sqrt r)/L$.)

In summary, for $n_f>1$, the one-loop potential will pick a unique vacuum value of $\f$.  If that $\f$ is in the interior of the gauge cell, then higher-order perturbative corrections can only move the position of the minimum by terms of order $g^2$, and so the vacuum will remain in the interior to all orders of perturbation theory and the gauge group will be fully abelianized,
\begin{align}\label{}
G \to U(1)^r, \qquad r = \text{rank}(G).
\end{align}
If, on the other hand, the value of $\f$ at the one-loop minimum is on some gauge cell walls, where the gauge group is not fully abelianized, 
\begin{align}\label{Gbreak}
G \to U(1)^n \times H,
\qquad \text{$H$ nonabelian,}\ \ 
n = \text{rank}(G)-\text{rank}(H),
\end{align}
then higher-order perturbative corrections will not move it off those walls, and the unbroken gauge group will remain as in (\ref{Gbreak}) to all orders in perturbation theory.

Since the masses of the $W$ and $\Psi$ states charged under the $U(1)^n$ abelian gauge factors are $\gtrsim L^{-1}$ while the neutral scalar masses are $m_\f \sim g/L$, then below the cut-off  scale $\m$ such that $g/L \ll \m \ll 1/L$, the effective $U(1)^n$ gauge theory can be dualized to $n$ scalars $\s$ governed by the action (\ref{3dmea}) plus the perturbative effective potential (\ref{pertpotl}) for $\f$.  At this scale any massive KK modes charged under the nonabelian gauge factor, $H$, are weakly coupled and can be classically integrated out to give an effective 3-d QCD(adj) for gauge group $H$.  Its 3-d gauge coupling only becomes strong at scales $\lesssim g^2 L^{-1}$.  Thus, the theory is weakly coupled $U(1)^n\times H$ 3-d QCD(adj) at the scale $\m\sim g/L$.   We will have nothing further to say about the non-abelian gauge factors in what follows (beyond the discussion given in the introduction), and will concentrate only on the semi-classical expansion of the effective action for the $U(1)^r$ gauge factors in the rest of the paper.

\section{1-loop potential minimization\label{sec2}}

\subsection{1-loop potential and summary of results\label{sec2.1}}

For a microscopic 4-d theory with massless complex scalars and Weyl fermions in representations $R_b$ and $R_f$, the 3-d one-loop effective potential for $\f$ is 
\begin{align}
V_\text{pert}(\f) = -\frac{1}{\cV} \ln\left(\frac{\prod_f \det(-D^2_{R_f})}{\det(-D^2_\ad) \prod_b \det(-D^2_{R_b})}\right)
\end{align}
where $\cV$ is the volume of $\R^3$.  The covariant derivative in representation $R$ acting on a field $\psi_\l$ in a basis labelled by the weights $\{\l\}$ of $R$ is
\begin{align}
\left(D^\m_R \psi\right)_\l
= \left( \del^\m \d_{\l \l'} + \frac{2\pi i}{L} \d^{\m 4} R(\f)_{\l\l'}\right) \psi_{\l'}
= \left( \del^\m + \frac{2\pi i}{L} \d^{\m 4} \l(\f) \right) \psi_\l
\end{align}
since, by definition, $R(\f)$ is diagonal in this basis with eigenvalues given by $\l(\f)$, the weight vectors evaluated on the Cartan subalgebra element.  Since all the $\phi_\l$'s are independent,
\begin{align}
\ln\det(-D^2_R) = \sum_{\l\in R} \ln\det\left[
-\vec\del^2 - (\del_4+\tfrac{2\pi i}{L} \l(\f) )^2\right]
= \frac{4\pi^2\cV}{3L^3} \sum_{\l\in R} B_4(\l(\f))
\end{align}
where the second equality comes from \cite{Gross:1980br, Unsal:2006pj} for periodic $S^1$, and $B_4(x)$ is the shifted 4th Bernoulli polynomial, which can be defined as
\begin{align}\label{Bernoulli}
B_4(x) &:= [x]^2[-x]^2 = [x]^2 (1-[x])^2, \\
&= x^4 - 2 |x|^3 + x^2 
\qquad \text{for}\ -\tfrac12\le x\le\tfrac12\ \text{and periodically extended,}
\nonumber
\end{align}
where $[x]$ is the fractional part of $x$, that is, $[x]:= x$ mod 1 so that $0\le[x]<1$ for all $x$.  Note that $B_4$ is non-analytic at $x\in\Z$ due to the $|x|^3$ term, but is analytic everywhere else.  
So the effective potential is
\begin{align}\label{veff}
V_{\rm pert}(\f) = \frac{4\pi^2}{3L^3} \left(
\sum_{\l\in{\rm ad}} +\sum_b \sum_{\l\in R_b}-\sum_f \sum_{\l\in R_f} \right) B_4(\l(\f)) .
\end{align}

For QCD(adj) where there are only $n_f$ massless (or light) adjoint Weyl fermions, then
\begin{align}\label{veffad}
V_\text{pert}(\f) = \frac{8\pi^2}{3L^3}(1-n_f) 
\sum_{\a\in\Phi_+} B_4(\a(\f)),
\end{align}
where $\Phi_+$ are the positive roots of $\gf$.  The roots are the non-vanishing weights of the adjoint representation; the exclusion of the zero weights is justified since $B_4(0)=0$. Also, the restriction to positive roots together with an extra factor of 2 is justified since $B_4(-\l (\f))=B_4(\l(\f) )$.

The periodicity of the Bernoulli polynomial (\ref{Bernoulli}) under $x\to x+1$ implies the potential (\ref{veffad}) is periodic under shifts $\f \to \f+\m$ such that $\a(\m)\in\Z$ for all roots $\a$.  Since the $\a$ integrally span the root lattice $\G_r$, this means that $\m\in\G^\v_w$ (since $\G^\v_w$ is integrally dual to $\G_r$).  A fortiori $V_\text{pert}$ is therefore periodic under shifts by $\m$ in the coarser lattice $\G^\v_r$.  Also, the roots are permuted by the Weyl group making the potential invariant under Weyl transformations, so $\tf$ can be restricted to a gauge cell $\hT = \tf/(W\ltimes\G^\v_r)$.  Furthermore, the invariance of $V_\text{pert}$ under shifts in $\G^\v_w$ which are not in $\G^\v_r$ implies the finite group $\G^\v_w/\G^\v_r \simeq Z(\tG)$ acts as a symmetry.  This shows how the restriction of $\f$ to $\hT$ and the action of the global discrete center symmetry, deduced earlier from gauge invariance, emerges explicitly in perturbation theory.

Minimizing this quartic potential directly is often difficult.  Instead, we rewrite it using the identity
\begin{align}
B_4(x) &= -48\sum_{n=1}^\infty \frac{\cos(2\pi n x)}{(2\pi n)^4} + \frac{1}{30}.
\end{align}
Thus, defining the shorthands
\begin{align}
g(x) := \sum_{n=1}^\infty \frac{\cos(2\pi n x)}{n^4},\qquad
\tV := \frac{\pi^2 L^3}{8(n_f-1)} V_\text{pert},
\end{align}
we have, dropping a constant term,
\begin{align}\label{vsum}
\tV =  \sum_{\a\in\Phi_+} g(\a (\f))
= \sum_{n=1}^\infty \frac{1}{n^4} \sum_{\a\in\Phi_+} \cos(2\pi n \a (\f)).
\end{align}
This shows that the potential is an infinite sum over $n$ of terms bounded by dim$(\gf)\cdot n^{-4}$, which therefore rapidly decrease with increasing $n$.  Thus a trial minimum of the potential can be found by minimizing these terms individually for low values of $n$.  We carry this out in section \ref{sec3p3} below.  We then have to check that the trial minimum is indeed a local and global minimum of the potential.  Some of these checks we do numerically.  

Table \ref{tab-1loop} in the introduction and figure \ref{fig-phases} below summarize the main properties of the 1-loop minima.  The figure plots the gauge holonomy eigenvalues for the rank-9 classical Lie algebras.  We have slightly horizontally offset the degenerate eigenvalues for the $B_N$ and $D_N$ theories so that they are apparent.
\begin{figure}
\begin{center}
\hspace{1.5cm}\includegraphics[width=8em]{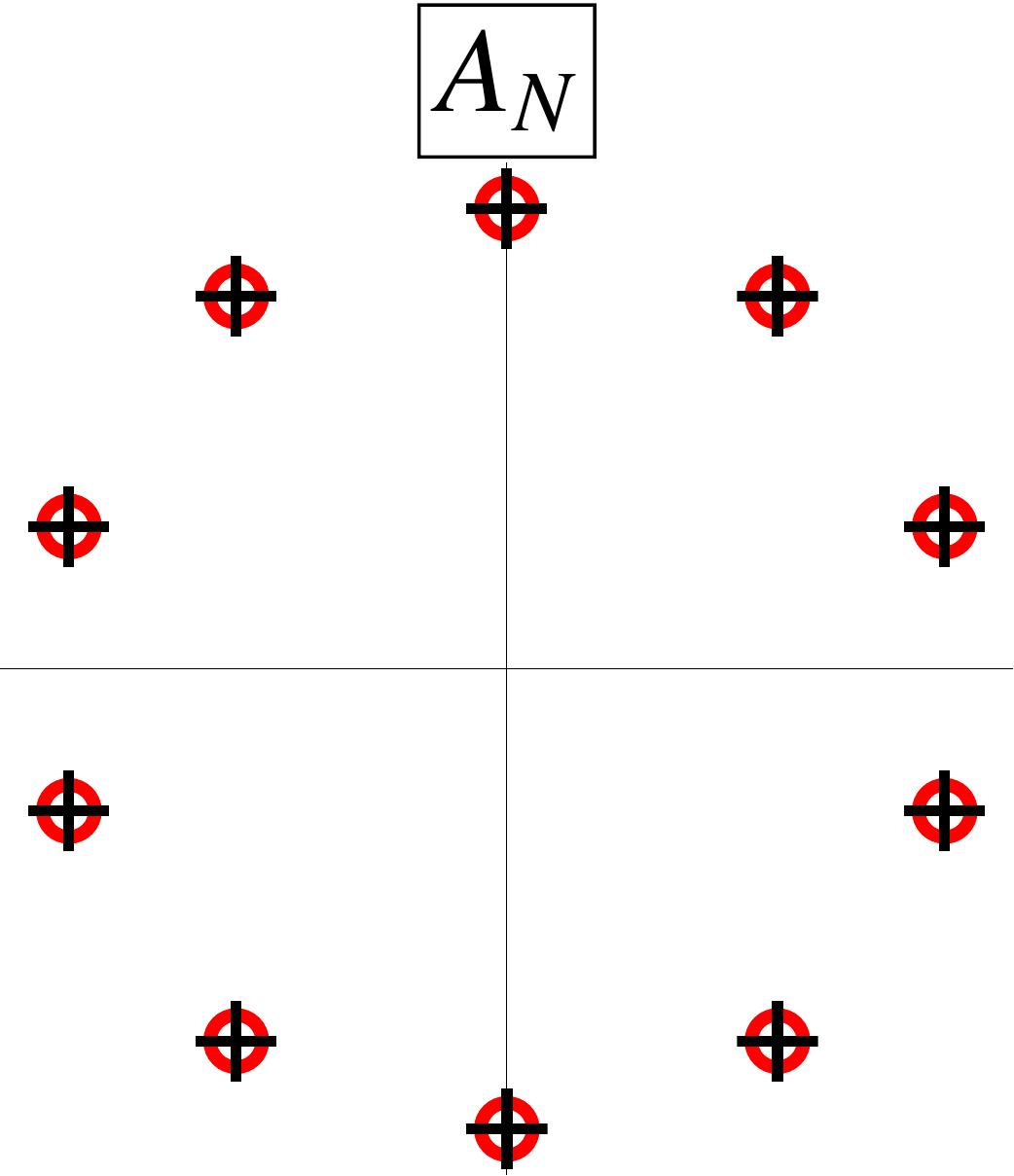}
\hspace{2.5cm}
\includegraphics[width=15em]{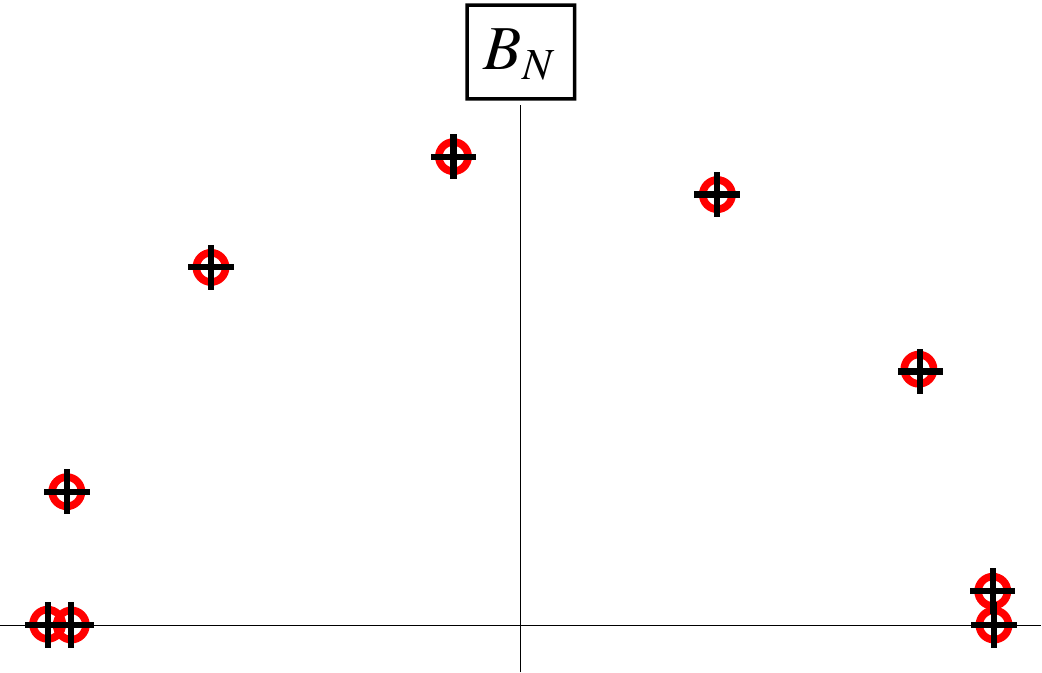}\\[1cm]
\includegraphics[width=15em]{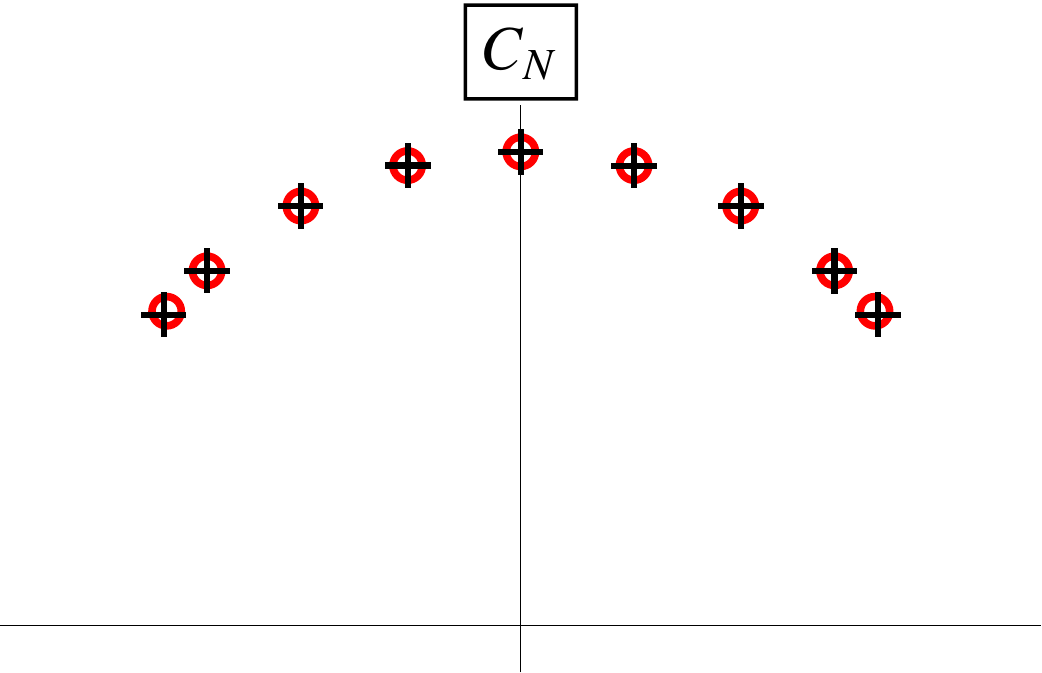}
\hspace{1cm}
\includegraphics[width=15em]{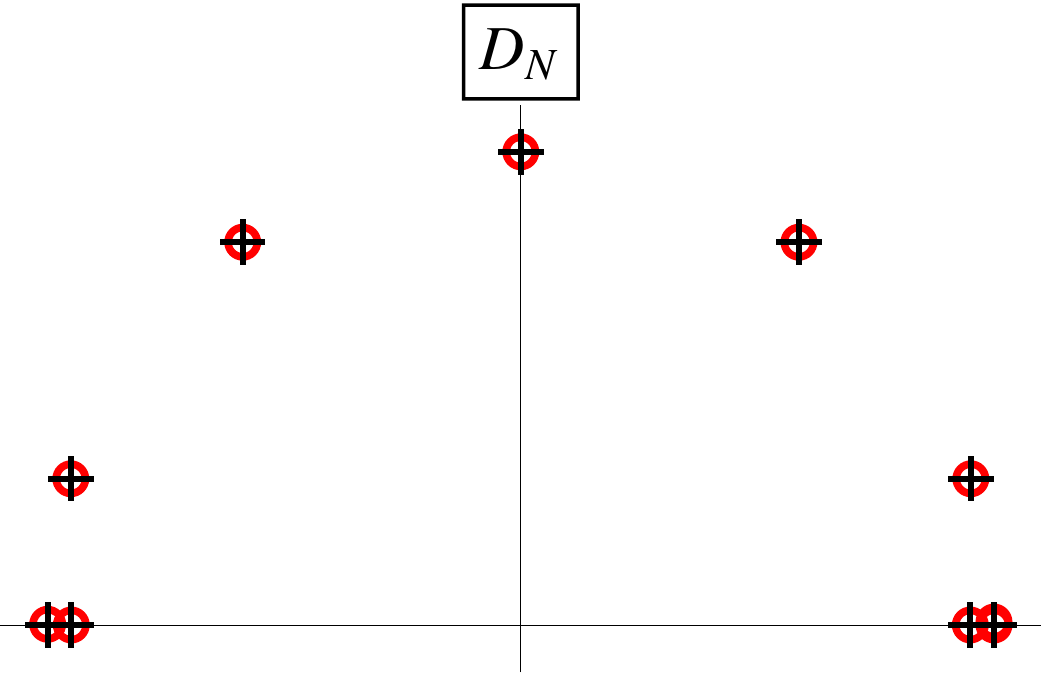}
\caption{Gauge holonomy eigenvalues $\exp\{2\pi i \f_j\}$ for the classical Lie algebras at rank $N=9$.  The red circles are the $\f^*$ predicted minima and the black ``+"'s mark the values found numerically.  The predicted minima are exact for $A_N$ and $D_N$, and thought to be correct only in the large-$N$ limit for $B_N$ and $C_N$.\label{fig-phases}}
\end{center}
\end{figure}

The center symmetry action on the holonomy eigenvalues can be read off from the results of appendix \ref{secB}.  For $A_N$ the $\Z_{N+1}$ center symmetry rotates the eigenvalues by $2\pi/(N+1)$; for $B_N$ the $\Z_2$ center symmetry reflects the eigenvalue closest to $-1$ through the $x$-axis and leaves the other eigenvalues unchanged; for $C_N$ the $\Z_2$ center symmetry reflects all the eigenvalues through the $y$-axis; and for $D_N$ ($N$ odd) the $\Z_4$ reflects the eigenvalue closest to $+1$ through the origin and reflects the rest through the $y$-axis.  All the distributions in the figure are center-symmetric.

Another way of visualising the holonomy eigenvalues is as a point in the gauge cell, which for a rank $r$ gauge group is an $r$-dimensional simplex, a region bounded by $r+1$ faces (which are themselves $(r{-}1)$-dimensional simplices).  The faces are defined by eigenvalue distributions fixed by a Weyl group element (e.g., a pair of eigenvalues coincide) and thus correspond to enhanced gauge symmetries.  The pattern of the gauge symmetry enhancement is described in appendix \ref{secB}.  For rank-2 gauge groups the gauge cells are just triangles, and are plotted in figure \ref{fig-cells} in the coordinates used in appendix \ref{secB}.  In this figure we also show the sub-simplices of center-symmetric holonomies,  fundamental domains for the center action, as well as the locations of the minima of the 1-loop potentials.
\begin{figure}[ht]
\begin{center}
\includegraphics[width=12em]{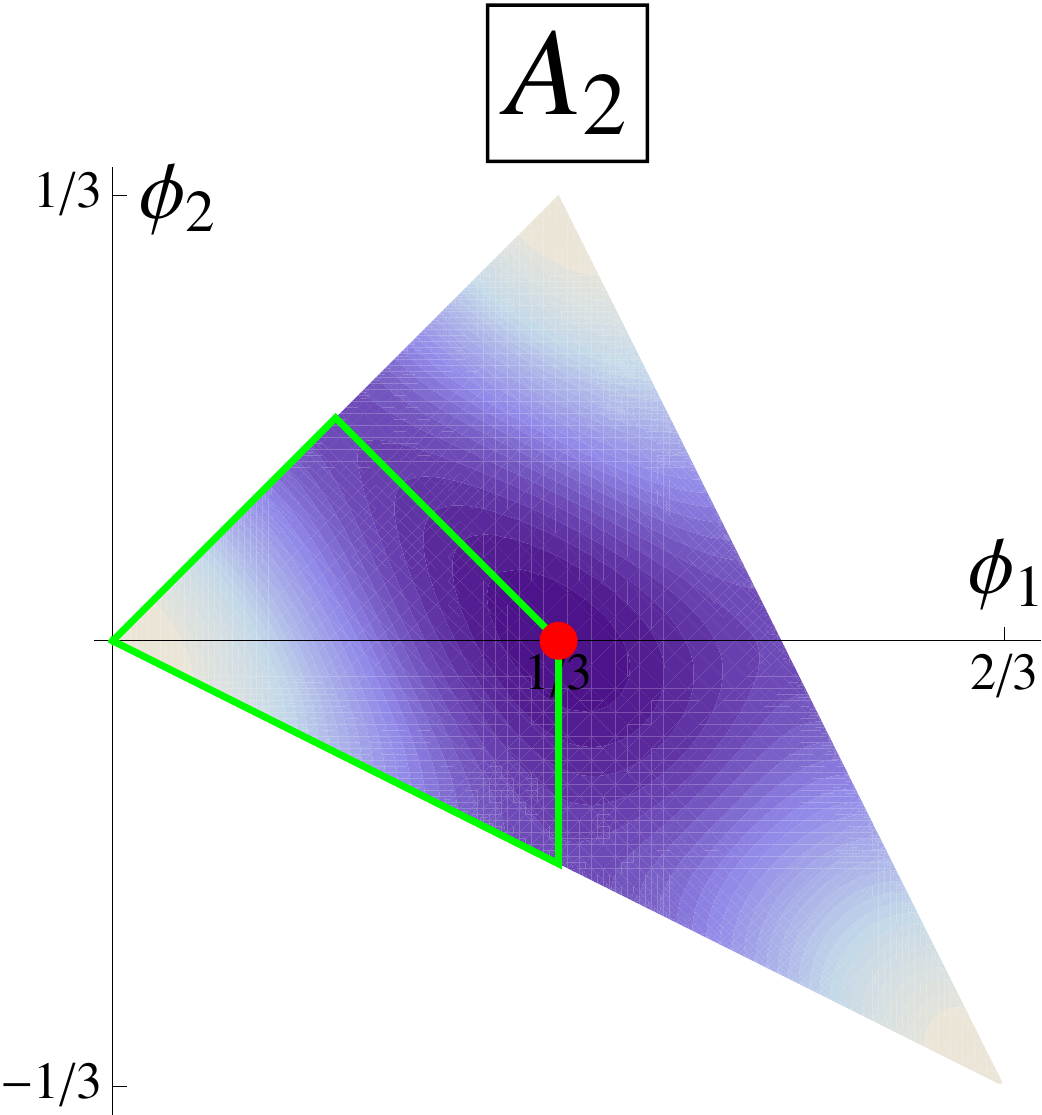}
\hspace{1cm}
\includegraphics[width=12em]{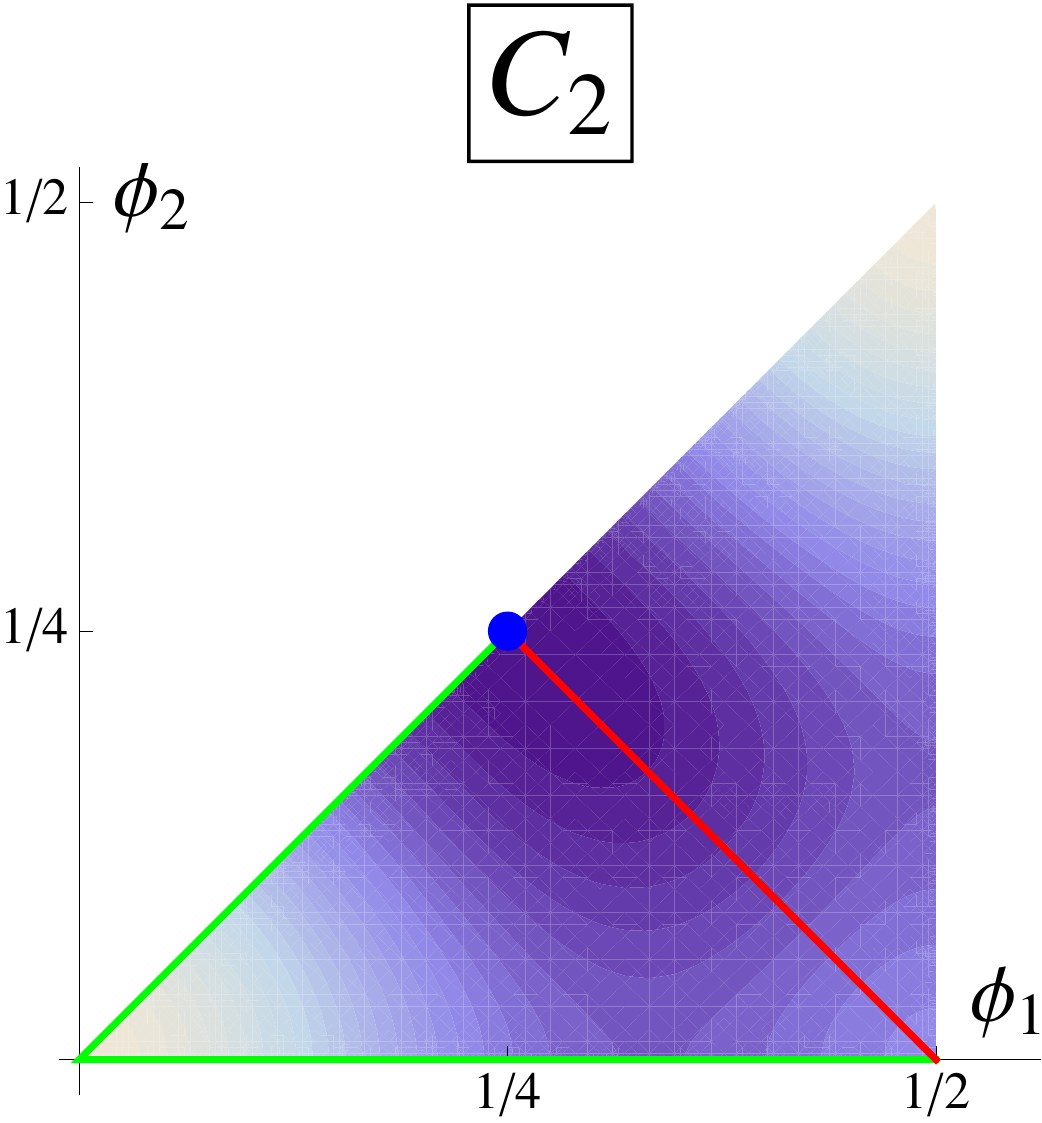}\\[1cm]
\includegraphics[width=12em]{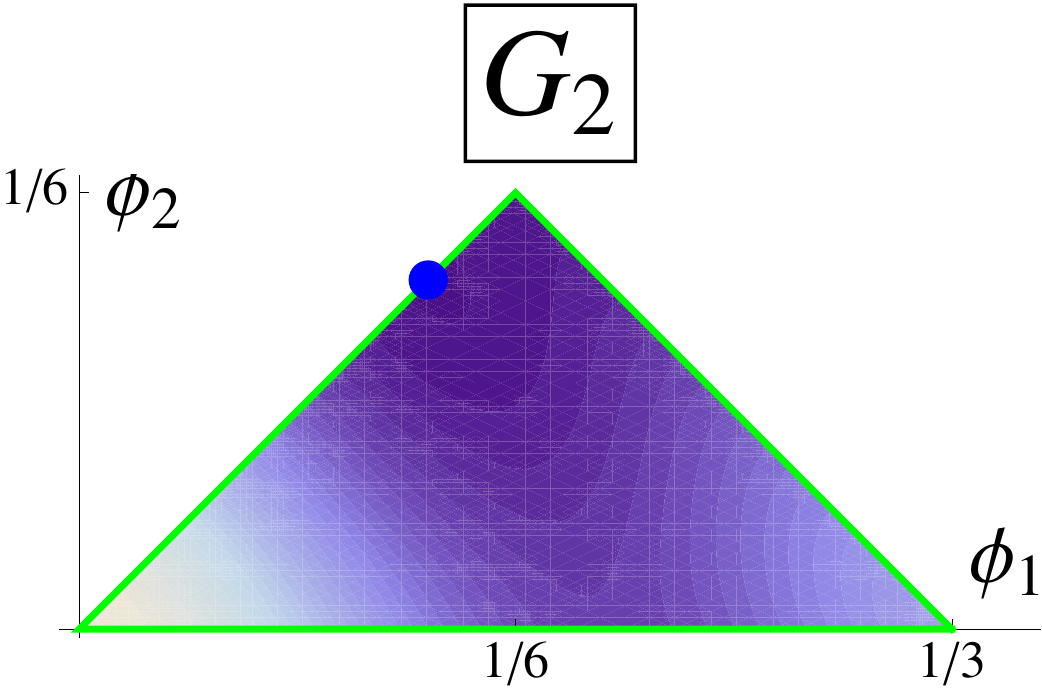}
\hspace{1cm}
\includegraphics[width=12em]{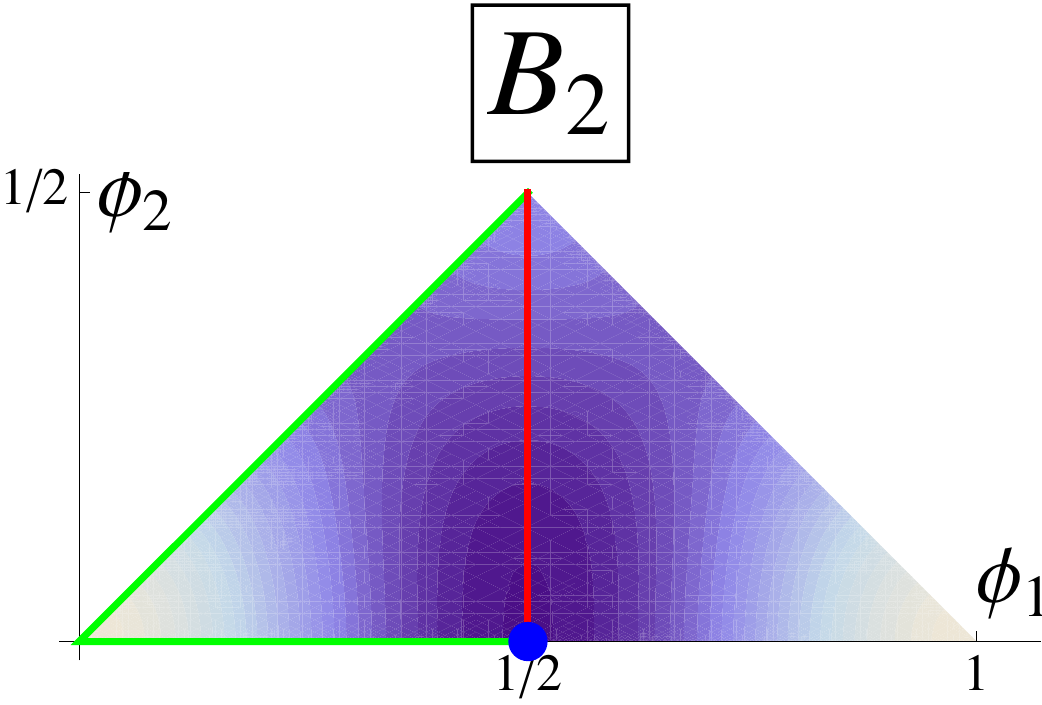}
\caption{Gauge cells for the rank-2 Lie algebras in the coordinates of appendix \ref{secB}, shaded according to the values of the 1-loop potential.  Green and red lines enclose fundamental domains for the action of the center $Z(\tG)$ on $\hT$, red lines or dots are points of unbroken center symmetry, and blue dots are the minima of the 1-loop potential.  The $B_2$ and $C_2$ cases are equivalent, but are expressed in different coordinate systems.\label{fig-cells}}
\end{center}
\end{figure}

\subsection{1PI versus Wilsonian 1-loop potential}

The 1-loop potential (\ref{veffad}) found above is not always the correct effective potential for the light fields (i.e., those with masses less than $\sim 1/L$).  The reason is that (\ref{veffad}) is the 1PI effective potential found from integrating \emph{all} the fields in the loops in the presence of a constant background $\vev{\f}$.  But to compute a consistent (Wilsonian) effective potential for the light modes at a generic $\vev\f$ we should only integrate out the massive degrees of freedom.  

Field components with non-zero weights, $\a$, are charged under the $U(1)^r$ low energy gauge group and have masses $\sim |\a(\f)|/L$ as found in (\ref{Wmass2}).  For $\f$ in the interior of the gauge cell $|\a(\f)|\sim 1$, and all these modes are massive.  The rest of the field components have zero weights in the adjoint representation are so are neutral under the $U(1)^r$ low energy gauge group and have masses at most $\sim g/L$ (from 1-loop effects).   The 1-loop potential (\ref{veffad}) was computed as a 1PI effective potential, in which both the light neutral as well as the heavy charged fields were integrated in the loop.  But, since this is just a 1 loop computation with no internal vertices, neutral fields do not contribute to the $\f$-dependence of $V_\text{pert}$; they only give a constant term, which is subtracted.   Indeed, this is reflected in the fact that in the expression (\ref{veffad}) for $V_\text{pert}$ only a sum over the roots (and not the zero weights) appears.   Thus the inclusion of the light neutral fields at 1 loop does not invalidate the potential.

But at the boundaries of the gauge cell, some of the massive charged modes become light (and are responsible for enlarging the low energy gauge group to contain nonabelian factors).  So, parametrically close to or at the boundaries, these light charged modes should not be integrated in loops.  Explicitly, when $\f$ is near the boundary of the gauge cell associated to the root $\a$, the two 3-d gauge bosons $W^{\pm\a}_m$ and the $2n_f$ adjoint fermions $\Psi_{\pm\a}$ associated to the roots $\pm\a$ become light with a common mass $m_\a = (2\pi/L)|\a(\f)|$.  Their contribution to the 1-loop effective potential is 
\begin{align}
V_\a(\f) &
= \frac{2-2n_f}{\cV}\ln\det\left[-\vec\del^2 + m_\a^2\right]
= -\frac{16\pi^2}{3L^3} (1-n_f)\, |\a(\f)|^3 .
\end{align}
Subtracting this from (\ref{veffad}) therefore increases the attraction to the $\a(\f)=0$ boundary of the gauge cell.  Thus, if the minimum of the 1PI $V_\text{pert}$ is on a gauge cell wall, then correcting to the Wilsonian effective potential does not move the minimum off the wall.  Thus using the 1PI potential does not lead to an incorrect location of the potential minimum.  Furthermore, since the difference between the two is a cubic term, the masses computed in the 1PI and Wilsonian potentials also agree at the minimum.  

Finally, note that subtracting $V_\a$ precisely cancels the $-2|x|^3$ term for $x=\a(\f)$ in $B_4$, so removing the non-analytic term from (\ref{veffad}) at the boundary.  Thus the 1-loop Wilsonian effective potential is never non-analytic, but is also not well-defined (single-valued) over the whole gauge cell.  The analytic Wilsonian expression $V_\text{Wilsonian}=V_\text{pert}-V_\a$ must be used whenever the $W^\a$ and $\Psi_\a$ masses are as light as the heaviest $\f$-mass.  We will see in the next subsection that $(m_\f)_\text{max} \sim \sqrt{N} g/L$ where $N$ is the rank of the gauge group.  Thus the effective 3d action with non-abelian gauge factors and the Wilsonian form of the potential should be used whenever $|\a(\f)| \lesssim \sqrt{N} g$.

\subsection{1-loop potential minima for $n_f>1$ adjoint fermions\label{sec3p3}}

In all of what follows $\{e_i\}$ is an orthonormal basis of $\R^N\supset\tf^*$ and $\{e^i\}$ is a basis of $(\R^N)^*\supset\tf$ dual to the $\{e_i\}$ so that $e_i(e^j)=\d^j_i$ and the $e^i$ are also orthonormal.   A general point $\f\in\tf$ will then have the coordinate expansion
\begin{align}
\f = \sum_i \f_i e^i .
\end{align}
Details of the coordinate systems that we use for the CSAs of the simple Lie algebras are given in appendix \ref{secB}.

\subsubsection{A$_{\bf N-1}$}

The potential (\ref{vsum}) is then given by
\begin{align}
\tV_{A_{N-1}} = \sum_{1\le i<j\le N} g(\f_i - \f_j) \qquad\mbox{with}\quad 
\sum_{1\le i\le N} \f_i = 0,
\end{align}
which can be rewritten by expanding out the cosines as
\begin{align}
\tV = \sum_{n=1}^\infty \frac{1}{4n^4}\left(|x_n|^2 - N\right)
\quad\text{where} \quad
x_n := \sum_j (e^{2\pi i \f_j})^n .
\end{align}
So the potential should be minimized if $x_n = 0$
for as many low values of $n$ as possible.  The general solution for $1\le n<N$ is that $e^{2\pi i \f_j}$ are the $N$-th roots of unity shifted by a phase to satisfy the $\sum_i \f_i=0$ constraint (which implies $\sum_i [\f_i] \in\Z$ for the fractional parts), a simple solution of which is
\begin{align}
\f_j = \frac{N+1-2j}{2N}  := \f^\star_j ,
\end{align}
defining the (trial) minimum point $\f^\star = \sum_j \f_j^\star e^j \in\tf$.  This has actually only determined the fractional parts of the $\f_j$.  Shifts by the co-weight lattice can be used to make arbitrary integer shifts of the $\f_j$ (preserving $\sum_j \f_j=0$) which can be used to put $\f^\star_j$ in the affine Weyl chamber.  The solution given above is already in this chamber, so no further shifts need be made.

To check that $\f^\star$ is a local minimum of the potential, evaluate the exact $\tV(H) = -(\pi^4/3) \sum_{i<j} [\f_i-\f_j]^2(1-[\f_i-\f_j])^2$ near $\f^\star$.  Take $\f_j = \f^\star_j - (\e_j/N)$ with $\sum_j \e_j=0$ for $\e_j$ small.  Then, since $\f_i - \f_j = (j-i+\e_j-\e_i)/N$ is between 0 and 1 for $i<j$, we can drop the fractional part $[\cdot]$ brackets to find
\begin{align}
\tV &= -\frac{\pi^4}{3N^4}\sum_{i<j} (i-j+\e_i-\e_j)^2(N+i-j+\e_i-\e_j)^2
\nonumber\\
&= -\frac{\pi^4(N^4-1)}{180N^2}
+ \frac{\pi^4}{3N^2}\sum_{i,j=1}^{N-1} M_{ij}\e_i \e_j
+ \cO(\e^3)
\nonumber
\end{align}
where $M_{ij}$ is the $(N{-}1)\times(N{-}1)$ symmetric matrix with $M_{ij}=12i(N-j)-N(1+\d_{ij})$ for $i\le j$.  Since the $\cO(\e)$ terms vanish, it is an extremum, and since all the entries of $M_{ij}$ are positive the $\e^2$ term is positive-definite, so $\f^\star$ is a local minimum of $V$.

A numerical search for $N\le 20$ supports that $\f^\star$ is also the global minimum; see figure \ref{fig-phases}.

Since $\f^\star$ is not at a boundary of the affine Weyl cell, the low energy gauge group is completely abelianized to $U(1)^{N-1}$.  Center symmetry is also unbroken, since $\f^\star$ is the unique center-symmetric vacuum derived in appendix \ref{secB}.

The eigenvalues $\{\l_i\}$ of $M_{ij}$ have the approximate distribution $\l_j \simeq \frac{5}{4} N^3j^{-2} + \frac{3}{4} j^2N^{-1}$, for $1\le j\le N-1$, implying a spectrum of $\f$ masses (squared)
\begin{align}\label{mphispec}
m^2_\f \simeq \frac{(n_f-1)g^2}{6L^2}
\left(5 \frac{N}{j^2} + 3 \frac{j^2}{N^3} \right) 
\qquad 1\le j\le N-1
\end{align}
which range from $\cO(g^2N)$ down to $\cO(g^2/N)$.

\subsubsection{B$_{\bf N}$}

The potential is
\begin{align}
\tV_{B_N} = \sum_{1\le i<j\le N} \left[ g(\f_i - \f_j) + g(\f_i+\f_j) \right] 
+\sum_{1\le i\le N} g(\f_i),
\end{align}
and can be rewritten by expanding out the cosines as
\begin{align}
\tV 
&= \sum_{n=1}^\infty \frac{1}{4n^4}\left\{
(x_n+x_n^*)^2 - (x_{2n}+x_{2n}^*) + 2(x_n + x_n^*) - 2N
\right\}
\nonumber\\
&=  \sum_{n\text{ odd}} \frac{1}{4n^4} \left\{
(x_n{+}x_n^*{+}1)^2 - (2N{+}1) \right\}
+\sum_{n\text{ even}} \frac{1}{4n^4} \left\{
(x_n{+}x_n^*{-}7)^2 - (2N{+}49) \right\}
\nonumber
\end{align}
where $x_n := \sum_j (e^{2\pi i \f_j})^n$,
and in the second line we have collected terms invloving $x_n$'s of like $n$ and completed squares.  This makes it plausible that the potential will be minimized if
\begin{align}\label{Bconds}
x_n + x_n^* =
\begin{cases}
-1 & \text{for $n$ odd,}\\
+7 & \text{for $n$ even,}\\
\end{cases}
\end{align}
for as many low values of $n$ as possible.  But $|x_n|\le N$, so the $+7$ value for even $n$ cannot be achieved for small values of $N$ ($N=2,3$).  For large $N$ the set of phases entering in $x_1+x_1^*$ should be unions of sets of all $q$ distinct $q$th-roots-of-unity with each set possibly shifted by an independent overall phase (since sums of their $n$th powers vanish for all $n$ up to $q$) plus the set of seven additional phases $\{-1,-1,-1,-1,+1,+1,+1\}$ (since for $n$ odd they contribute a total of $-1$ to $x_n+x_n^*$, while for $n$ even they contribute $+7$).  So, to satisfy (\ref{Bconds}) for as many $n$ as possible, we should take $q=2N-7$ with overall phase 1, giving the (trial) solution $\f^?:=\f_j^? e^j\in\tf$ with
\begin{align}\label{Bguess}
\{\f^?_j\} = \left\{ \frac{1}{2}, \frac{1}{2}, 
\frac{N-4}{2N-7}, 
\frac{N-5}{2N-7}, \ldots, 
\frac{2}{2N-7}, 
\frac{1}{2N-7}, 0, 0 \right\}.
\end{align}
This solution only makes sense for $N\ge4$.  (In any case, for $2\le N\le6$ the exact minimum can be found by brute force; see below.)

We check whether $\f^?$ is a local minimum of the potential by evaluating at $\f=\f^?$ the first and second derivatives of the exact potential,
\begin{align}\label{}
\frac{3}{\pi^4}\tV &= - \sum_{i<j} \Bigl\{ [\f_i{-}\f_j]^2(1{-}[\f_i{-}\f_j])^2 + [\f_i{+}\f_j]^2(1{-}[\f_i{+}\f_j])^2 \Bigr\} 
- \sum_i [\f_i]^2(1{-}[\f_i])^2.
\nonumber
\end{align}
First, label the $\f^?_j$ in decreasing order as in (\ref{Bguess}).  Then, for nearby points $\f=\f^?+\e$ (with certain choice of signs and relative sizes of the $\e_j$),\footnote{There is no loss in generality in assuming the $\e_j$ have definite signs since the first and second derivatives of $[x]^2(1-[x])^2$ are continuous across the jump from $[x]=1$ to $[x]=0$.  These derivatives are all that are needed to assess whether $\f^?$ is a local minimum.  (The third derivative, on the other hand, has a discontinuity across the jump.)} $[\f_j]=\f_j$ for all $j$ and $[\f_i\pm \f_j]=\f_i\pm \f_j$ for all $i<j$, so 
\begin{align}
\frac{3}{\pi^4}\tV
&= - {\tsum_{i<j}} \left((\f_i{-}\f_j)^2(1{-}\f_i{+}\f_j)^2
+(\f_i{+}\f_j)^2(1{-}\f_i{-}\f_j)^2\right) 
-\tsum_i (\f_i)^2(1{-}\f_i)^2
\nonumber\\
&= \tsum_i \left((7{-}2N) \f_i^4 + (4N{-}4i{+}2) \f_i^3 
+ (1{-}2N) \f_i^2\right)
- 6 \bigl({\tsum_i} \f_i^2 \bigr)^2 
+ 12 \tsum_{i<j} \f_i \f_j^2 .
\nonumber
\end{align}
Then the first derivatives of the potential are
\begin{align}\label{BddV}
\frac{3}{\pi^4}\del_k\tV &= 
4(7-2N)\f_k^3 +6(2N-2k+1)\f_k^2 +2(1-2N)\f_k
\nonumber\\
&\qquad\qquad\mbox{}
+ 12 (\tsum_{i>k} \f_i^2) + 24\f_k (\tsum_{i<k} \f_i) 
-24\f_k (\tsum_i \f_i^2) 
\end{align}
which implies that $\del_k \tV|_{\f=\f^?}=0$ (for $N\ge4$) and shows that the trial minimum is an extremum.  But the second derivatives of the potential are
\begin{align}
\frac{3}{\pi^4}\del_k\del_l \tV
&= \Bigl[12(3-2N)\f_k^2-24(\tsum_i \f_i^2)
+12(2N-2k+1)\f_k +24(\tsum_{i<k}\f_i ) 
\nonumber\\
&\qquad\qquad\mbox{}
+2(1-2N)\Bigr]\d_{kl}
+ 24 \f_l (1-2\f_k)\th_{l>k} + 24\f_k (1-2\f_l)\th_{l<k}
\end{align}
which evaluates at $\f=\f^?$ to
\begin{align}
\del_k\del_l \tV
&\propto
\begin{cases}
\tfrac12(2N-7) \d_{kl} & k,l\in\{1,2\} \\
12(2k{-}5)(N{-}1{-}l)-(24N^2{-}118N{+}149)\d_{kl} & 
k\le l\in\{3,...,N{-}1\}\\
-(24N^2-118 N+149) & k=l=N \\
0& \text{otherwise}\\
\end{cases}
\nonumber
\end{align}
where the proportionality factor is $\frac{2\pi^4}{3}(2N-7)^{-2}$.  
This matrix has only positive eigenvalues in the first $2\times2$ block (i.e., for the $\f^?_1=\f^?_2=\tfrac12$ values) and negative for the remaining $N-2$ eigenvalues.  Thus $\f^?$ is not a minimum, but only a saddle point where the $\f_{1,2}$ coordinates are stable, but the rest are not.

To see where the actual minimum of the potential is, we did a numerical search for global minima for $N\le 25$.  This gives the following global minima with coordinates $\hat\f_j$ of $V$ in the gauge cell:
\begin{align}
N=1: &\quad \{\hat\f_j\}=\{\tfrac12\}.
\quad\text{(exact)}\nonumber\\
N=2: &\quad \{\hat\f_j\} = \{\tfrac12,0\}.
\quad\text{(exact)}
\nonumber\\
N=3: &\quad \{\hat\f_j\} = \{\tfrac12,\tfrac25,0\}
\quad\text{(exact)}
\nonumber\\
N=4: &\quad \{\hat\f_j\} = \{\tfrac12,\tfrac12,\tfrac17,0\}
\quad\text{(exact)}
\nonumber\\
N=5: &\quad \{\hat\f_j\} = \{\tfrac12,\tfrac12,0.3297,0.0422,0\}
\nonumber\\
N=6: &\quad \{\hat\f_j\} = \{\tfrac12,\tfrac12,0.4002,0.1980,0.0253,0\}
\nonumber\\
N=7: &\quad \{\hat\f_j\} = \{\tfrac12,\tfrac12,0.4286,0.2859,0.1415,0.0181,0\}
\nonumber\\
N=8: &\quad \{\hat\f_j\} = \{\tfrac12,\tfrac12,0.4444,0.3333,0.2223,0.1100,0.0141,0\}
\nonumber
\end{align}
For $2\le N\le6$ {\sl Mathematica}\texttrademark\ can find exact algebraic expressions for these values.  For $N=5,6$ the decimal values shown are approximations to irrational numbers (roots of cubics or quartics).  None of these agree with the $\f^?_j$ given in (\ref{Bguess}), but as $N$ increases they rapidly approach $\f^\star_j$ where
\begin{align}\label{Bapprox}
\{\f^\star_j\} \approx 
\left\{ \frac12, \frac12, \frac{N{-}4}{2N{-}7}, 
\frac{N{-}5}{2N{-}7}, \ldots, \frac{2}{2N{-}7}, 
\frac{1}{2N{-}7}, \frac{1}{8(2N{-}7)}, 0 \right\} ;
\end{align}
see figure \ref{fig-phases}.  More accurately, $\f^\star$ is given to six significant figures by
\begin{align}\label{Baccurate}
\{\f^\star_j\} &=
\bigl\{ \frac12, \frac12, \ldots, 
\frac{3(0.999966)}{2N{-}7}, 
\frac{2(1.000500)}{2N{-}7}, 
\frac{(0.990203)}{2N{-}7}, 
\frac{(1.012080)}{8(2N{-}7)}, 0 \bigr\} .
\end{align}
(The first $N-5$ $\f^\star_j$ equal $\f^?_j$ to this accuracy.)

Thus the evidence from the exact solutions for $N\le6$ and the numerical solutions for larger $N$ is that the minima $\f^\star$ have two $\f^\star_j=\tfrac12$ and one $\f^\star_j=0$, exactly.  This implies that the $\f^\star$ vacua are invariant under an $SO(4)\times U(1)^{N-3}\times SO(3)$ nonabelian gauge group.   Also, because there is a $\f^\star_j=\frac12$, the $\Z_2$ center symmetry is not broken (see appendix \ref{secB}).

Evaluating (\ref{BddV}) at the actual minima and diagonalizing gives an approximate spectrum of $\f$ masses at large $N$
which is the same as (\ref{mphispec}) found in the $A_N$ case for $1\le j \le N-4$, plus four masses lighter by about a factor of 4 than the lightest of the above spectrum.  (More accurately, these four have masses about
\begin{align}
m^2_\f = \l \frac{2(n_f-1)g^2}{3(2N-7)L^2}.
\end{align}
with $\l\in\{ 1.6669 , 1.0000 , 1.0000 , 0.8231 \}$.  The two with equal masses are associated to the unbroken $SO(4)\simeq SU(2)^2$ factors, while the lightest is associated with the unbroken $SO(3)$ factor.) 

\subsubsection{D$_{\bf N}$}

In the coordinates of appendix \ref{secB}, the potential is 
\begin{align}
\tV_{D_N} = \sum_{1\le i<j\le N} \left[ g(\f_i - \f_j) + g(\f_i+\f_j) \right] 
\end{align}
which can be rewritten by expanding out the cosines and completing squares as
\begin{align}
\tV 
&= \sum_{n=1}^\infty \frac{1}{4n^4}\left\{
(x_n+x_n^*)^2 - (x_{2n}+x_{2n}^*) - 2N
\right\}
\nonumber\\
&=  \sum_{n\text{ odd}} \frac{1}{4n^4} \left\{
(x_n{+}x_n^*)^2 - 2N \right\}
+ \sum_{n\text{ even}} \frac{1}{4n^4} \left\{
(x_n{+}x_n^*-8)^2 - (2N{+}64) \right\}
\nonumber
\end{align}
where $x_n := \sum_j (e^{2\pi i \f_j})^n$.  This is clearly minimized if
\begin{align}\label{Dconds}
x_n + x_n^* =
\begin{cases}
0 & \text{for $n$ odd,}\\
8 & \text{for $n$ even,}\\
\end{cases}
\end{align}
for as many low $n$ as possible.  But $|x_n|\le N$, so the $+8$ value for even $n$ cannot be achieved for $N=2,3$.  For large $N$ the set of phases entering in $x_1+x_1^*$ should be the union of the set of all $q$ distinct $q$th-roots-of-unity (possibly shifted by an overall phase) plus the set of eight additional phases $\{-1,-1,-1,-1,+1,+1,+1,+1\}$ (since for $n$ odd they contribute a total of $0$ to $x_n+x_n^*$, while for $n$ even they contribute $+8$).  So, to satisfy (\ref{Dconds}) for as many $n$ as possible, we should take $q=2N-8$ with overall phase $\exp\{2\pi i /(4N-16)\}$ (so that the set is invariant under complex conjugation), giving the (trial) solution $\f^?=\f_j^? e^j\in\tf$ with (for $N\ge5$)
\begin{align}\label{Dguess}
\{\f^?_j\} = \left\{
\frac{1}{2},
\frac{1}{2},
\frac{2N-9}{4(N-4)},
\frac{2N-11}{4(N-4)},\ldots,
\frac{1}{4(N-4)},0,0\right\}.
\end{align}

It remains to see whether this trial solution is a minimum of the exact potential.  To check whether $\f^?$ is a local minimum of the potential, evaluate the exact potential $\tV(\f)= - \frac{\pi^4}{3}\tsum_{i<j} \left( [\f_i{-}\f_j]^2 (1{-}[\f_i{-}\f_j])^2 + [\f_i{+}\f_j]^2 (1{-}[\f_i{+}\f_j])^2 \right)$.  Taking the $\f_j$ in the Weyl cell implies $[\f_i\pm \f_j]=\f_i\pm \f_j$ for all $i<j$, so
\begin{align}
\tV & =
- \frac{\pi^4}{3}
\tsum_{i<j} \left((\f_i{-}\f_j)^2(1{-}\f_i{+}\f_j)^2
+(\f_i{+}\f_j)^2(1{-}\f_i{-}\f_j)^2\right)
\nonumber\\
&= -\frac{2\pi^4}{3}
{\tsum_i} \left((N{-}4) \f_i^4 - 2(N{-}i) \f_i^3 
+ (N{-}1) \f_i^2\right)
- \frac{6\pi^4}{3} \bigl({\tsum_i} \f_i^2 \bigr)^2 
+ \frac{12\pi^4}{3} {\tsum_{i<j}} \f_i \f_j^2 .
\nonumber
\end{align}
Then
\begin{align}
\frac{3}{\pi^4} \del_k \tV 
&= -8(N-4)\f_k^3 +12(N-k)\f_k^2 -4(N-1)\f_k
\nonumber\\
&\qquad\qquad\qquad\mbox{}
-24 \f_k (\tsum_i \f_i^2) + 12 (\tsum_{i>k} \f_i^2) + 24\f_k (\tsum_{i<k} \f_i),
\end{align}
which implies that $\del_k\tV\big|_{\f^?}= 0$ (for $N\ge5$) and shows that the trial minimum is an extremum.  Also, 
\begin{align}
\frac{3}{\pi^4}\del_k\del_j \tV
&= \Bigl[-24(N-2)\f_k^2 - 24 (\tsum_i \f_i^2) 
+24(N-k)\f_k +24 (\tsum_{i<k}\f_i )
\nonumber\\
&\qquad\qquad\mbox{}
- 4(N-1)\Bigr]\d_{kj}
+ 24 \f_j(1-2\f_k) \th_{j>k} + 24\f_k(1-2\f_j) \th_{j<k} ,
\nonumber
\end{align}
so
\begin{align}\label{DddV}
\del_k\del_j \tV\Big|_{\f=\f^?}
&\propto 
\begin{cases}
\tfrac12(N{-}4) \d_{kj} & k,j\in\{1,2,N{-}1,N\} \\
12(k{-}\tfrac52)(N{+}1{-}j{-}\tfrac52)-(N{-}4)\d_{kj} & 
k\le j\in\{3,...,N{-}2\}\\
0& \text{otherwise}\\
\end{cases}
\end{align}
where the proportionality factor is $\frac{\pi^4}{3} (N-4)^{-2}$.  This matrix is positive definite for all $N\ge5$.  Thus $\f^?$ is a local minimum.

To see where the global minimum of the potential is, we did a numerical search for $N\le 20$.  This gives to within numerical accuracy that the global minimum equals the trial minimum given in (\ref{Dguess}), 
\begin{align}
\f^\star =\f^? \quad\text{for}\ N\ge5 ;
\end{align} 
see figure \ref{fig-phases}.
For the other values of $N$, we can determine the exact minima algebraically to be
\begin{align}
N=2: &\quad \{\f^\star_j\} = \{\tfrac12,0\}.
\nonumber\\
N=3: &\quad \{\f^\star_j\} =  \{\tfrac12,\tfrac14,0\}
\nonumber\\
N=4: &\quad \{\f^\star_j\}  = \{\tfrac12,\tfrac12,0,0\}
\nonumber
\end{align}
The $N=2$ and $N=3$ cases show no gauge enhancement, and indeed coincide with the results for $SO(4)\simeq SU(2)\times SU(2)$ and $SO(6)\simeq SU(4)$, as expected.

This evidence implies that for $N\ge4$ the minima $\f^\star$ have two $\f^\star_j=\tfrac12$ and two $\f^\star_j=0$.  This implies that the $\f^\star$ vacua are invariant under an $SO(4)\times U(1)^{N-4}\times SO(4)$ nonabelian gauge group. 
Also, the full center symmetry (either $\Z_4$ or $\Z_2\times\Z_2$) is not broken by $\f^\star$.

Diagonalizing (\ref{DddV}) gives an approximate spectrum of $\f$ masses at large $N$ which is the same as (\ref{mphispec}) found in the $A_N$ case for $1\le j \le N-4$, plus four equal masses lighter by a factor of 4 than the lightest of the above spectrum.  More precisely, these four have the 1-loop exact mass
\begin{align}\label{DNlightmass}
m^2_\f = \frac{(n_f-1)g^2}{3(N-4)L^2},
\end{align}
and are associated to the unbroken $SO(4)^2\simeq SU(2)^4$ gauge factors.

\subsubsection{C$_{\bf N}$}

The potential is 
\begin{align}
\tV_{C_N} = \sum_{1\le i<j\le N} \left[ g(\f_i - \f_j) + g(\f_i+\f_j) \right] 
+\sum_{1\le i\le N} g(2\f_i),
\end{align}
which can be rewritten by expanding out the cosines as
\begin{align}
\tV &= \sum_{n=1}^\infty \frac{1}{4n^4}\left\{
(x_n+x_n^*)^2 - (x_{2n}+x_{2n}^*) 
+ 2(x_{2n} + x_{2n}^*) - 2N  \right\} \\
&=  \sum_{n\text{ odd}} \frac{1}{4n^4} \left\{
(x_n+x_n^*)^2 - 2N \right\}
+ \sum_{n\text{ even}} \frac{1}{4n^4} \left\{
(x_n+x_n^*+8)^2 - (2N{+}64) \right\}
\nonumber
\end{align}
where $x_n := \sum_j (e^{2\pi i \f_j})^n$, and in the second line we have collected terms invloving $x_n$'s of like $n$ and completed squares.  This is clearly minimized if
\begin{align}\label{Cconds}
x_n + x_n^* =
\begin{cases}
0 & \text{for $n$ odd,}\\
-8 & \text{for $n$ even,}\\
\end{cases}
\end{align}
for as many low $n$ as possible.  But $|x_n|\le N$, so the $-8$ value for even $n$ cannot be achieved for small values of $N$ ($N=2,3$).  As in the $B_N$ and $D_N$ cases, we can try to satisfy these constraints for $n\lesssim N$ by choosing the set of phases entering in $x_1+x_1^*$ as the union of the set of all $q$ distinct $q$th-roots-of-unity (possibly shifted by an overall phase) plus the set of additional phases to account for the $-8$ for even $n$.  This can be done with $q=2N-8$ with an overall phase shift of $\exp\{2\pi i /(4N-16)\}$ together with the eight additional phases $\{-i,-i,-i,-i,+i,+i,+i,+i\}$ (since for $n$ odd they contribute a total of $0$ to $x_n+x_n^*$, while for $n$ even they contribute $-8$).  This will then satisfy (\ref{Cconds}) up to about $n=2N-8$.

But because $x_n+x^*_n$ is negative and even for even $n$, there is another way of (approximately) satisfying (\ref{Cconds}) but for higher values of $n$:  instead of adding additional phases, choose a larger value of $q$, $q=2N+8$, (with an overall phase shift of $\exp\{2\pi i /(4N+16)\}$ to keep the set invariant under complex conjugation) and {\it remove} the four phases closest to $+1$ and the four closest to $-1$.  In this way, for odd $n<2N+8$ (\ref{Cconds}) will be exactly satisfied, and for even $n$ we will have subtracted approximately $(4)^2+(-4)^2=8$, thus closely satisfying (\ref{Cconds}) for even $n<2N+8$.  This gives the (trial) solution $\f^?=\f_j^? e^j\in\tf$ with
\begin{align}\label{Cguess}
\{\f^?_j\} = \left\{
\frac{2N+3}{4(N+4)}, \ldots,
\frac{2(N-j)+5}{4(N+4)}, \ldots,
\frac{5}{4(N+4)}
\right\}.
\end{align}

It remains to see whether this trial solution is a minimum of the exact potential.  Evaluate the exact potential, 
\begin{align}\label{}
V(\f)\sim - {\tsum_{i<j}} \left([\f_i{-}\f_j]^2(1{-}[\f_i{-}\f_j])^2+[\f_i{+}\f_j]^2(1{-}[\f_i{+}\f_j])^2\right)+{\tsum_i} [2\f_i]^2(1{-}[2\f_i])^2,\nonumber
\end{align} 
by taking the $\f_j$ in the gauge cell determined in appendix \ref{secB}, so that $[2\f_j]=2\f_j$ for all $j$ and $[\f_i\pm \f_j]=\f_i\pm \f_j$ for all $i<j$.  
Minimizing this numerically for $N\le20$ gives the following global minima $\f^\star$:
\begin{align}
N=1: &\quad \{\f^\star_j\} = \{\tfrac14\}.
\quad\text{(exact)}
\nonumber\\
N=2: &\quad \{\f^\star_j\} = \{\tfrac14,\tfrac14\}.
\quad\text{(exact)}
\nonumber\\
N=3: &\quad \{\f^\star_j\} = \{0.2885,\tfrac14,0.2115\} 
\nonumber\\
N=4: &\quad \{\f^\star_j\}  = \{0.3149,0.2815,0.2185,0.1851,\} 
\nonumber\\
N=5: &\quad \{\f^\star_j\} = \{0.3354,0.3058,\tfrac14,0.1942,0.1646,\} 
\nonumber\\
N=6: &\quad \{\f^\star_j\} = \{0.3519,0.3252,0.2750,0.2250,0.1748,0.1481\}
\nonumber\\
N=7: &\quad \{\f^\star_j\} = \{0.3654,0.3411,0.2954,\tfrac14,0.2046,0.1589,0.1346\}
\nonumber\\
N=8: &\quad \{\f^\star_j\} = \{0.3766,0.3544,0.3125,0.2709,0.2291,0.1875,0.1457,
0.1234\}.
\nonumber
\end{align}
None of these agree with the $\f^?_j$ given in (\ref{Cguess}), but as $N$ increases they approach the $\f^?_j$ more closely except for $j=1$ and $j=N$ which are consistently pushed away from $\tfrac12$ and $0$:
\begin{align}\label{Capprox}
\{\f^\star_j\} \underset{N\gg1}{\approx} \left\{ 
\frac{N+1}{2(N+4)}, 
\frac{2N+1}{4(N+4)},
\ldots,
\frac{2(N-j)+5}{4(N+4)},\ldots, 
\frac{7}{4(N+4)},
\frac{3}{2(N+4)} \right\} ;
\end{align}
see figure \ref{fig-phases}.

For $N>2$ there is no gauge enhancement, though the $\Z_2$ center symmetry is unbroken at these minima.

\subsubsection{G$_{\bf 2}$}

In the coordinates of appendix \ref{secB}, the potential in the gauge cell is 
\begin{align}
\tV 
&= g[3 \f_1]+g[3 \f_2]+g[3\f_1+3\f_2]+g[\f_1-\f_2]+g[2\f_1+\f_2]+g[\f_1+2\f_2]
\nonumber\\
&= -180 (\f_1^4 + 2 \f_1^3 \f_2 + 3 \f_1^2 \f_2^2
+ 2 \f_1 \f_2^3 + \f_2^4) 
\nonumber\\
& \qquad \text{} + 4( 32 \f_1^3  + 48 \f_1^2 \f_2  + 51 \f_1 \f_2^2 + 28 \f_2^3)
- 24 (\f_2^2 + \f_1 \f_2 + \f_1^2).
\nonumber
\end{align}
The exact minimum is at 
\begin{align}
\f^\star = \{\f_1=2/15\ ,\ \f_2=2/15\}
\end{align} 
which is at a boundary of the gauge cell, and so has an enhanced $SU(2)\times U(1)$ gauge invariance.

\subsubsection{F$_{\bf 4}$}

In the coordinates of appendix \ref{secB}, the potential in the gauge cell is
\begin{align}
\tV &= -\sum_i g[\f_i] - \sum_{i<j}\left(g[\f_i{-}\f_j]
+g[\f_i{+}\f_j]\right)
- \sum_{a,b,c} g\left[\tfrac12\left(\f_1{+}(-)^a \f_2 
{+} (-)^b \f_3 {+} (-)^c \f_4 \right)\right]
\nonumber\\
&= -\frac{15}2 (\f_1^2+\f_2^2+\f_3^2+\f_4^2)^2
-9 (\f_1^2 +\f_2^2 +\f_3^2 +\f_4^2)
+16 \f_1^3 +10 \f_2^3 +6 \f_3^3 +2 \f_4^3
\nonumber\\
&\ \ \text{}
+18 \f_1 (\f_2^2+ \f_3^2+ \f_4^2)
+12 \f_2 (\f_3^2+\f_4^2)
+12 \f_3 \f_4^2 .
\nonumber
\end{align}
The global minimum appears to be 
\begin{align}
\f^\star := \{\f_1=3/5, \f_2=2/5, \f_3=1/5, \f_4=0\},
\end{align}
to within numerical accuracy.  It is at a boundary of the gauge cell since it saturates $\f_1=\f_2+\f_3+\f_4$, $\f_4=0$, and $\f_1+\f_2=1$ corresponding to vanishing vevs for the two short simple roots and one long root (orthogonal to the short roots), implying an $SU(3)\times SU(2)\times U(1)$ enhanced gauge symmetry.

\subsubsection{E$_{\bf N}$}

For the $E_N$ exceptional groups the global minima of the potentials (whose expressions are too long to reproduce here) are found to be, within numerical precision,
\begin{align}\label{ENmin}
E_6 : &\quad \f^\star = \{ 
\f_1 = 1/2\ ,\ 
\f_2 = 
\f_3 = 
\f_4 = -1/6\ ,\ 
\f_5 = 
\f_6 = 1/6\ \}  \nonumber \\
E_7 : &\quad \f^\star = \{ 
\f_1 = 
\f_2 = 
\f_3 = 1/4\ ,\ 
\f_4 =  
\f_5 = 
\f_6 = 
\f_7 = 0\ \}   \\
E_8 : &\quad \f^\star = \{ 
\f_1 = 5/6\ ,\ 
\f_2 =  
\f_3 = 
\f_4 =  
\f_5 = 
\f_6 = 1/6\ ,\ 
\f_7 = 
\f_8 = 0\ \} \nonumber
\end{align}
in the coordinates described in appendix \ref{secB}.  These minima are easily checked to correspond to the minimal breakings
\begin{align}
E_8 &\supset SU(2)\times SU(3)\times SU(6),
\nonumber\\
E_7 &\supset SU(2)\times SU(4)\times SU(4),
\nonumber\\
E_6 &\supset SU(3)\times SU(3)\times SU(3).
\end{align}
These vacua all preserve the center symmetry (though the center symmetry is trivial for $E_8$).

\section{Self-dual topological configurations  on $\R^3 \times S^1$} 

So far we have argued that in QCD(adj) on $\R^3\times S^1$ the  gauge holonomy, $\f$, at interior points of the gauge cell Higgses $G\to U(1)^r$ at a scale $m_{W^\a}\sim L^{-1}$, where $L$ is the size of the $S^1$.  The 3-d effective action with a cutoff scale $\m$ such that $g/L\ll\m\ll L^{-1}$ and in the interior of the gauge cell is given in perturbation theory by
\begin{align}\label{L0}
\cL_0 = \tfrac{g^2}{4 L} (\del_m\s,\del_m\s) 
+ \tfrac{4\pi^2}{g^2L} \left( \del_m\f \,,\del_m\f \right) 
+ i \tfrac{2L}{g^2} \left( \bar\psi_f , \slashed{\del} \psi_f \right)
+ V_\text{pert}(\f),
\end{align}
where $V_\text{pert}$ is given by (\ref{veffad}) plus corrections smaller by powers of $g^2$ which do not shift the minimum of $V_\text{pert}$ qualitatively.  Interactions involving the dual photon, $\s$, are not generated at any order in perturbation theory.

To understand whether and what effective interactions for $\s$ are generated, we must go beyond perturbation theory.  So we now turn to computing the semi-classical expansion,
\begin{align}\label{scexp}
\cL = \cL_0 + \cL_1 + \cL_2 + \cdots,
\end{align}
of the 3-d effective action, where we will see that the typical size of the $n$th order term in this expansion is, very approximately,
\begin{align}\label{}
\cL_n \approx \exp\{-n S_I \n(\f)\},
\end{align}
where $S_I := 8\pi^2/g^2$ is the 4-d instanton action, and $\n(\f)$ are fractional instanton charges which depend on the vacuum value of $\f$.  For generic values of $\f$ in the interior of the gauge cell, 
\begin{align}\label{}
\n(\f) \sim \frac{1}{h^\v},
\end{align}
where $h^\v$ is the dual Coxeter number of the gauge algebra (a number on the order of the rank of the algebra).  In particular, in this case the semi-classical expansion will be dominated by contributions with fractional instanton number.  

In cases where $\f$ is on a boundary of the gauge cell, some $\n(\f)$ vanish, and the semi-classical expansion becomes invalid, as does the abelian effective action (\ref{L0}) itself.  As computed in the previous section, this actually occurs in QCD(adj) for all gauge groups except $SU(N)$ and $Sp(2N)$.  So the following discussion of the semi-classical expansion is only strictly valid for those groups.

The remainder of this section reviews, following \cite{Lee:1997vp,Davies:2000nw}, the elementary semi-classical configurations, and derives the first order corrections to the effective Lagrangian.  The next section will be devoted to higher-order corrections, which, though much smaller, lead to qualitatively new effects.

Finite action field configurations on $\R^3 \times S^1$ are classified according to two pseudo quantum numbers, the magnetic charge (vector) $\m\in\tf$ and the topological charge (or instanton number) $\n$:
\begin{align}\label{}
\m &:= \frac{1}{2\pi} \int_{S^2_{\infty}} f  ,
\qquad\qquad
\n := \frac{1}{16 \pi^2} \int_{\R^3 \times S^1} ( F_{\m \n},   \til  F_{\m \n} )  .
\end{align}
The magnetic charge is defined here in terms of the low energy 3-d effective $U(1)^r$ 2-form field strength, $f$, while the topological charge is given in terms of the microscopic 4-d Yang-Mills field strength.  The Killing form is normalized so that the smallest instanton number on $\R^4$ is 1, and corresponds to the normalization where the lengths-squared of long roots are 2.  These quantum numbers are protected to all orders in perturbation theory, but their conservation can be violated non-perturbatively.  The perturbative vacuum has $\m=\n=0$. 

\subsection{Monopole-instantons\label{sec4.1}}  

A subclass of the finite action topological configurations on $\R^3 \times S^1$ arises as solutions to the self-duality equation
\begin{align}\label{4dins}
F_{\m\n} = \til F_{\m\n} := \tfrac12 \e_{\m\n\r\s} F_{\r\s}.
\end{align}
The Higgsing of the gauge group by a compact adjoint Higgs field---in our case the gauge holonomy $\f$ around the $S^1$---implies the existence of $r+1$ types of elementary monopole-instantons.  These are solutions to the Bogomolny equation,
\begin{align}
F_{mn}= \e_{mnp} D_p A_4 = \frac{2 \pi}{L}\e_{mnp} D_p \f ,
\label{monins}
\end{align}
which is the dimensional reduction of the self-duality equation (\ref{4dins}), found by assuming the gauge fields are $x_4$-independent.  Ordinarily, one expects only $r$ elementary monopoles due to the Higgsing to $U(1)^r$, each associated with a simple root $\a_j$,  $j=1, \ldots, r$ of the gauge algebra.  But since the adjoint Higgs field is compact, there is an extra monopole associated with the affine (or lowest) root $\a_0$.  The magnetic charge, $\m^{(j)}$, of the monopole-instanton of type $j$ is, in the normalization of section \ref{sec2.4},
\begin{align}
\m^{(j)}= \a_j^\v, 
\qquad  j = 0, \ldots, r ,
\label{magcharge}
\end{align}
where $\a_j^\v$ are the affine or simple co-roots, defined in (\ref{corootmap}).  Since the electric charge of a $W$-boson of type $j$ is the affine or simple root $\a_j$,  $j=0, \ldots, r$, and since $\a_i(\a_j^\v) = \hat A_{i,j}$ is the integer-valued extended Cartan matrix---defined in appendix \ref{secA3}---one checks that the Dirac quantization condition is satisfied.  Thus in the long distance 3-d theory, the 3-d instantons are monopoles.  As described around (\ref{disorderop}),  a magnetic source at point $x$ in the dual path integral is accompanied by the insertion of the disorder or monopole operator, $\exp\{2\pi i \s(\a_j^\v)\}$.

In the microscopic 4-d theory, these monopole-instantons are semi-classical field configurations with finite action,
\begin{align}\label{fracinst1}
S_j(\f) = S_I \cdot |\n^{(j)}| ,
\end{align}
where $S_I := 8\pi^2/g^2$ is the action of a single 4-d instanton, and where the monopole-instanton fractional topological charges are
\begin{align}\label{fracinst2}
\n^{(j)} = (\a_j^\v,\f) + \d_{j,0}
= \begin{cases}
\frac{2\,\a_j(\f)}{(\a_j,\a_j)} & \quad j = 1, \ldots, r \\
\a_0(\f)+1 & \quad j=0
\end{cases}
\end{align}
for $\f$ in the fundamental $\tG$-cell, where they are all positive.    In the second equality we have used the normalization of the Killing form mentioned above, for which long roots---and so in particular $\a_0$---have length-squared 2.  Recall that the fundamental $\tG$-cell is defined (\ref{cellwalls}) by the inequalities $\d_{j,0}+\a_j(\f)\ge0$.  These functions on the $\tG$-cell will appear often in what follows, so we define the special notation,
\begin{align}\label{}
\ba_j(\f) := \a_j(\f) + \d_{j,0} .
\end{align}
Then the monopole-instanton topological charges are
\begin{align}\label{fracinst3}
\n^{(j)} = \frac{2}{(\a_j,\a_j)} \, \ba_j(\f) .
\end{align}
Also, the masses of the lightest massive W-bosons and fermions in the $\tG$-cell (\ref{Wmass2}) are given by
\begin{align}\label{Wmass3}
m_{W^{\a_j}} = m_{\Psi_{\a_j}} = \frac{2\pi}{L} \, \ba_j(\f).
\end{align}

At the specific vacua found in section \ref{sec2} by minimizing the 1-loop potential for $\f$ in QCD(adj), one finds for the classical groups at large rank (and exactly for $SU(N)$ and $SO(2N)$) the values shown in table \ref{tab4.1}.  The vanishing entries for $SO(N)$ correspond to the unbroken nonabelian $SU(2)$ gauge factors at the perturbative vacuum.
\begin{table}[ht]
\begin{center}
\begin{tabular}{|c||c c c c|} \hline
& $A_r=SU(r+1)$ & $C_r=Sp(2r)$ & $B_r=SO(2r+1)$ & $D_r=SO(2r)$ \\ \hline\hline
$h^\v$ & $r+1$ & $r+1$ & $2r-1$ & $2r-2$ 
\\ \hline
$\n^{(0)}$ & $(h^\v)^{-1}$ & $3 (h^\v+3)^{-1}$ &
0 & 0 
\\
$\n^{(1)}$ & $(h^\v)^{-1}$ & $\frac12(h^\v+3)^{-1}$ &
0 & 0
\\
$\n^{(2)}$ & $(h^\v)^{-1}$ & $(h^\v+3)^{-1}$ &
$\frac12 (h^\v-6)^{-1}$ & $\frac12 (h^\v-6)^{-1}$
\\
$\vdots$ & $\vdots$ & $\vdots$ &$\vdots$ &$\vdots$
\\
$\n^{(j)}$ & $(h^\v)^{-1}$ & $(h^\v+3)^{-1}$ & 
$(h^\v-6)^{-1}$ & $(h^\v-6)^{-1}$ 
\\
$\vdots$ & $\vdots$ & $\vdots$ &$\vdots$ &$\vdots$
\\
$\n^{(r-2)}$ & $(h^\v)^{-1}$ & $(h^\v+3)^{-1}$ & 
$\frac78 (h^\v-6)^{-1}$ & $\frac12 (h^\v-6)^{-1}$
\\
$\n^{(r-1)}$ & $(h^\v)^{-1}$ & $\frac12(h^\v+3)^{-1}$ &
$\frac18 (h^\v-6)^{-1}$ & 0
\\
$\n^{(r)}$ & $(h^\v)^{-1}$ & $3 (h^\v+3)^{-1}$ &
0 & 0
\\ \hline
\end{tabular}
\caption{Fractional instanton numbers of fundamental monopole-instantons for QCD(adj) with classical gauge groups.  These are exact for $A_r$ and $D_r$, but only approximate for large $r$ for $B_r$ and $C_r$.  The dual Coxeter number for each group is also shown.\label{tab4.1}}
\end{center}
\end{table}

Each monopole-instanton also has four bosonic zero modes, $a\in\R^3$ is its position and $\phi\in U(1)$ is the internal angle of the monopole.  Global electric $U(1)$ gauge transformations (in the $U(1)$ subgroup associated with the type-$j$ monopole-instanton) shift $\phi$.  Since the monopole-instanton is electrically neutral, its $\phi$-dependence is trivial.

Since 3-d monopole-instantons have finite action, they will have finite space-time density in the vacuum as in the Polyakov model:  in a given three-volume $V_3$ in $\R^3$ there will be approximately $V_3 L^{-3} e^{-S_j}$ instantons.  But, unlike what happens in the Polyakov model, a dilute gas of monopole-instantons does not cause a mass gap for gauge fluctuations.  The reason is that they carry a certain number of fermion zero modes given by the Nye-Singer index theorem  \cite{Nye:2000eg, Poppitz:2008hr} (which is a generalization of the Atiyah-Singer index theorem to a manifold with boundary, and thus applicable to $\R^3 \times S^1$).  In QCD(adj) each monopole-instanton has $2n_f$ fermionic zero modes.

Putting these ingredients together one expects the gas of type-$j$ monopole-instantons to induce an operator
\begin{align}
\cM_j &= \cC_j \ 
\exp \left[ -S_j(\f) + 2 \pi i \s(\a_j^\v) \right] 
\ \det_{f,f'} \left[\a_j(\psi_f) \cdot \a_j(\psi_{f'}) \right]
\end{align}
in the effective 3-d theory in the interior of the gauge cell, where the light fields are the holonomy $\f\in\tf$, the $r$ dual photons $\s\in\tf^*$, and the $n_f$ fermions $\psi_f\in\tf$.  This form of $\cM_j$, as well as its $\f$-dependent coefficient $\cC_j$ will, be determined below from a careful analysis of the path integral zero-mode measure.  We will refer to $\cM_j$ as the type-$j$ monopole operator in what follows.  Note that $\cM_j$ preserves a global $SU(n_f)$ symmetry.  An anti-monopole-instanton, $\bar\cM_j$, has the opposite magnetic and topological charges as a monopole-instanton, and its operator is the complex conjugate of the monopole operator.

\subsection{4-d instanton as a composite at long distances}

Since the theory at short distances is a 4-d gauge theory, it also has 4-d instantons obeying the self-duality equation (\ref{4dins}) which carry topological charge one and zero magnetic charge.  The action of a single ($\n=1$) 4-d instanton is
\begin{align}\label{}
S_I := \frac{1}{2g^2} \int (F_{\m \n},  F_{\m \n}) 
= \frac{1}{2g^2}\int ( F_{\m \n},  \til F_{\m \n} )
= \frac{8 \pi^2}{g^2} .
\end{align}
This self-dual field configuration is not independent of the monopole-instantons described above.

It is instructive to see how this defect arises as a composite of the elementary monopole-instantons.  There exists a unique positive integral linear relation among the simple and affine co-roots, $\sum_{j=0}^{r} k_j^\v \a_j^\v = 0$, with $k_0^\v=1$.   The $k_j^\v$ are the co-marks or dual Kac labels, and are described in Appendix \ref{secA3}.  Thus, the smallest magnetically neutral combination of the monopole-instantons is given by combining $k_j^\v$ monopole-instantons of type $\cM_j$ for $j=0,\ldots,r$.  Schematically, if the instanton-induced operator is $I$, then $I \sim \prod_{j=0}^{r} [\cM_j]^{k_j^\v}$.  Since $\sum_{j=0}^r k_j^\v = h^\v$, the dual Coxeter number, this presents the 4-d instanton as a combination of $h^\v$ monopole-instantons.  The values of $h^\v$ for the simple Lie algebras are given in table \ref{tabA2} in appendix \ref{secA3}.  It follows from (\ref{fracinst3}) that the instanton number of this combination is then $\n = \sum_{j=0}^r k^\v_j \n^{(j)} = 1$, irrespective of the vacuum value of $\f$.
   
The 4-d instanton has $4 h^\v$ bosonic zero modes which matches the counting of the zero modes of the $h^\v$ monopole-instantons.  
The 4-d instanton zero modes are associated with the classical symmetries of the self-duality equation:  4 are the position of the instanton $({\bf a}_I \in \R^4)$ and arise due to translation invariance, one is the size modulus $(\r \in \R^+)$ and is associated with invariance under dilatations, and the remaining $4h^\v-5$ are angular coordinates in the gauge group $(U \in G_\text{stability})$ associated with new solutions obtained under the action of the stability group, see \cite{Vandoren:2008xg} for a review:
\begin{align}
4h^\v & \ \xrightarrow{\text{short-distance}}\ 
4+1+(4h^\v -5)  =  ({\bf a}_I \in \R^4) + (\r \in \R^+) 
+ ( U \in G_\text{stability}).  
\end{align}

In unHiggsed gauge theories the existence of the size modulus $\r$ implies that the instanton comes in arbitrarily large sizes at no cost in action, and prevents a meaningful long-wavelength description of a dilute instanton gas from first principles.   But since the small $\R^3 \times S^1$ regime of QCD(adj) is in a Higgs phase, instantons have a maximal size and an effective coupling associated with the scale of the Higgsing.  At long distances where the 4-d instanton is described as a composite of $h^\v$ 3-d monopole-instantons, we have
\begin{align}  
4h^\v & \ \xrightarrow{\text{long-distance}}\ 
h^\v [3 +1] = h^\v [({\bf a} \in \R^3) + (\phi \in U(1))]  .    
\end{align}
In particular the 4-d instanton size modulus is no longer present in the long distance description of QCD(adj) on small $\R^3 \times S^1$.  This permits a meaningful dilute gas expansion; however, the 4-d instanton plays a negligible role in the semi-classical expansion since the constituent monopole-instantons have smaller action.  

We can also easily check that the counting of the fermionic zero modes match.  A 4-d instanton has $2h^\v n_f$ fermionic zero modes and an associated instanton operator, $I \sim e^{-S_I} [\det_{f, f'} (\psi_f, \psi_{f'})]^{h^\v}$, which is invariant under an $SU(n_f)$ continuous symmetry.  Alternatively, since $\sum_{j=0}^r k_j^\v = h^\v$, the $2n_f$ fermionic zero modes of each monopole-instanton give the same total number as for a 4-d instanton.

Finally, the 4-d instantons reduce the classical $U(1)_A$ symmetry down to a $\Z_{2 n_f h^\v}$ discrete chiral symmetry of the quantum theory. We will discuss the realization of this symmetry in section \ref{sec6} after we construct the low-energy effective Lagrangian. 
 
\subsection{Collective coordinates of monopole-instantons}
\label{sec:B}

The appropriate one-loop measure for integrating over configurations of a single type-$j$ monopole-instanton is\footnote{The following summary is an adaptation of the appendix of \cite{Davies:2000nw}, which treats the monopole measure in supersymmetric Yang-Mills theory.}
\begin{align}
d\m_{\rm B}  d\m_{\rm F} =
e^{-S_j} \cdot
\frac{d^3 a\, d\phi}{(2 \pi)^2} \prod_{f=1}^{n_f} d^2 \xi_f \cdot
\m^{4-n_f} \cdot 
J_a J_\phi (J_\xi)^{-n_f} \cdot 
\left[{\det}'(-D^2)_\text{adj} \right]^{n_f-1} .
\nonumber
\end{align}
\begin{list}{$\bullet$}{\itemsep=0pt \parsep=0pt \topsep=2pt}
\item $a\in\R^3$ is the monopole-instanton position, $\phi\in U(1)$ is the global electric angle of the monopole, $\xi_f$ are the Grassmann-valued fermionic zero modes.  Since all the 3-d effective fields and defects in QCD(adj) are electrically neutral, there is no $\phi$-dependence in the integrand and so the integral over $\phi$ just gives a factor of $2\pi$.
\item $\m$ is the (Pauli-Villars) renormalization scale.  The factor of $\m^4$ can be viewed as the contribution of the Pauli-Villars regulator fields associated with the 4 bosonic zero modes. Similarly, $\m^{-n_f}$ can be viewed as the contribution of the Pauli-Villars regulator fields associated with the $2n_f$ fermionic  zero modes.
\item The $J$'s are the collective coordinate Jacobians, $J_a = S_j^{3/2}$, $J_{\phi} = L S_j^{1/2} [2\pi\ba_j(\f)]^{-1}$, and $J_\xi = 2S_j$.  (Our value for $J_\phi$ differs from that given in \cite{Davies:2000nw} by the substitution $\a_j(\f) \to \bar\a_j(\f)$.)
\item The primed determinant comes from integrating over the Gaussian fluctuations of the non-zero modes.  Because in a self-dual background $[{\det}'(-D^2\d_{\m\n}-2F_{\m\n})_\text{adj} ]^{-1/2}$ $=\, [{\det}'(-D^2)_\text{adj}]^{-2}$ and ${\det}'(\slashed{D})_\text{adj} = {\det}'(-D^2)_\text{adj}$, the contributions from the Gaussian integrals over all bosonic and fermionic fluctuations other than zero modes combine to give
\begin{align}
\underbrace{
[{\det}'(-D^2\d_{\m\n}-2F_{\m\n})_\text{adj}]^{-1/2} 
}_\text{gauge bosons}
\times \underbrace{ 
{\det}'(-D^2)_\text{adj} 
}_\text{ghosts}   
\times \underbrace{ 
[{\det}'(\slashed{D})_\text{adj}]^{n_f}
}_\text{fermions} 
=  [{\det}'(-D^2)_\text{adj}]^{n_f-1} \, .
\nonumber 
\end{align}
Note that when $n_f=1$, the bosonic and fermionic primed  determinants cancel precisely due to supersymmetry and absence of non-compact scalars.\footnote{In supersymmetric theories with non-compact scalars the fluctuation determinants may not cancel due to possible differing continuum state densities.}
\end{list}

The dependence of the regularized scalar determinant on the renormalization scale is determined by the counterterm for the gauge action due to the scalar field fluctuations, which has the form $\d\cL = - (8\pi^2)^{-1} (T(R)/12) (F_{\m\n},F_{\m\n}) \ln\m$ for complex scalars in the representation $R$ (as in the 1-loop beta function (\ref{b0})).  In the adjoint representation $T(\ad)=2 h^\v$.  Exponentiating this in a Euclidean type-$j$ monopole-instanton background gives $\exp\{ \ln(\m)\, h^\v S_j(\f) /(3 S_I) \} = \exp\{\ln(\m)\, h^\v \n^{(j)} /3\}$.  Thus $\det'(-D^2)_\text{adj} \sim \m^{h^\v \n^{(j)} /3}$.  For $SU(N)$ gauge group, $h^\v \n^{(j)} = 1$ for all $j$, but for the other simple groups the exponent will vary with $j$ according to table \ref{tab4.1}.

The fields of a type-$j$ monopole-instanton are embedded entirely within the regular $SU(2)$ subgroup of $G$ associated with the root $\a_j$.  The only scale which appears in the classical equations for the type-$j$ monopole-instanton is $2\pi \ba_j(\f)/L$, the mass of the $W$-boson associated with $\a_j$.  Since the determinant is dimensionless, it must therefore have the form
\begin{align}\label{fluctdet}
[{\det}'(-D^2)_\text{adj}]^{n_f-1}= 
\left(\tfrac12(\a_j,\a_j)C_j\right)^{n_f-1} \, 
\left(\frac{\m L}{\ba_j(\f)} \right)^{(n_f-1)h^\v \n^{(j)} /3} \,,
\end{align}
where $C_j$ is a pure number presumably of order one.  (The factor of $(\a_j,\a_j)/2$ is to simplify some later formulas.)  $C_j$ may have some $N$- and $\f$-dependence.  It could, in principle, be computed along the lines of \cite{'tHooft:1976fv}, but we will not attempt that calculation here.

Putting this all together, the one-loop type-$j$ monopole-instanton measure becomes
\begin{align}
d\m_{\rm B} d\m_{\rm F} =
\frac{C_j^{n_f-1}}{32\pi^2} 
\left(\frac{\m L}{\ba_j(\f)} \right)^{\b^{(j)}}
\left(\frac{L}{\ba_j(\f)} \right)^{n_f-3}
\frac{(4S_j)^{2-n_f}}{(\a_j,\a_j)^{1-n_f}} e^{-S_j}\,
d^3a \, \prod_{f=1}^{n_f} d^2  \xi_f ,
\label{Eq:measure}
\end{align}
where
\begin{align}\label{}
\b^{(j)} := \frac13 [(12-h^\v\n^{(j)})-(3-h^\v\n^{(j)})n_f].
\end{align}
Note that $\sum_{j=0}^r k^\v_j \b^{(j)} = \b_0$, the 1-loop beta function (\ref{b0adj}).  This was expected since the 4-d instanton is a combination of $h^\v$ monopole-instantons ($k^\v_j$ of type $j$), and the 4-d instanton measure is proportional to $\m^{\b_0}$.

\subsubsection*{Monopole operator induced in the 3-d effective Lagrangian}

The long-distance asymptotics of the fermionic zero mode profile for a type-$j$ monopole-instanton located at $a\in\R^3$ is 
\begin{align}
\psi^{(j)}_f (x)  &= F_{mn}(x-a)\,\s^{mn}\xi_f     
\ \xrightarrow{\text{long-distance}}\ 
4 \pi S_F(x-a) \xi_f\, \a_j^\v,
\end{align}
where $S_F(x)= \s^m x_m/(4\pi |x|^3)$ is the free fermion propagator.  We deduce that in the long wavelength effective theory
\begin{align}
\left\langle\prod_{f=1}^{n_f}\psi^{(j)}_f\cdot\psi^{(j)}_f 
\right\rangle 
=  \int d\m_B d\m_F \,
(\a^\v_j)^{\otimes 2 n_f}
\prod_{f=1}^{n_f}  
\left(4\pi S_F(x-a) \xi_f\right) \cdot  
\left(4\pi S_F(x-a) \xi_f\right),
\nonumber
\end{align}
where the dot denotes spinor index contraction.  The integration over the Grassmann-valued collective coordinates, $\x_f$, gives a product of free fermion Green's functions and factors involving the co-roots.  Such a correlator is reproduced by adding to the perturbative 3-d effective Lagrangian, $\cL_0$ (\ref{L0}), the interactions
\begin{align}\label{L1}
\cL_1 = \sum_{j=0}^r \left( \cM_j + \text{h.c.}  \right),
\qquad
\cM_j := \til\cA_j \, e^{-S_j(\f)+2\pi i\s(\a_j^\v)}
\prod_{f=1}^{n_f} (\a_j^\v,\psi_f)^2, 
\end{align} 
where
\begin{align}
\til\cA_j  :=  
\left(\frac{2L}{g^2}\right)^{2n_f} 
(4\pi)^{2n_f}
\frac{C_j^{n_f-1} }{32\pi^2}
\left(\frac{\m L}{\ba_j(\f)}\right)^{\b^{(j)}}   
\left(\frac{L}{\ba_j(\f)}\right)^{n_f-3} 
\frac{(4S_j(\f))^{2-n_f}}{(\a_j,\a_j)^{n_f-1}}.
\end{align}
The $(2L/g^2)^{2n_f}$ factor reflects our normalization of the kinetic term in (\ref{L0}).  The dual photon field, $\s$, dependence follows from the long-distance coupling (\ref{disorderop}) to a point magnetic charge $\a_j^\v$.

Using Fierz identities, the fermion product in $\cM_j$ can be rewritten as
\begin{align}\label{fierz}
\prod_{f=1}^{n_f} (\a_j^\v,\psi_f)^2 = \frac{2^{n_f}}{(n+1)!} 
\det_{f,f'} \left[ (\a_j^\v,\psi_f) \cdot (\a_j^\v,\psi_{f'})\right], 
\end{align}
which makes apparent the fact that $\cM_j$ is invariant under an $SU(n_f)$ global symmetry.  Since $(\a_j^\v,\psi_f) = 2\a_j(\psi_f)/(\a_j,\a_j)$, and recalling the expressions (\ref{fracinst1}) and (\ref{fracinst3}) for the monopole-instanton action $S_j(\f)$, we can rewrite (\ref{L1}) as
\begin{align}\label{mon1a}
\cM_j &= \cA_j \, e^{-S_j(\f)+2\pi i\s(\a_j^\v)} 
\prod_{f=1}^{n_f} [\a_j(\psi_f)]^2, 
\qquad \text{with} \\
\cA_j &= 
L^{n_f-3} \left(\frac{2L}{g^2}\right)^{2n_f} (g^2)^{n_f-2}
\frac{128\pi^2}{(\a_j,\a_j)^3}  
C_j^{n_f-1}
(\ba_j(\f))^{5-2n_f}
\left(\frac{\m L}{\ba_j(\f)}\right)^{\b^{(j)}},
\nonumber
\end{align} 
to make the $\f$- and $g^2$-dependence more explicit.  We will refer to $\cM_j$ as the type-$j$ monopole operator.
 
Alternatively, (\ref{L1}) can be obtained by considering a dilute gas of monopole-instantons and by summing over all such events.  Treating the scalars as background fields, the grand-canonical ensemble of a dilute gas of 3-d instantons can be recast into a Lagrangian, as was shown by 't Hooft in the context of 4-d-instantons \cite{'tHooft:1986nc}.

\section{Topological molecules (non-self-dual configurations)\label{sec5}}
 
Since the fundamental (self-dual) monopole-instantons have fermionic zero modes, they cannot generate a mass gap for gauge fluctuations \cite{Unsal:2007vu, Unsal:2007jx}.  Instead, they generate multi-fermion dual photon interactions as shown in (\ref{L1}). 

In order to generate a mass gap for gauge fluctuations, we need a potential purely in terms of dual photon fields, similar to the Polyakov model where an $e^{-S_I} \cos\s$ term induces a mass gap and, equivalently, confinement of electric charge.  Such a bosonic potential is induced at second order---$\cL_2$ in the semi-classical expansion (\ref{scexp})---from semi-classical configurations involving a monopole-instanton and an anti-monopole-instanton.  In the language of the Euclidean dilute monopole--anti-monopole gas, these appear as topological ``molecules", since the second-order terms in the semi-classical expansion arise from the interactions of between the monopoles and anti-monopoles. 

There are two types of topological molecules.  One type, the ``magnetic bion", has been discussed in \cite{Unsal:2007vu, Unsal:2007jx} for $SU(N)$ gauge group.  Here, we generalize that discussion to all gauge groups.  We also emphasize the existence and properties of a second type of bion, which has non-trivial implication for the Wilson line dynamics.  

The topological molecules appearing at second order in the semi-classical expansion are in one-to-one correspondence with the non-vanishing entries of the extended Cartan matrix, $\hat A_{ij} := \a_i(\a_j^\v) \propto (\a_i,\a_j)$.  Its diagonal elements are all positive, and its off-diagonal elements are either negative or vanish.  In particular, for $i\neq j$, $(\a_i,\a_j) \neq 0$ whenever the $i$ and $j$th nodes of the extended Dynkin diagram are connected by a link.  The extended Dynkin diagrams for the simple Lie algebras are shown in figure \ref{fig-dynkin} in appendix \ref{secB}.  The key properties of the two types of bions are as follows.
\begin{itemize}
\item  {\it Magnetic bions:} For each pair $(i,j)$ such that $(\a_i,\a_j)<0$,  there exists a magnetic bion $[\cM_i \bar\cM_j]$ with magnetic and topological charges 
\begin{align}
(\m,\n) = 
\left(\a_i^\v-\a_j^\v  \, , \,  \n^{(i)}-\n^{(j)}\right),
\end{align} 
associated with an operator in the effective action proportional to 
\begin{align}
\cB_{ij} \sim 
e^{-S_i(\f)-S_j(\f)} e^{2\pi i \s(\a_i^\v-\a_j^\v)},
\end{align}  
\item {\it Neutral bions:}  For each $i$  there exists a bion $ [\cM_i\bar\cM_i]$ with magnetic and topological charges
\begin{align}
(\m,\n) = (0,0),
\end{align}
associated with an operator proportional to 
\begin{align}
\cB_{ii} \sim e^{-2 S_i(\f)}.
\end{align}  
\end{itemize}
The magnetic bions carry non-zero magnetic (and possibly also topological) charge, so are distinguishable from the perturbative vacuum.   The neutral bions, on the other hand, are indistinguishable from the perturbative vacuum in that sense.

Since these topological molecules are not solutions to the first order  Bogomol'nyi-Prasad-Sommerfield equations in a simple way, we need to show their stability due to dynamics.\footnote{The analogous  instanton--anti-instanton molecules in quantum mechanics are the (complex) analytic continuation of bounce solutions.  Perhaps there is a generalization of this to quantum field theory; see \cite{Balitsky:1986qn}.}  This requires a careful study of the zero and quasi-zero modes of the molecules.  The magnetic bions provide an example of stable semi-classically calculable bound states of a monopole-instanton and an anti-monopole-instanton.

\subsection{Zero and quasi-zero modes of the topological molecules} 
\label{sec:zeromode}

In the path-integral formalism one sums over fluctuations around the topological defect field configuration.  This requires the study of the eigen-spectrum of the small fluctuation operator (corresponding to the second derivative of the action in the background of the defect).  The eigenvalues are of two types: {\it i)} Zero modes, reflecting  the symmetries of the system, which do not cost extra action; the corresponding integrals are trivial.  {\it ii)} Non-zero modes or small fluctuations in a semi-classical analysis; the corresponding integrals can be dealt with within a Gaussian approximation.  A review of this material can be found in, e.g., \cite{Coleman}.  

\begin{figure}[t]
\begin{center}
\includegraphics[angle=-90, width=5.0in]{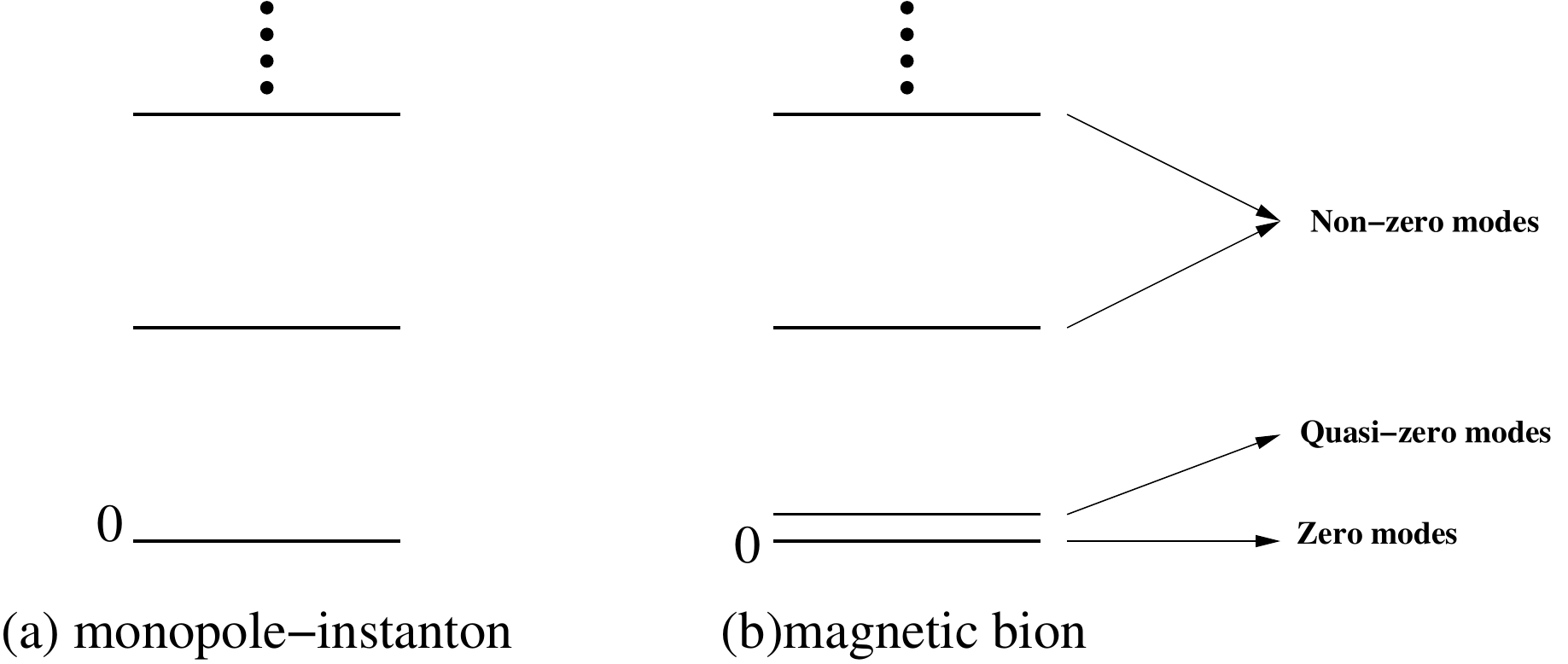}
\caption{Typical eigen-spectrum of the small-fluctuation operator (a) for a monopole-instanton, and (b) for a topological molecule, e.g., a magnetic bion.  To get the correct prefactor for the magnetic bions,  the quasi-zero modes integrals need to be done exactly.}
\label {fig:fluc}
\end{center}
\end{figure}
If the eigen-spectrum of the small fluctuation operator involves a mode parametrically separated from the non-zero modes, the situation is more subtle.  Such modes cannot be treated in the Gaussian approximation as they are not normal Gaussian modes, and they are not exact zero modes either.  The integrals over them need to be done exactly in the path integral formalism to get the correct results.  Therefore, it is appropriate to consider in the eigen-spectrum a third type of eigenvalue in the above classification:  {\it iii)} Quasi-zero modes.  

Quasi-zero modes are typical when one considers topological molecules such as instanton--anti-instanton pairs.  In such examples, the separation between the defects is a quasi-zero mode.  The way to see this is to consider these defects at asymptotically large separation, where they interact only weakly.  For example, consider an instanton $I(t-\t/2)$ and anti-instanton $\bar I(t+\t/2)$ pair where $\t$ is the separation between the two in the quantum mechanical double-well problem.  The action of the pair is $S=2 S_I - c e^{-\o\t}$ where $\o^{-1}$ is the instanton size and $c$ is numerical factor.  In the regime where $\o\t\gg1$, changing $\t$ has a very small impact on the action, and hence it is a quasi-zero mode.  On the other hand, the ``center of mass" position $t$ is an exact zero mode as the action does not depend on it.  

Similarly, in QCD(adj) a change in the separation between a monopole-instanton and an anti-monopole-instanton corresponds to a quasi-zero mode.  Long range interactions between the two defects induced by the light fields ($\f$, $\s$, and $\psi_f$) lift this mode slightly to become a quasi-zero mode.  When these long-range interactions are attractive, they indicate the existence of new, higher-order terms in the semi-classical expansion of the effective action.

The path integral of the effective 3-d theory to first order in the semi-classical expansion,
\begin{align}\label{}
Z = \int [D\f D\s D\psi_f] e^{-\int d^3x (\cL_0 + \cL_1)},
\end{align}
is the partition function for a grand-canonical ensemble describing a dilute monopole plasma.  Expanding the exponential of the first-order terms,
\begin{align}\label{}
e^{-\int d^3x \cL_1} = 1 - \int d^3x \,\cL_1 + \frac12 \left(\int d^3x \,\cL_1\right)^2 + \cdots
\end{align}
induces terms at second order including the terms
\begin{align}\label{}
+ \sum_{ij} \int d^3x \int d^3y \, \cM_i(\vec x) \bar\cM_j(\vec y) 
= \sum_{ij} \int d^3R \int d^3r \, \cM_i(\vec R +\tfrac12\vec r) \bar\cM_j(\vec R -\tfrac12\vec r) ,
\nonumber
\end{align}
where we have pulled out the integration over the exact ``center of mass" zero mode $\vec R$.  This induces an effective second-order term in the semi-classical expansion of the effective action,
\begin{align}\label{bion0}
\cL_2 \supset \sum_{ij} \cB_{ij} := - \sum_{ij} \int d^3r \, 
\Big\langle \cM_i(\vec R +\tfrac12\vec r) 
\bar\cM_j(\vec R -\tfrac12\vec r) \Big\rangle ,
\end{align}
where the brackets denote a connected correlator in the perturbative vacuum.  If the correlator is mainly supported at separations $r<r_b$ for some length scale $r_b$, then it is consistent to treat $\cB_{ij}$ as independent operators in an effective action valid on length scales much larger than $r_b$.  We will call the $\cB_{ij}$ ``bion operators".

From the explicit form of the monopole operators $\cM_i$ given in (\ref{mon1a}) and the connected correlators
\begin{align}
& \Big\langle  
e^{-S_i(\f) + 2\pi i \s(\a_i^\v)} (\tfrac12\vec r) \, 
e^{-S_j(\f) -  2\pi i \s(\a_j^\v)} (-\tfrac12\vec r)
\Big\rangle
=  \exp \left[ (2\pi)^2\frac{2L}{g^2} 
\frac{ (\a_i^\v,\a_j^\v) (1+ e^{-m_\f r}) } {4 \pi r}\right] ,
\nonumber\\ 
& \Big\langle
\prod_{f=1}^{n_f} [\a_i(\psi_f)]^2(\vec x) \,
\prod_{f=1}^{n_f} [\a_j(\bar\psi_f)]^2(\vec y) \Big\rangle 
= \left( \frac{g^2}{2L}\right)^{2n_f}\frac{(\a_i,\a_j)^{2n_f}}{(2\pi)^{2n_f} r^{4n_f}},
\nonumber
\end{align}
we obtain 
\begin{align}\label{bion1}
\cB_{ij} = -\cA_{ij} e^{-S_i(\f)-S_j(\f)} 
e^{2\pi i \s(\a_i^\v-\a_j^\v)},
\end{align}
where
\begin{align}\label{bioncoeff}
\cA_{ij} =
\cA_i\cA_j\, \left( \frac{g^2}{2L}\right)^{2n_f}
\frac{(\a_i,\a_j)^{2n_f}}{(2\pi)^{2n_f}}
\int d^3r \, e^{-V_\text{eff}^{ij} (r)},
\end{align}
and
\begin{align}\label{mmbar}
V_\text{eff}^{ij}(r) =  
- (\a_i^\v,\a_j^\v) \frac{2 \pi}{g^2} 
(1+ e^{-m_\f r}) \frac{L}{r}+ 4n_f {\ln}(r).
\end{align}
Here $m_\f$ is the mass of $\f$ in the perturbative vacuum and the $\cA_i$ are given in (\ref{mon1a}).

Note, first of all, that by virtue of the factors of $(\a_i,\a_j)$ in (\ref{bion1}), no bion operator is generated if $(\a_i,\a_j)=0$.  Secondly, the sign of the first term in (\ref{mmbar}) depends on the sign of $(\a_i,\a_j)$, while the second term does not.

$V^{ij}_\text{eff}$ has a straightforward physical interpretation as a monopole--anti-monopole effective potential in the Euclidean monopole plasma picture.  The second term in (\ref{mmbar}) is an attractive force induced by fermion zero mode exchange, while the first term in (\ref{mmbar}) is the Coulomb interaction between a monopole and anti-monopole, which is repulsive for $(\a_i,\a_j) <0$ and attractive for $(\a_i,\a_j)>0$.  The $1/r$ part of the first term is due to exchange of the dual photon scalar $\s$.  Recall that $\s$ remains massless to all orders in perturbation theory.  The $e^{-m_\f r}/r$ term is due to the exchange of the $\f$-scalar.  Since $m_\f \sim g/L$ at one loop in perturbation theory for $n_f>1$ (as we computed in section \ref{sec2}), this force is short range.  When $n_f=1$, however, it is massless to all orders in perturbation theory.  (This is because for $n_f=1$ QCD(adj) is supersymmetric and $\f$ and $\s$ are in the same supermultiplet.)  In other words, for the purpose of a long distance effective theory, $\f$ decouples for $n_f>1$, whereas it should be kept when $n_f=1$.  For this reason, we introduce 
\begin{align}
\z := 
\begin{cases}
1 &\text{for $n_f=1$,}\\ 
0 &\text{for $2 \leq n_f \leq 5$,}
\end{cases}
\end{align}
and replace
\begin{align}\label{}
e^{-m_\f r} \to \z
\end{align}
in $V^{ij}_\text{eff}$.  

\subsection{Magnetic bions}

The previous discussion makes it clear that there will be qualitative differences between the $\cB_{ij}$ bions with $i\neq j$, which we call magnetic bions, and the $\cB_{ii}$ which we call neutral bions.  We start with the magnetic bions.

\begin{figure}[t]
\begin{center}
\includegraphics[angle=0, width=3.0in]{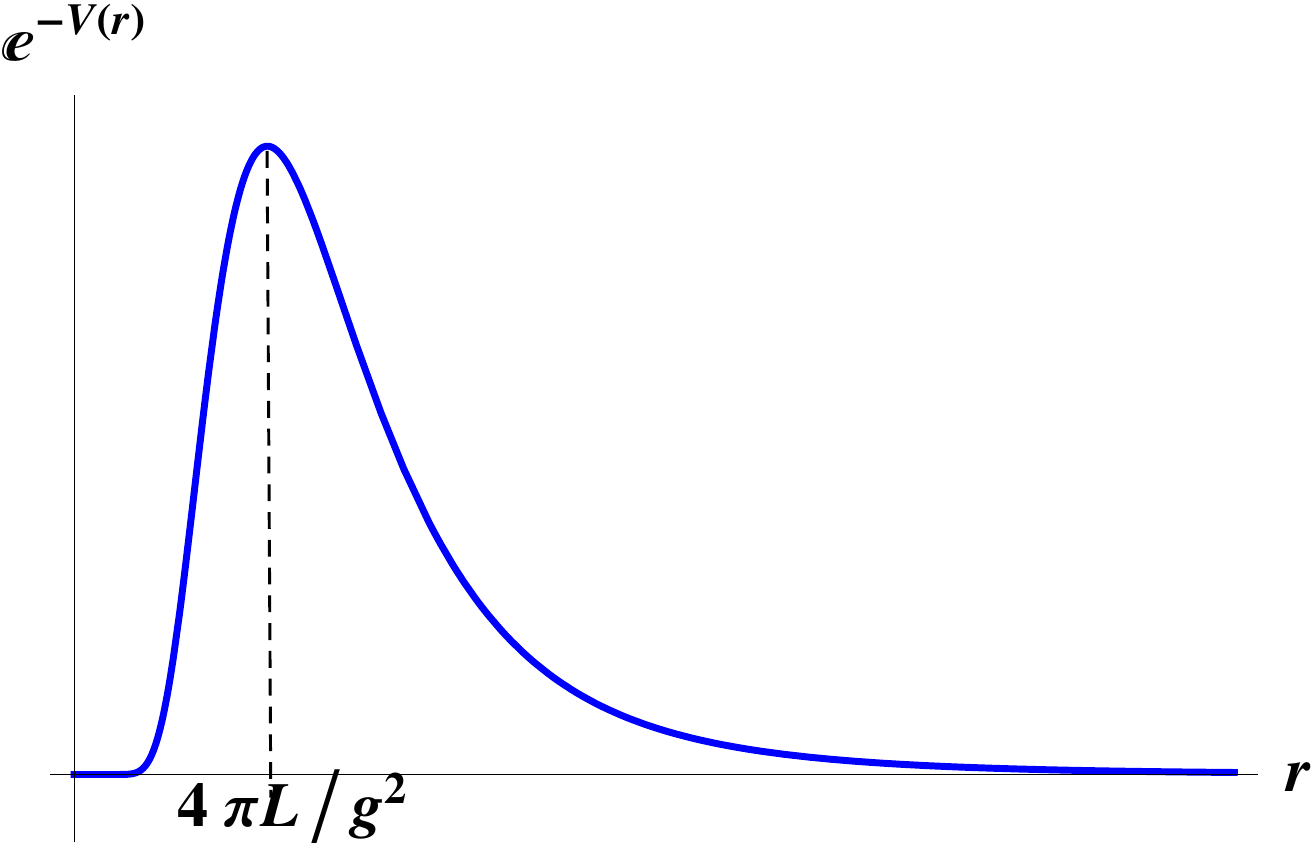}
\caption{The integral over the quasi-zero mode---the separation between $\cM_i$ and $\bar\cM_j$---is dominated by separations $r \sim L/g^2$.   In Euclidean space for $i\neq j$ the interaction between the two monopole-instantons is repulsive at short distances due to Coulomb repulsion and attractive at long distances due to fermion zero-mode exchange, leading to the stable saddle. 
}
\label {fig:quasi}
\end{center}
\end{figure}

For $i\neq j$ such that $(\a_i,\a_j) <0$ (which correspond to linked nodes of the extended Dynkin diagram), the prefactor of the magnetic bion amplitude (\ref{bion1}) evaluates to
\begin{align}
\label{bionamp}
\cA_{ij} 
&= 
-\; \frac{(\a_i,\a_j)^{3-2n_f}}{g^8 L^3} \cdot
\frac{2^{13}\pi^2}{1+\z} \cdot \til C_i\til C_j \cdot
I(g^2, n_f) 
\end{align}
where
\begin{align}\label{int2}
I(g^2, n_f) &= 
\int_0^\infty \! dz  \; 
\exp\left[-\frac{1}{g^2 z} - (4n_f-2) \ln z \right] 
=  \left( \frac{1}{g^2} \right)^{3-4n_f} \G(4n_f-3) 
\end{align} 
and
\begin{align}
\til C_j &:= \left[\frac{(\a_j,\a_j)^4 C_j}{(4\pi)^3(1+\z)^2}\right]^{n_f-1}
\frac{\ba_j(\f)^{5-2n_f}}{(\a_j,\a_j)^2} 
\left(\frac{\m L}{\ba_j(\f)}\right)^{\b^{(j)}}.
\end{align}
Note that $\cA_{ij}$ is positive since $(\a_i,\a_j)<0$.  The $I(g^2,n_f)$ factor arises as the integral over $\exp( - V_\text{eff})$ in (\ref{bioncoeff}) in rescaled variables.  The short-distance Coulomb repulsion and the long-distance fermion-induced attraction in $V_\text{eff}$ means that the integrand of $I(g^2,n_f)$ is peaked as shown in the physical units in Fig.\ref{fig:quasi}.  
The integral is over the quasi-zero mode and is dominated by the scale $r_{\rm b}\sim L/g^2$.  Separations between an instanton and anti-instanton less than $L/g^2$ are virtually forbidden by a Coulomb blockade, $e^{-r_{\rm b}/r}$.  At large separation, the integral is cut off by the fermion zero mode exchange in a power law manner.  (See \cite{Anber:2011de} for an alternative derivation). 
 
The existence of magnetic bions is reliable within the region of validity of semi-classical analysis because of the clear separation of all the scales involved:
\begin{align}\label{hierarchy2}
\begin{matrix}
r_{\rm m}&\ll&r_{\rm b}&\ll&d_{\rm m-m}&\ll&d_{\rm b-b},\\ 
\downarrow   &&\downarrow&&\downarrow && \downarrow  \\
L&\ll&\frac{L}{g^2}&\ll&Le^{S_0/3}&\ll& Le^{2S_0/3}.
\end{matrix}
\end{align}
At first order in the semi-classical expansion, we have monopole-instantons with typical size $r_{\rm m} \sim L$ set by the scale of Higgsing of the microscopic gauge group.  These monopoles are rare because of their large action, $S_0 \sim (g^{2}N)^{-1}$.  Their mean separation is $d_{\rm m-m} \sim n_{\rm m}^{-1/3} = L e^{S_0/3}$ where $n_{\rm m}$ is the monopole density.  At second order in the semi-classical expansion are magnetic bions which we have just shown have typical size $r_{\rm b} \sim L/g^2$.  Thus $r_{\rm m} \ll r_{\rm b} \ll d_{\rm m-m}$, which allows us to consistently interpret magnetic bions as a second-order effect in a semi-classical expansion which are clearly distinct from the first-order dilute monopole plasma.  The density of bions is $n_{\rm b} \sim  e^{-2S_0}$ and the mean separation between these molecules is $d_{\rm b-b} \sim n_{\rm b}^{-1/3} = e^{2S_0/3}$.  Evidently, bions are much rarer than monopoles, but, as we explain in section \ref{sec6}, they are the {\it leading} topological defects to give rise to a non-perturbative mass term to gauge fluctuations.

There are a few basic consistency checks on the form of the magnetic bion induced terms in the action.  Keeping only the parametric dependence of the bion amplitude on the coupling, compactification scale $L$, and cut off $\m$, we have
\begin{align}
\cB_{ij} \sim L^{-3} g^{8n_f-14} (\m L)^{2\b^{(j)}}.
\end{align}
The factor of $1/L^3$ means that our analysis is dimensionally correct.  The power of $\m$ leads to the correct appearance of the leading order beta function coefficient for instanton operators, as explained after (\ref{Eq:measure}).  Finally, the power of the coupling for $n_f=1$ is $g^{-6}$, agreeing with power of $g$ appearing in the bosonic potential of $\cN{=}1$ superYang-Mills \cite{Davies:2000nw}.  For the non-supersymmetric theory, the same power has recently been obtained in \cite{Anber:2011de} through a different method.

\subsection{Neutral bions and the BZJ prescription}\label{sec5.3}

When $i=j$, both terms in (\ref{mmbar}) induce an attractive interaction since $(\a_i,\a_i)>0$.  Since the magnetic charges of the monopole and the anti-monopole are opposite in this case, we call such configurations neutral bions.  The contribution of the neutral bion operator to the effective action is, formally, 
\begin{align}\label{type-two}
\int d^3x  \, \cB_{ii} (\vec x) &= -\int d^3x\, 
\cA_{ii} \,  e^{-2S_i(\f)} 
\end{align}
where the integral over the quasi-zero mode gives
\begin{align} \label{neutamp}
\cA_{ii} &=   
+\; \frac{(\a_i,\a_i)^{3-2n_f}}{g^8 L^3} \cdot
\frac{2^{13}\pi^2}{1+\z} \cdot (\til C_i)^2 \cdot
\til I(g^2, n_f) 
\end{align}
where  
\begin{align}\label{int3}
\tilde I(g^2, n_f) &=  
\int_0^\infty dz  \, \exp\left( +\frac{1}{g^2 z} 
- (4n_f-2) \log(z) \right) .
\end{align}
The main differences from the magnetic bion induced term are that: i) the neutral bion operator (\ref{type-two}) has no $\s$-dependence so contributes only to the effective potential for the gauge holonomy, $\f$; and ii) the sign of the Coulomb interaction term in the quasi-zero mode integral (\ref{int3}) changes.   Note that since $(\a_i,\a_i)>0$ the overall sign of the prefactors of $\til I$ in $\cA_{ii}$ is positive, just as in the magnetic bion case.

But an apparent problem is that the quasi-zero mode integral (\ref{int3}) is badly divergent at small $z$.  Even worse, the small-separation region, $z \ll 1/g^2$ (or  $r \ll r_{\rm b}$ in physical units), which dominates the integral is the region where the effective monopole--anti-monopole interaction (\ref{mmbar}) is actually incorrect as it becomes strong and there are other strong  corrections which we cannot control.  Therefore, in this regime the notion of a $[\cM_i \bar \cM_i]$ molecular configuration seems meaningless.

A second, apparently unrelated, problem is that since the $[\cM_i \bar \cM_i]$ configuration has both vanishing magnetic and topological charges, $\m=\n=0$, it is indistinguishable from the perturbative vacuum.  This raises the question of whether a well-defined semi-classical expansion even exists in this sector.  In particular, the perturbative $U(1)_\s^r$ symmetry mentioned in section \ref{sec2.6} prohibits the appearance of magnetic bion-like operators $\cB_{ij} \sim e^{2\pi i \s(\m)}$ which violate magnetic charge conservation, but not neutral bion-like ones.  Indeed, the neutral bion operator (\ref{type-two}) induces a potential for $\f$ qualitatively similar to the perturbatively induced potential (\ref{vsum}).

We claim that these two problems are, in fact, intimately related and are related to the large-order behavior and IR divergences of gauge theory perturbation theory.  Understanding these relations leads to a quantitatively precise definition of the neutral  bion contribution to the semi-classical expansion.

These problems we are encountering with neutral bions are not new;  in fact, the analog of this field theory obstacle has already been met and understood in quantum mechanics  \cite{Bogomolny:1980ur, ZinnJustin:1981dx}.  But a generalization to general field theories has not yet been achieved, and this is a necessary step to make sense out of neutral bions and other neutral molecule configurations.  We will undertake this step below. 
 
The analog of the neutral bion problem was first discussed by Bogomolny \cite{Bogomolny:1980ur} for double-well quantum mechanics, and Zinn-Justin realized the relation of Bogomolny's prescription to the large-order behavior of perturbation theory and Borel summability \cite{ZinnJustin:1981dx}.  Because of the combined deep insights that these two authors brought to this problem, we will refer to their procedure as the {\it Bogomolny--Zinn-Justin (BZJ) prescription}.  The BZJ prescription was applied by Balitsky and Yung to supersymmetric quantum mechanics and a few supersymmetric field theories \cite{Balitsky:1985in, Balitsky:1986qn,Yung:1987zp}.

There are a few cases where the result of the BZJ prescription can be cross-checked by other reliable methods.  For example, for bosonic non-supersymmetric quantum mechanics Bogomolny and Zinn-Justin provided evidence for the correctness of this prescription by comparing the results with the WKB approximation.  Yung \cite{Yung:1987zp} evaluates the bosonic potential which is induced by a 4-d instanton--anti-instanton pair---unlike the superpotential which is induced by an instanton---directly using the BZJ prescription giving a result identical to the bosonic potential derived from the superpotential.  On $\R^3 \times S^1$ Poppitz and one of us (M.\"U) were able to provide a consistency check for the prescription for $\cN{=}1$ superYang-Mills \cite{Poppitz:2011wy}.

In what follows we will use the same prescription for non-supersymmetric quantum field theory.  Currently, we do not know how to cross-check our results with another technique.  It is desirable to find such an alternative technique, i.e., a generalization of the WKB approximation to the Hamiltonian formulation of gauge theory, or a new method.

\paragraph{The BZJ prescription:} 

Bogomolny proposes to do integrals over the quasi-zero modes of instanton--anti-instanton molecules as follows.  Deform the contour of integration over the complexified quasi-zero mode so that the instanton--anti-instanton interaction becomes repulsive.  Then evaluate the integral by using the steepest descent path exactly.  In practice this is equivalent to changing the sign of the coupling $g^2$ in the instanton--anti-instanton interaction.  This turns the attractive Coulomb force into a repulsive one.  One then calculates the resulting integral exactly, without any gaussian approximations as emphasized in section \ref{sec:zeromode}.  Finally, analytically continue the final result back to positive $g^2$.  We will describe Zinn-Justin's important insights in connection with large orders in perturbation theory and Borel resummation in the next subsection. 

Following this prescription, we modify $\til I(g^2, n_f) \to \til I(-g^2,n_f)$ so that the Coulomb interaction becomes repulsive and the integral converges.  Note that $\til I(-g^2,n_f) = I(g^2,n_f)$, the quasi-zero mode integral (\ref{int2}) that we already evaluated for the magnetic bion.  Next, we substitute $g^2\to -g^2$ giving \begin{align}\label{sign}
\til I(g^2,n_f) \to I(-g^2,n_f) = \left(- \frac{1}{g^2} \right)^{3-4n_f} \G(4n_f-3) = -I(g^2,n_f). 
\end{align}  
The last equality is only valid for integer $n_f$.  Thus the BZJ prescription makes the neutral bion quasi-zero mode integral the same as for the magnetic bion integral, but gives an overall relative sign between the magnetic and neutral bion amplitudes.

This predicted relative sign is physically relevant.  In the $n_f=1$ theory which is supersymmetric and for which no perturbative potential is generated, the effective potential for the $\f$ and $\s$ scalars are due to both the magnetic and neutral bion amplitudes $\cB_{ij}$ and $\cB_{ii}$.  The relative sign between these terms from the BZJ prescription accounts for the vanishing vacuum energy in the supersymmetric theory.  A more detailed comparison of our result for $n_f=1$ with the bosonic potential obtained through the superpotential in supersymmetric theory \cite{Davies:2000nw} shows that they coincide.

The power and importance of the BZJ prescription for our purposes is that it transcends supersymmetry.  It can be applied to non-supersymmetric theories, and it yields correct results for supersymmetric theories without recourse to supersymmetric selection rules and non-renormalization theorems.

\subsection{High orders in perturbation theory, Borel summation and neutral molecules}\label{sec5.4}

Bogomolny's directive to analytically continue quasi-zero mode integrals from negative to positive $g^2$ gives convergent answers when applied to instanton--anti-instanton pairs, but would render the already convergent integrals for instanton--instanton pairs divergent.  Zinn-Justin \cite{ZinnJustin:1981dx} gives a justification for applying Bogomolny's prescription only to instanton--anti-instanton pairs, and improves upon it when Bogomolny's analytic continuation gives complex (as opposed to real) answers which depend on the choice of path of analytic continuation in the complex $g^2$-plane.

As Zinn-Justin's argument depends on the structure of the high-order behavior of perturbation theory, let us review that briefly.  There are other equivalent descriptions of what we will outline below; for a review, see \cite{Beneke:1998ui}.

It is well known that in theories with degenerate minima perturbation theory gives an asymptotic expansion, and hence is divergent.  In such theories, the perturbation series (even after being regularized and renormalized properly) is not even Borel resummable.  There are cases in which perturbation series become Borel resummable if the expansion parameter in the sum is taken to be negative, $g^2<0$.  This occurs, for example, in double-well quantum mechanics.  Let us call the resulting Borel resummed series $\B_0(g^2)$.  We then define the perturbative sum as the analytic continuation of $\B_0(g^2)$ in the $g^2$ complex plane from negative coupling, $g^2<0$, to  the the positive real axis, $g^2>0$.  The fact that the original ($g^2>0$) series was not Borel resummable implies that the function $\B_0(g^2)$ has a branch point at $g^2=0$.  Upon analytically continuing from $g^2<0$ to the positive real axis $\B_0(g^2)$ develops an imaginary part whose sign is ambiguous, depending on whether one approaches the real axis from below or above,
\begin{equation}
\label{Borel}
\B_0(|g^2| \pm i \e)  = \Re\B_0(|g^2|) \pm i \Im\B_0(|g^2|)  
\end{equation}
where $ \Im\B_0(|g^2|) \sim \pi e^{-2 S_0}$, and is inherently non-perturbative.  Thus the Borel resummation prescription for perturbation theory, i) produces a two-fold ambiguous result, and ii) produces complex results for what should be real observables.

The Bogomolny prescription for the semi-classical expansion has similar problems:  for instanton--anti-instanton amplitudes it also induces a complex answer with a branch point at $g^2=0$.  This structure is to some extent shown for the neutral bion molecule in (\ref{sign}) for non-integer $n_f$.  Of course, for QCD(adj), $n_f$ is an integer, in which case the analytic continuation gives a real and unambiguous answer.  But this is an exception to a general rule: as we discuss in the next subsection, a branch point at $g^2=0$ is encountered for general neutral topological molecules so that an imaginary part with ambiguous sign is generated upon continuation to positive real $g^2$.  The size of this imaginary part is $\sim e^{-2S_0}$, just as in the Borel resummed perturbative series.

Zinn-Justin states that these two ambiguous imaginary contributions---one from the perturbative Borel resummation prescription and one from the semi-classical (non-perturbative) Bogomolny prescription for quasi-zero mode integration---cancel.
This can be checked explicitly in some quantum mechanical examples, but also makes sense on more general grounds:  both are contributions to the same physical quantity, so only their sum need be real and unambiguous.  So Zinn-Justin's prescription is that, for $g^2$ small and negative, we should calculate {\it both} the sum of the perturbation series and the relevant instanton--anti-instanton contributions, and perform an analytic continuation to positive $g^2$ for both quantities in the same way.  Therefore, from this point of view, Bogomolny's prescription is required for the consistency of the Borel resummation prescription.  

How do we decide to which topological defects this BZJ prescription should be applied?  In the double-well quantum mechanics example, instanton--anti-instanton amplitudes have vanishing topological charge and so can contribute to the same quantities as the perturbation series.  In more general quantum mechanical examples where there is only one topological quantum number, vanishing of the topological charge is a sufficient condition for selecting the appropriate topological defects to include in the BZJ prescription.  But in gauge theories on $\R^3\times S^1$ in a vacuum in which the gauge group is Higgsed $G\to U(1)^N$, the topological defects carry two types of quantum number, magnetic and topological charge $(\m,\n)$, instead of just a single topological charge (instanton number).  We have seen that the semi-classical expansion of QCD(adj) on $\R^3\times S^1$ is organized in powers of $e^{-S_0}$, the fugacity or diluteness of the monopole-instanton, and incorporates effects from topological defects of all different combinations of charges, e.g.,
\begin{list}{$\bullet$}{\itemsep=0pt \parsep=0pt \topsep=2pt}
\item $e^{-S_0}$: monopole-instantons with $\m\neq0$ and $\n\neq0$,
\item $e^{-2S_0}$: magnetic bions with $\m\neq0$ and $\n\approx0$,
\item $e^{-2S_0}$: neutral bions with $\m=0$ and $\n=0$,
\item $e^{-NS_0}$: 4-d instantons with $\m=0$ and $\n\neq0$,
\item $e^{-2NS_0}$: 4-d instanton--anti-instanton pairs with $\m=0$ and $\n=0$.
\end{list}
It only makes sense to combine a perturbation series around the vacuum with semi-classical contributions from topological defects, such as neutral bions or instanton--anti-instanton pairs, with \emph{all} topological charges vanishing, i.e., $(\m,\n)=(0,0)$.  So we propose the following slight sharpening of the Bogomolny-Zinn-Justin prescription which applies, in particular, to topological defects on $\R^3 \times S^1$.

\begin{quote}
{\bf Refined BZJ prescription}: For $g^2$ small and negative, one should calculate {\it both} the sum of the perturbation series and the sum of all neutral topological molecule and multi-instanton contributions with quantum numbers the same as those of perturbative vacuum, and perform an analytic continuation to positive $g^2$ of the sum of these two quantities. 
\end{quote}

Furthermore, we suggest a sectorial dynamics in gauge theory.  The imaginary part that arises from the analytic continuation of a perturbation series around the vacuum (\ref{Borel}) can never be related to a magnetic bion or any other object which has a non-vanishing topological charge, but can be cancelled by neutral molecular defects.  Likewise, the magnetic bion, $\cM_i$, which may have zero topological charge but has non-vanishing magnetic charge, already gave a sensible answer at positive $g^2$ by itself.  It gives the leading contribution to quantities in this topological charge sector.  There can be perturbative corrections to these quantities whose Borel resummation may give imaginary parts upon continuation which should be cancelled by higher-action topological defects in the same charge sector, such as $[\cM_i\cM_j\bar\cM_j]$ or more complicated molecules.

\subsection{High orders in perturbation theory and exotic topological molecules}
\label{etm}

The key point of the above discussion was that, based on general arguments about perturbation theory for theories with degenerate minima, one expects the contribution of neutral molecules to be complex so that they will cancel the imaginary part of Borel resummed perturbation theory.  But the amplitude that we obtained for a neutral bion through the BZJ prescription, $I(-g^2) \sim (-1/g^2)^{3-4n_f}$, is real for integer $n_f$ and complex otherwise.  And, of course, non-integer $n_f$ is unphysical.  This is not a contradiction as long as the imaginary part of the Borel resummed perturbation series is of order $e^{-4S_0}$ or smaller so that they can be cancelled by neutral topological molecules at higher order in the semi-classical expansion.

A study of various examples shows a connection between whether or not a given type of neutral topological molecule induces an imaginary part through the BZJ prescription and the occurrence of fermion zero modes in its constituent topological defects.  The following pattern holds for all quantum mechanical and quantum field theories we have examined, although we state our observations in a language appropriate for gauge theories on $\R^3\times S^1$. 

\begin{list}{$\bullet$}{\itemsep=0pt \parsep=0pt \topsep=2pt}
\item[\bf 1.] 
In purely bosonic theories with topological defects (instantons, monopole-instantons, etc.), the topologically neutral molecules induce an imaginary part proportional to the 2-defect fugacity, $\pm e^{-2S_0}$.  
\item[\bf 2.] 
If the theory has fermions, there are two cases depending on whether a given defect has a fermionic zero mode or not. 
\begin{list}{$\bullet$}{\itemsep=0pt \parsep=0pt \topsep=2pt}
\item[\bf a.] 
If it has a zero mode, the associated topologically neutral defect--anti-defect molecule does not induce an imaginary part for integer number of fermion flavors.\footnote{Ref.~\cite{Balitsky:1985in} has an example which at first sight seems to contradict to this claim.  They deform the Yukawa term in supersymmetric quantum mechanics into $p W'' \bar\psi\psi$ where $W$ is the superpotential and the theory is supersymmetric for $p=1$, and they find that the quasi-zero mode integral is proportional to $(-1)^p$.  However, one can show rigorously that this system describes the ground state properties of a multi-fermion flavor (non-supersymmetric) quantum mechanics where $p$ acquires an interpretation as $n_f$.}
\item[\bf b.] 
If it does not have a zero mode, then its associated topologically neutral molecule will induce an imaginary part as in case 1. 
\end{list}
\item[\bf 3.]
If the theory has fermions, and if all defects have fermionic zero modes, then there will be topologically non-neutral molecular events without any zero modes, which we can call 2-defects.  Then there are topologically neutral molecules made out of these 2-defects as in case 2b which induce an imaginary part as in case 1, but now proportional to $\pm e^{-4S_0}$. 
\item[\bf 4.]
Cases 1 and 3 generalize to higher molecules, with induced imaginary parts $e^{-2nS_0}$, $n=1,2,\ldots$ and  $e^{-4nS_0}$, $n=1,2,\ldots$, respectively. 
\end{list}

Examples of some of these cases are: the 3-d Polyakov model for case 1, where the defects are monopole-instantons; and QCD(adj) on $\R^3\times S^1$ for case 2a, where the defects are again monopole-instantons.  We can illustrate cases 3 (and 2b) in QCD(adj) by considering a neutral molecule composed of two magnetic bions.
Denote a magnetic bion by $\cB_{ij} = [\cM_i\bar\cM_j]$.  Then at 4th order in the semi-classical expansion there can be amplitudes of the form 
\begin{align}\label{bibion}
[\cB_{ij} \cB_{ji}] &:= [\cB\bar\cB],
\qquad\text{and}\qquad
[\cB_{ij} \cB_{ij}] := [\cB\cB],
\end{align}
both giving contributions $\sim e^{-4S_0}$.  Since the bions have no fermion zero modes the associated amplitudes only involve bosonic fields.  These are permitted by the symmetries of the effective Lagrangian and there is no reason for them not to be generated.  Note, however, that for $i\neq j$ the $[\cB\cB]$ configuration is not magnetically (or topologically) neutral while $[\cB\bar\cB]$ always is.  Thus these will contribute to different ``charge sectors" in the sense of the discussion at the end of section \ref{sec5.4}.

(We focus on the two 4th-order configurations in (\ref{bibion}) just for illustrative purposes.  There are more general molecules at 4th order, such as $[\cB_{ij} \cB_{kl}]$ with all indices different.  Note that if there is no interaction between, say, $\cB_{ij}$ and $\cB_{kl}$, as determined by the inner product of their associated root vectors, they cannot form correlated molecular instanton events.   The following discussion of the quasi-zero mode integrals can in principle be generalized to arbitrary topological molecules.) 

According to our general discussion in the previous section, the integral over the quasi-zero modes  between these molecules should not yield an imaginary part for $[\cB\cB]$ and should yield an imaginary part for $[\cB\bar\cB]$.  The quasi-zero mode integrals are of the form 
\begin{align}
&\label{bb}   
I (g^2) =  \int d^3r\, \exp\left(- V(r)  \right)    
\qquad \text{for $[\cB\cB]$, and}\\
\label{bbbar}  
& \til I(g^2) = \int d^3r\, \exp\left( +V(r) \right) 
\qquad \text{for $[\cB\bar\cB]$},
\end{align}
where 
\begin{align}\label{bbpot}
V(r) = (\m_{\cB} , \m_{\cB}) \frac{2 \pi}{g^2} \frac{L}{r} 
\end{align}
and $\m_\cB = \a^\v_i - \a^\v_j$ is the magnetic charge of the magnetic bion $\cB_{ij}$.  There are two problems with these integrals: first, both integrals diverge at large separation; and second, the $[\cB\bar\cB]$ integral diverges at small $r$.

The first problem appears for bosonic molecules because the integrals are no longer cut off by fermion zero mode exchange.  Such an effect is also seen in quantum mechanics by Bogomolny \cite{Bogomolny:1980ur}, who instructs us that if the separation between pairs is asymptotically large, we should count them as independent (uncorrelated) events, not as composites.  In our case, if the bions are distant, their effects are already accounted for in the dilute plasma of bions.  Therefore, we should subtract the large-separation divergence to prevent double counting.  In fact, we have already calculated in (\ref{int2}) the integral for general $n_f$.  All we need to do is to take the $n_f =\e\to 0$ limit in a meaningful way.  

For the $[\cB\cB]$ integral we have
\begin{align}
I(g^2) = 4\pi\left[ (\m_\cB,\m_\cB)2\pi L  \right]^3   I(g^2,\e)
\equiv C\, I(g^2,\e) 
\end{align}
where we recall that
\begin{align}\label{intep} 
I(g^2,\e) 
= \int_0^\infty dz\, \exp\left(-\frac{1}{g^2 z}-(4\e-2)\ln z\right) 
= g^{8\e-6}\, \G(4\e-3).
\end{align} 
Expanding around the pole at $\e=0$, we obtain
\begin{equation}
g^6 I (g^2, \e) 
= g^{-8\e} \G(4\e-3)
= - \frac{1}{24\e}
+ \frac{1}{6} \left[\ln(g^2) + \g - \frac{11}{6} \right] + O(\e).
\end{equation}
Our subtraction scheme, which gets rid of the double counting of independent bion events, is to drop the $1/\e$ pole term, and leads to 
\begin{align}
I (g^2) = \frac{C}{6}  \left(\frac{1}{g^2}\right)^3  
\left[\ln(g^2) + \g -\frac{11}{6}\right] ,
\end{align}
a real and finite answer.

Now consider the $[\cB\bar\cB]$ case.  Since the constituents of the molecule are attractive at short distances and the composite is topologically neutral, we have to follow the BZJ prescription.  Hence, as a first step, we take $g^2 \to -g^2$, leading to $\til I(g^2) \to \til I(-g^2) =  I(g^2)$.  Now the interaction is repulsive at short distances, and the resulting integral is the one we just did above.  Finally, we have to continue back to positive $g^2$ in $\til I(g^2)$ which gives 
\begin{align}\label{BBbarcont}
\til I (g^2) 
&= - \frac{C}{6} \left(\frac{1}{g^2}\right)^3  
\left[\ln(-g^2) +\g -\frac{11}{6} \right]   
= - I(g^2) \mp i \pi \frac{C}{6}
\left(\frac{1}{g^2}\right)^3 .
\end{align}
Thus the BZJ prescription gives an imaginary part to the $[\cB\bar\cB]$ amplitude of the form $\pm i \pi e^{-4S_0}$.  The sign ambiguity arises because the logarithm is multi-valued.  

Since it is topologically neutral, the $[\cB\bar\cB]$ amplitude gives a contribution to the vacuum energy density (times the circumference $L$) of the theory.  Previously we have argued that the Borel resummation and analytic continuation prescription for perturbation theory gives a result, $\B_0 (g^2)$, which also has an imaginary part of ambiguous sign.  This result, therefore, is meaningless by itself, because the vacuum energy density is real.  Let us write $g^2 = |g^2| e^{i \th}$, where $\th$ is the phase of the complexified coupling.  The imaginary parts on the two sides must cancel in order for the theory to make sense, 
\begin{align}\label{borelbion}
\Im\B_{0, \th=0^\pm}  
+ \Im [\cB\bar \cB]_{\th=0^\pm}
=0.
\end{align}
As $\th$ goes from $0^-$ to $0^+$, there is a ``jump" in $\B_{0, \th}$.  The interesting thing is that the $[\cB\bar\cB]_\th$ amplitude also undergoes a similar jump, in the opposite direction, so that the physical observable, which ought to be real, remains real as $\th \to 0$.  From our calculation of $\Im[\cB\bar\cB]$ above, this implies
\begin{align}\label{borelbion2}
\Im \B_{0, \th=0^\pm} 
\pm \pi \frac{C}{6} \left(\frac{1}{g^2}\right)^3  
\cA_{ij}^2  e^{-4 S_0} =0,
\end{align} 
where the prefactor of the magnetic bion amplitude, $\cA_{ij}$, is calculated in (\ref{bionamp}) and $S_0$ is the typical size of the monopole-instanton action.  Recall that the monopole-instanton action actually depends on its magnetic charge as shown for example in table \ref{tab4.1} in section \ref{sec4.1}.  To keep the discussion simple, we will just use the average monopole-instanton action $S_0 = S_I / h^\v$ where $S_I = 8\pi^2/g^2$ is the 4-d instanton action and $h^\v$ is the dual Coxeter number of the gauge group; for $SU(N)$, $h^\v = N$.

We thus get a prediction for the size of the imaginary part of the Borel resummed perturbation series, which in turn determines the size of the large-order terms in the original perturbation series.  This prediction could, in principle, be checked by studying infinite sequences of Feynman diagrams to give estimates of the size of large-order terms in perturbation theory.  The large-order behavior of the perturbation series determines the location of the singularities (branch points) of the Borel transform of the series.  Recall that the Borel transform of a perturbative series, $
G(g^2) = \sum_{n=0}^\infty a_n g^{2n}$,
 is $B G(t) = \sum_{n=0}^\infty (a_n/n!) t^n$, and the Borel resummation of $G$ is 
\begin{equation} 
\B(g^2) = \int_0^\infty B G(t g^2) e^{-t} dt \; .
\end{equation}
  The complex $t$-plane is called the Borel plane.  The Borel transform has singularities at values of $t$ corresponding to $g^2$ times the action of classical Euclidean topologically neutral solutions, and can have singularities at other places as well.  A Borel-plane singularity at positive real $t=t_0$ contributes to a branch point in $\B(g^2)$ at the origin with a resulting branch cut along the positive real $g^2$ axis across which $\Im\B$ is discontinuous by $\exp\{-t_0/g^2\}$ (typically times some analytic function of $g^2$).  See \cite{Weinbergv2} section 20.7 and \cite{'tHooft:1977am} for lucid explanations of these facts.

For $SU(N)$ gauge theory, for example, since $S_0 = S_I/N = 8\pi^2 /(N g^2)$, the $e^{-4S_0}$ term in (\ref{borelbion2}) implies a singularity in the Borel plane at $t=32\pi^2/N$.  By contrast, a 4-d instanton--anti-instanton configuration has action $S_{I{-}\bar I} = 2 S_I = 2N S_0$, and so gives a Borel plane singularity $2N$ times further from the origin.  It should be noted that this prediction of the position of the Borel-plane singularity from (\ref{borelbion2}) only reflects the cancellation of the leading imaginary part of Borel resummed perturbation theory.  Sub-leading ambiguities in perturbation theory must cancel with neutral topological molecules with higher action.

We also note that in a bosonic center-symmetric theory on small $S^1 \times\R^3$ (e.g., pure Yang-Mills appropriately deformed by holonomy double trace operators), the counterpart of the above cancellation occurs at order $e^{-2 S_0}$, and the counterpart of the relation (\ref{borelbion}) reads 
\begin{align}
\label{borelmon}
\Im\B_{0, \th=0^{\pm} }  +   \Im [\cM_{i}  \bar \cM_{i} ]_{\th=0^{\pm}}=0 .
\end{align}
We comment on the implications of this in the next subsection. 

\subsection{Neutral bions as the semi-classical realization of renormalons?}

We now argue that the neutral bion molecules discussed above are intimately related to 't Hooft's renormalons on $\R^4$. They are, very plausibly, their weak coupling incarnation in a sense we will make precise.  We will illustrate our arguments just using $SU(N)$ QCD(adj) for simplicity.

\begin{figure}[ht]
\begin{center}
\includegraphics[angle=0, width=4.0in]{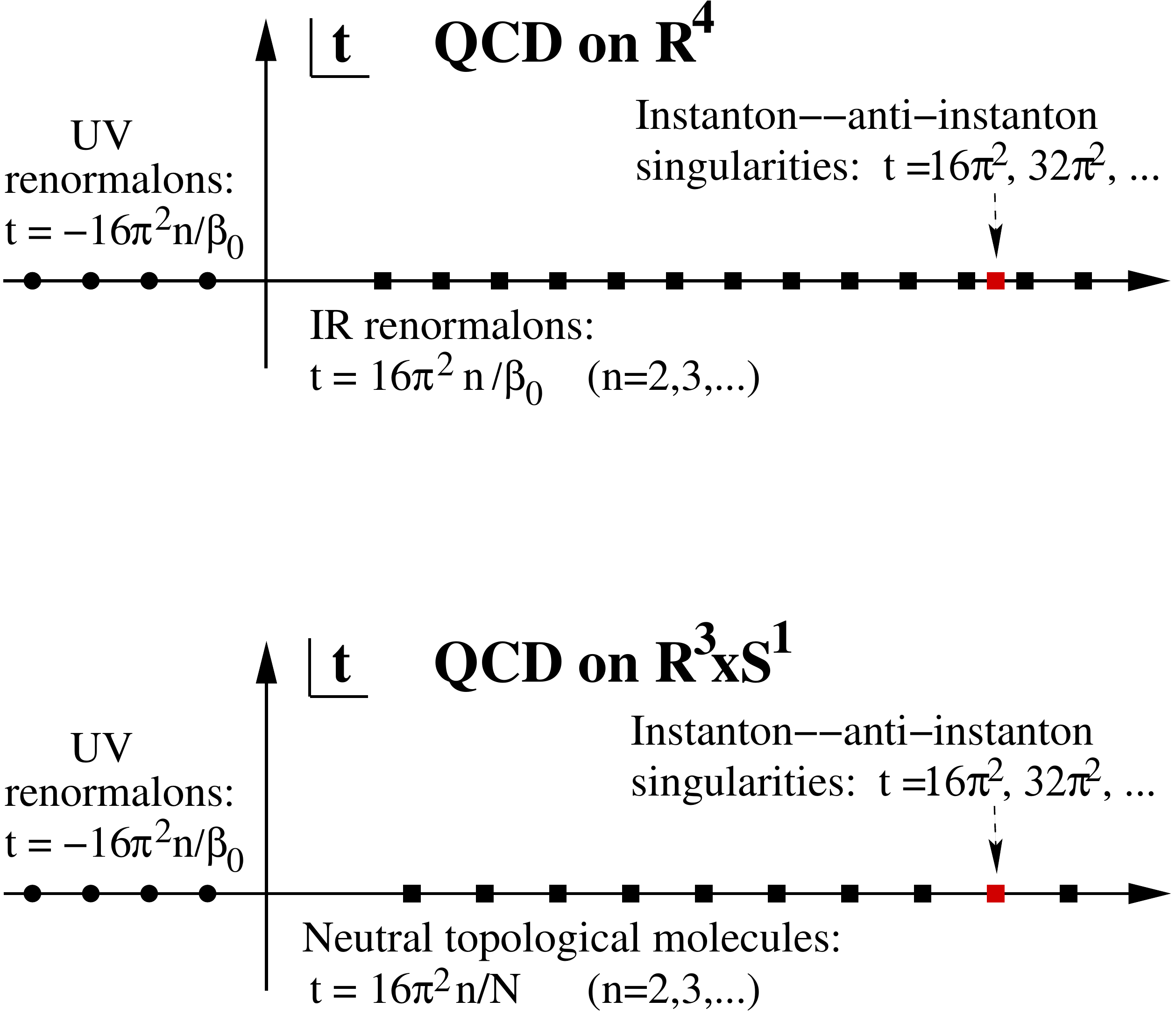}
\caption{Upper figure: The  conjectural structure of the Borel plane for  a QCD-like gauge theory on $\R^4$.  Lower figure: The  semi-classical poles associated with QCD(adj) with massless fermions
on small $S^1\times\R^3$.}\label {fig:Borel}
\end{center}
\end{figure}

Let us review the (conjectural) distribution of Borel plane singularities for QCD-like theories on $\R^4$, shown in the upper figure in fig.~\ref{fig:Borel}.  4-d instanton--anti-instanton molecules are known to produce singularities at \cite{Bogomolny:1977ty}
\begin{equation}\label{R4singsIIbar}
t_{\R^4} = n S_{I{-}\bar I}g^2  = 2n S_Ig^2= 16\pi^2 n,
\qquad n\in \Z^+.
\end{equation}
These give the leading Borel-plane singularities (i.e., those closest to the origin on the positive real axis) associated to semi-classical configurations.  But in renormalizable asymptotically free gauge theories, the large-order behavior of perturbation theory seems to be dominated by what are called renormalon divergences \cite{'tHooft:1977am} which are associated to singularities closer to the origin of the Borel plane.  For example, for $SU(N)$ QCD(adj) on $\R^4$ the IR renormalon singularities are at
\begin{equation}\label{R4singsIRrenorm}
t_{\R^4} = \frac{16\pi^2}{\b_0} n
= \frac{48\pi^2}{N(11-2n_f)} n
\qquad n=2,3, \ldots \ , 
\end{equation}
which are closer to the origin by a factor of order $N$.  

The Borel plane IR renormalon singularities are associated with divergent sub-series in perturbation theory whose terms get their main contribution from processes at the strong-coupling scale of the theory, therefore at a much lower energy than the cut-off scale.  They render the theory non-Borel summable.  They induce a branch cut on the positive $g^2$ axis and associated imaginary parts $\Im\B_0(g^2) \sim \pm \exp\{-16\pi^2 n/\b_0\}$, just like the sub-leading singularities (\ref{R4singsIIbar}) induce small imaginary parts $\Im\B_0 \sim \pm \exp\{-16\pi^2 n \}$.  But a crucial difference between the two is that there are semi-classical solutions with action $16\pi^2 n/g^2$ (namely, $n$ instanton--anti-instanton pairs) whereas there are no semi-classical solutions with action $16\pi^2 n/(\b_0 g^2)$.  This means that the BZJ prescription can be used to cancel the ambiguous imaginary parts associated with the former, but no precise prescription is known to cancel the IR renormalon divergences.

By contrast, our analysis of gauge theory on $\R^3 \times S^1$ explicitly demonstrates that there are semi-classically calculable poles in the Borel plane much closer to the origin than the 4-d instanton--anti-instanton poles.  Perturbation theory mixes with molecules such as $[\cB_{ij}\cB_{ji}]$, $[ \cB_{ij} \cB_{jk}\cB_{ki}]$  and related molecules with action $S= n S_{I{-}\bar I}/N$, $n=2, 3, \ldots$ in QCD(adj) and correspond to Borel-plane singularities at
\begin{eqnarray}\label{R3S1singsQCDadj}
t_{\R^3 \times S^1} = \frac{16\pi^2}{N} n, \qquad n=2,3, \ldots, 
\;\; \text{for massless or small-mass QCD(adj).} 
\end{eqnarray}
The resulting distribution of singularities in the Borel plane is shown in the lower figure of figure \ref{fig:Borel}.   This picture of the Borel plane for gauge theories on small $S^1\times\R^3$ is new and is a result of our semi-classical analysis together with the BZJ prescription.

Also, from our discussion of exotic topological molecules for general theories in the last subsection, we can easily extend this picture to other gauge theories on small $S^1\times\R^3$ whose gauge group Higgses to abelian factors.  For example, we have already seen that the neutral bion molecule $[\cM_i\bar\cM_i]$ also has the same quantum numbers as the perturbative vacuum, but does not induce an imaginary part in the BZJ prescription for theories with massless fermions.  (More precisely, this type of molecule does not produce an imaginary part provided that each $\cM_i$ has fermionic zero modes.  In QCD(adj), this is always the case, whereas in QCD with fundamental fermions or in pure Yang-Mills this is not the case.  The situation for general matter representations is controlled by an index theorem \cite{Nye:2000eg, Poppitz:2008hr}.)  But, in a bosonic theory such as trace-deformed Yang-Mills (dYM) or in a theory in which fermions are heavy, the leading pole in the Borel plane is due to the mixing of perturbation theory with $[\cM_i\bar\cM_i]$ and related molecules in the semi-classical domain with action $S=n S_{I{-}\bar I}/N$.  These correspond to Borel-plane singularities at
\begin{eqnarray}
t_{\R^3 \times S^1} = \frac{16\pi^2}{N} n, 
\ \ n\in\Z^+  
\qquad \text {for large-mass QCD(adj) or dYM.}
\end{eqnarray}
These are again more relevant than 4-d BPST instanton--anti-instantons molecules.   They are twice as dense as the singularities (\ref{R3S1singsQCDadj}) of massless QCD(adj) theory on $\R^3\times S^1$.  

Clearly, the singularities in the Borel plane associated with these neutral gauge theory molecules on small $S^1\times \R^3$ are of the same order as the elusive renormalons on $\R^4$ in the sense of counting powers of $N$, the rank of the gauge group.  We conjecture that the neutral bion and related molecules are the weak coupling incarnation of IR renormalons.

Starting with an asymptotically free theory on $\R^4$ with IR renormalons, if we compactify it on $\R^3 \times S^1$ in such a way as to avoid phase transitions as the circle shrinks,\footnote{I.e.,  prevent center symmetry changing phase transitions by judicious choice of boundary conditions as in QCD(adj), or by using double-trace deformations as in dYM.} then we expect the positions of the Borel plane singularities to change continuously with radius.  When the radius of the circle, $L$, is much larger than the strong-coupling length scale, $\L^{-1}$, the location of the renormalon singularities will be independent of radius, and universal for a given theory.  When the theory reaches the semi-classical domain where $L\ll \L^{-1}$, by continuity the renormalon singularities must acquire a semi-classical interpretation.

For asymptotically free theories on $\R^4$, IR renormalons arise from processes which get a large contribution from energies (momentum transfers) of order the strong coupling scale, $\L$.  In these theories this is precisely the regime where perturbative and semi-classical methods break down.  By contrast, in the small $S^1\times \R^3$ limit in theories like QCD(adj) in which the gauge group is Higgsed to abelian factors at a scale well above $\L$, the IR dynamics is weakly coupled.  Thus, it must be possible to describe the remnant of the IR renormalon in these theories by semi-classical physics on $\R^3\times S^1$. 

The expectation that the positions of the Borel plane singularities will change continuously for our class of theories as the radius of compactification is changed is supported in the large-$N$ limit by the fact that these theories exhibit volume independence \cite{Kovtun:2007py} in the $N=\infty$ limit.  Large-$N$ volume independence (also called Eguchi-Kawai reduction) states that perturbation theory on a compact space, provided the theory does not break its center symmetry spontaneously, reproduces perturbation theory in infinite volume as $N\to\infty$.  Heuristically, in these theories it is $LN$ and not $L$ which provides the effective compactification volume.  Therefore, in the large-$N$ limit, both UV and IR renormalon singularities must be present on small $S^1\times \R^3$.

This identification of topologically neutral semi-classical configurations on $\R^3\times S^1$ as the origin of IR renormalons on $\R^4$ gives a new perspective on some old problems.  It suggests that for the class of field theories we are considering, it may be possible to give a complete non-perturbative continuum definition of the field theory, at least in the semi-classical domain, on $\R^3 \times S^1$.  

Furthermore, it suggests that by studying the large-order behavior of perturbation series for compactified center-symmetric theories, it may be possible to understand the IR renormalons of the theory on $\R^4$, i.e., to understand the 4-d prescription for how to remove the ambiguity in the Borel resummed perturbation series that IR renormalons induce.  This is an important issue which we leave for future work. 

Finally, we are led to a sharpening of an old conjecture.  Some time ago, 't Hooft speculated that IR renormalons may be related to the quark confinement mechanism; see for example \cite{'tHooft:1977am}.  In the semi-classical domain on $\R^3 \times S^1$ where confinement and mass gap are calculable,  't Hooft's speculation is not correct, but it is very close to being  correct.  For example, in QCD(adj) it is not the neutral $[\cB_{ij}\cB_{ji}]$ molecule (which is the realization of the IR renormalon in the semi-classical domain), but the proliferation of their constituent magnetic bions, $\cB_{ij}$, which leads to quark confinement \cite{Unsal:2007jx}. (This will be reviewed in section \ref{sec6} below).  Analogously, in QCD(adj) with heavy fermions or in deformed Yang-Mills theory, the realization of IR renormalons is $[\cM_i \bar\cM_i]$ while confinement is generated by the proliferation of monopole-instantons $\cM_i$ in the semi-classical domain.

\section{Effects of the neutral bion-induced potential}\label{sec7}

The bion operators, entering at second order in the semi-classical expansion (\ref{scexp}), give purely bosonic potential terms for the $\f$ and $\s$ scalars in the interior of the gauge cell.  Thus the leading terms in the 3-d bosonic effective lagrangian are
\begin{align}\label{Lbos}
\cL_\text{bosonic} = 
\tfrac{g^2}{4 L} (\del_m\s,\del_m\s) 
+ \tfrac{4\pi^2}{g^2L} \left( \del_m\f \,,\del_m\f \right) 
 + V_\text{pert.}(\f) +  V_\text{n.p.}(\f, \s).
\end{align}
Here $V_\text{pert}$ is the one-loop perturbative potential given by (\ref{veffad}) plus higher-order corrections in perturbation theory.  To all orders in perturbation theory, it has no $\s$-dependence, and the minimum in the gauge cell for $\f$ is given by the minima for the one-loop potential described in section \ref{sec2} up to small corrections which do not move the minimum off a gauge cell wall if it is there at one loop.
$V_\text{n.p.}$ is the semi-classical non-perturbative potential induced by a dilute gas of bion defects, and, from (\ref{bion0}), (\ref{bion1}), (\ref{bionamp}), and the BZJ prescription sign (\ref{sign}), is given by
\begin{align}\label{bion-np-pot}
V_{\rm n.p.} (\f,\s) 
&= \sum_i |\cA_{ii}| e^{-2 S_i(\f)} 
\ - \sum_{\{i,j | (\a_i,\a_j) < 0 \}} |\cA_{ij}|  
e^{-S_i(\f)- S_j(\f)} e^{2\pi i\s(\a_i^\v-\a_j^\v)}.
\end{align}
The positive terms in this sum come from neutral bions while the negative ones are from magnetic bions.  The magnetic bion terms induce a potential for $\s$, which will be discussed in the next section (along with the fermionic terms).

In this section we concentrate on the effect of the bion potential terms for $\f$, ignoring their $\s$-dependence.  The expectation is that in the semi-classical regime where this potential is calculable, the perturbative potential will control the location of the minimum and the non-perturbative terms will only give small corrections.  But there are theories where this is not the case.  

QCD(adj) with $n_f=1$ massless fermion has $V_\text{pert}\equiv 0$ to all orders in perturbation theory (enforced by the $\cN{=}1$ supersymmetry this model has).  The non-perturbative potential then gives the leading effect.  The bion-induced potential exactly reproduces the one derived from the $\cN{=}1$ superYang-Mills superpotential.  This potential has been analyzed in \cite{Davies:2000nw} who show that it is minimized at the geometric ``center" of the gauge cell, namely the point where $\ba_j(\f) = 1/h^\v$ for all $j=0,\ldots,r$.  

One can also deform this theory, breaking the supersymmetry slightly, by adding a bare fermion mass, $m$.  This generates a perturbative potential whose size is proportional to $m^2$.  In the small-$m$ limit, the vacuum is determined by a competition between the perturbative and non-perturbative contributions to the potential. 

We have seen in section \ref{sec2} that for massless QCD(adj) with $n_f>1$ the perturbative potential is not, in general, minimized at the center of the gauge cell (except for $SU(N)$ gauge group); see table \ref{tab4.1}.  Furthermore, except for $SU(N)$ and $Sp(2N)$ gauge groups, the minimum is on a boundary of the gauge cell, implying that the effective 3-d theory perturbative vacuum has a few unbroken non-abelian (typically $SU(2)$) gauge factors.  Since an arbitrarily small shift of the minimum off the gauge cell wall would lead to a qualitative change in the low energy behavior of the theory, it is interesting to ask whether in these non-abelian cases the non-perturbative bion-induced potential can abelianize the theory by shifting the minimum slightly.

In section \ref{sec2.6} we showed that in the cut-off theory analyticity in the background field $\f$ of the effective potential keeps higher-order perturbative effects from moving a minimum off the gauge cell wall.  
But in the semi-classical expansion, non-analytic dependence on $\f$ is introduced by the integral over the monopole-instanton collective coordinates and fluctuation determinants, as reviewed in section \ref{sec:B}.  In particular, the $\f$-dependence of the bion terms in (\ref{bion-np-pot}) is of the general form
\begin{align}\label{}
V_\text{n.p.} \sim \pm\, \ba_j(\f)^{-p}\, e^{-S_I \ba_k(\f)}
\end{align}
where $S_I = 8\pi^2/g^2$ is the 4-d BPST instanton action, and $p$ is some positive constant.  Recalling that in the gauge cell $\ba_j(\f)\ge 0$ and that the cell walls are where one or more $\ba_j(\f)=0$, we see that these terms, though suppressed by the $S_I$ in the exponent, nevertheless diverge at the cell walls.  In particular, the neutral bion terms which come with the positive sign give rise to a potential which is strongly repulsive from the cell walls.

Of course, the calculation of the bion-induced effective potential is not valid precisely at the gauge cell walls where it diverges: the semi-classical expansion breaks down since topologically-protected monopole-instanton solutions do not exist when the effective gauge group is non-abelian, and there are presumably no semi-classical saddle point solutions either.  Thus we look for a \emph{self-consistent} minimum of $V=V_\text{pert} + V_\text{n.p.}$.  This means that the minimum must be at $\f=\f_\text{min}$ such that
\begin{list}{$\bullet$}{\itemsep=0pt \parsep=0pt \topsep=2pt}
\item $\f_\text{min}$ is not at a gauge cell wall, so that $V_\text{n.p.}$ is well-defined, and
\item the value of the $\f$ mass, $m_\f$, and the W-boson mass, $m_W$, satisfy $m_\f < m_W$, so that there can exist an intermediate scale $m_\f<\m<m_W$ at which to define an effective  abelian theory where the W-bosons are integrated out, but the $\f$ fields are light.
\end{list}
The $\f$ mass is determined by the curvature of the potential at the minimum, $L^2m^2_\f \sim L^3V''(\f_\text{min})$, while the W-boson associated with root $\a$ has mass $Lm_W \sim \ba(\f_\text{min})$ by (\ref{Wmass3}), and is thus determined by the distance of $\f_\text{min}$ from the cell walls.

But, it is not too difficult to show that no such self-consistent abelianizing minimum exists, at least near to a gauge cell wall.  It is enough to keep track of the powers of the coupling, of the size of the gauge group, and of the number of fermion flavors to understand the behavior of the potential near a cell wall.  So define
\begin{align}\label{}
N := \text{rank}(G), \qquad
\l := \frac{N g^2}{8\pi^2} , \qquad
\n := \frac{2}{3}(n_f-1) .
\end{align}
Thus $\l$ is the 't Hooft coupling, and the large-$N$ limit should be taken holding $\l$ fixed.  Also, note that $N/\l = 8\pi^2/g^2$ is the BPST instanton action, and that $1/\l$ is approximately the monopole-instanton action for vacua not near any cell walls (where $\a(\f)\sim 1/N$).  But near the $\a$-wall, $\a(\f)\ll 1/N$, and the monopole instanton action is $\sim N\a(\f)/\l$.  Thus the dilute monopole-instanton gas approximation requires $\l\ll N\a(\f)$.

We are interested in the component of $\f$ perpendicular to the cell wall where the perturbative minimum is located.  This is $\f_\perp := \a(\f)$ where $\a$ is the simple root associated to that wall.  In particular, $\f_\perp\ge 0$ to be in the gauge cell, and vanishes at the cell wall.  The other components of $\f$ parallel to the cell wall, $\f_{||} \sim \b(\f)$ for other roots $\b$, must also be positive to be in the gauge cell.  Recall that the geometric center of the gauge cell is at $\a_j(\f) = 1/h^\v \sim 1/N$ for all $\a_j$.  Thus for the minimum of $V$ to be close to the original cell wall we must have
\begin{align}\label{}
0\le \f_\perp \ll 1/N \quad\text{and}\quad \f_{||} \sim 1/N .
\end{align}

The form of the leading quadratic part of perturbative potential is
\begin{align}\label{}
L^3 V_\text{pert} \sim \tfrac{1}{N}\f_\perp^2
+ \tfrac{1}{N}(\f_{||} - \tfrac{1}{N})^2, 
\end{align}
which tends to drive $\f_\perp \to 0$ and $\f_{||}\to 1/N$.
Its normalization corresponds to the perturbative value of the $\f$ mass, $L^2 m^2_{\f-\text{pert}} \sim \l/N^2$, found in sec 3.  (A factor of $g^2 \sim \l/N$ comes from canonically normalizing $\f$ as in (\ref{Lbos}).)

The form of a magnetic bion-induced potential term which involves $\f_\perp$ is (neglecting the $\s$-dependence)
\begin{align}\label{vmb}
L^3V_\text{mag. bion} &\sim  - (\l/N)^{6\n-3}  
\f_{||}^{-2\n} \f_\perp^{-2\n} 
e^{-N\f_{||}/\l} e^{-N\f_\perp/\l} .
\end{align}
The negative sign means it tends to push $\f_\perp\to0$ with an inverse power at short ranges $\f_\perp \lesssim \l/N$ due to the exponential cut off.  The prefactor and the exponential dependence mean that this term is very small compared to the perturbative potential for points in the interior of the gauge cell ($\f \sim 1/N$).  This justifies looking only near the cell wall for a minimum, and justifies neglecting the fluctuations of $\f_{||}$.  Thus (\ref{vmb}) becomes
\begin{align}\label{vmb2}
L^3V_\text{mag. bion} &\sim  - (\l/N)^{6\n-3} 
N^{2\n} \f_\perp^{-2\n} 
e^{-1/\l} e^{-N\f_\perp/\l} .
\end{align}

The form of the neutral bion-induced potential term for $\f_\perp$ is
\begin{align}\label{vnb}
L^3V_\text{neut. bion} &\sim  + (\l/N)^{6\n-3}  
\f_\perp^{-4\n} e^{-2N\f_\perp/\l} .
\end{align}
Its positive sign means it tends to push $\f_\perp$ away from the gauge cell wall.   Even though it has a shorter range than the magnetic bion term, it has a stronger power-law behavior, so dominates in most of the gauge cell.  Indeed, the magnetic bion term only becomes comparable to the neutral bion term for $\f_\perp \sim 1/N$ which is near the center of the gauge cell.  Thus we can safely ignore the magnetic bion terms.

The resulting potential with just the leading perturbative and neutral bion terms is
\begin{align}\label{pnpot}
L^3V
&= (1/N) \f_\perp^2 + (\l/N)^{6\n-3} 
\f_\perp^{-4\n} e^{-2N\f_\perp/\l} .
\end{align}
This always has a minimum for positive $\f_\perp$.  The only question is whether it simultaneously satisfies
\begin{align}\label{}
\l\ll N\f_\perp &\quad\text{(dilute monopole-instanton gas approximation)},
\label{con-dig}\\
0<\f_\perp\lesssim 1/N &\quad\text{(minimum inside gauge cell)},
\label{con-cell}\\
2\le n_f \le5 &\quad\text{(non-vanishing $V_\text{pert.}$ 
and asymptotic freedom)},\\
m_\f \ll 
m_W &\quad\text{(consistency of the effective action)}.
\label{con-ea}
\end{align}
The minimum of (\ref{pnpot}) satisfies 
$\f_\perp^{4\n+1} e^{2N\f_\perp/\l} \sim N^{5-6\n} \l^{6\n-4}$, at which point the $W$-mass from (\ref{Wmass3}) and the $\f$-mass from the curvature at the minimum are
\begin{align}\label{Wmass4}
L^2 m_W^2 \approx \f_\perp^2, \qquad
L^2 m_\f^2 \approx \f_\perp/N,
\end{align}
where we have used (\ref{con-dig}).  But then (\ref{con-ea}) cannot be satisfied for any $\f_\perp$ satisfying (\ref{con-cell}).

Thus there is no self-consistent abelianizing minimum induced by the bion potential.  Physically, the minima coming from the competition of the perturbative and non-perturbative pieces consistent with the semi-classical approximation can only occur so close to the cell walls that it gives a mass for the $W$-boson (which we were trying to integrate out) much smaller than the mass for $\f$ (which we were trying to keep in the effective action).

This discussion is an oversimplification for all the non-abelian minima except for the one with gauge group $G_2$.  The reason is that for all the others the perturbative minimum is not just at a wall of the gauge cell, but at a corner, where several walls meet.  Thus in these theories there are several relevant variables---an independent $\f_\perp$ for each wall that meets at the corner of interest.  The neutral bion terms do not couple these different variables, but the magnetic bion ones do.  

Furthermore, in the cases where the relevant walls are not orthogonal (the nodes associated to their roots are connected by a link in the extended Dynkin diagram), our argument above for the smallness of the magnetic bion terms relative to the neutral bion terms no longer holds.  This case, which only occurs for the exceptional groups $E_{6,7,8}$ and $F_4$, would require a truly multidimensional analysis, which we will not attempt here.

\section{Long-distance effective theory and confinement}\label{sec6}

We now set $\f$ to its perturbative minimum, and look at the physical implications of the effective action for the $\s$ scalar fields (which are the dual 3-d photons) and the $n_f$ light fermions, $\psi_f$.  The results of sections \ref{sec2} and \ref{sec7} imply that this is only valid for $SU(N)$ and $Sp(2N)$ QCD(adj) for which the vacuum Higgses  the gauge group completely to $U(1)$ factors.  For other gauge groups where there are unbroken nonabelian gauge group factors, some other method is needed to analyze the effective 3d dynamics. 

The Euclidean non-perturbative long distance effective theory in the case where the gauge group abelianizes, $G\to U(1)^r$, is governed by the proliferation of topological defects, as illustrated in figure \ref{fig:plasma}.  In particular, as we have discussed above in detail, the Euclidean vacuum may be seen as a grand-canonical ensemble of topological defects and molecules, which may be written as 
\begin{equation}
Z= \int [d \s] [d \phi] \left( \prod_{f=1}^{n_f} [d \psi_f]   [d \bar \psi_f] \right) 
 \exp \left[ - \int_{\R^3}  \cL \right]
\end{equation} 
where 
\begin{align}\label{Leff}
\cL = \cL_0  +  \cL_1 +  \cL_2 + \ldots  
\end{align}
with
\begin{align}\label{Leff0}
\cL_0 &= \tfrac{g^2}{4 L} (\del_m\s,\del_m\s) 
+ \tfrac{4\pi^2}{g^2L} \left( \del_m\f \,,\del_m\f \right) 
+ i \tfrac{2L}{g^2} \left( \bar\psi_f , \slashed{\del} \psi_f \right)
+ V_\text{pert}(\f),
\\
\label{Leff1}
\cL_1 &= \sum_{j=0}^r \left(    \cA_j \, e^{-S_j(\f)+2\pi i\s(\a_j^\v)} \prod_{f=1}^{n_f} [\a_j(\psi_f)]^2, + \text{h.c.}  \right),
\\
\label{Leff2}
\cL_2  
&= \sum_i |\cA_{ii}| e^{-2 S_i(\f)} 
- \!\!\!\!\!\!\sum_{\{i,j | (\a_i,\a_j) < 0 \}}\!\!\!\!\!\! |\cA_{ij}|  
e^{-S_i(\f)- S_j(\f)} e^{2\pi i\s(\a_i^\v-\a_j^\v)},
\end{align}
where the $\cA_i$ are given in (\ref{mon1a}), the $\cA_{ij}$ in (\ref{bionamp}), and the $\cA_{ii}$ in (\ref{neutamp}).

The proliferation of the monopole-instanton events corresponds to   operators in (\ref{Leff1}), while neutral bion events and magnetic bions are associated with, respectively, the first and second classes of operators in (\ref{Leff2}).  The dual description (\ref{Leff}) is valid for distances large compared to the inverse $W$-boson mass $m_W^{-1}/g^2$.  The theory exhibits a mass gap for gauge fluctuations and confinement via the magnetic bion mechanism. 

\subsection{Mass gap for gauge fluctuations}\label{sec7.1}

Consider the bosonic part of the  long-distance effective theory (\ref{Leff}).  In the small-$S^1$ regime, bosonic fluctuations are $\f$ and $\s$ associated with the gauge holonomy and the dual photons.  These two types of fluctuations have different masses at weak coupling for $n_f \neq 1$ QCD(adj).  $m^2_\f$ receives contributions both from perturbation theory around the perturbative vacuum, and non-perturbative contributions due to neutral and magnetic bions.  In contradistinction, the mass gap for $\s$ fluctuations is zero to all orders in perturbation theory, and is induced at $m^2_\s L^2 \sim e^{-2S_0}$ order due to magnetic bions, where $S_0 \sim 8\pi^2/(g^2 N)$ is the typical bion action.  The leading $\cO(g^2)$ one-loop result for the mass of $\f$-fluctuations is given in (\ref{mphispec}), and there are also perturbative corrections from all orders in perturbation theory.  The leading non-perturbative contribution to the $m^2_\f$ appears at order $\cO(e^{-2S_0})$ and is due to bions. 

Thus we may write, schematically, the following mass formula 
\begin{align}\label{masssquare1}
m^2_\f &=   m^2_{\rm pert.} +   m^2_{\rm n.p.} 
\hspace{-3cm}&&= \left[ (n_f-1) \cO(g^2) + \cO(e^{-2S_0})  \right]L^{-2} ,
\nonumber\\
 m^2_\s &= m^2_{\rm n.p.}  
 \hspace{-3cm}&&= \cO(e^{-2S_0}) L^{-2} .
\end{align}
For example, for $SU(2)$ gauge theory, using the one-loop beta function and dimensional transmutation, and ignoring logarithmic corrections momentarily, the mass spectrum for $\f$ and $\s$ fluctuations takes the form
\begin{align}\label{masssquare2}
m^2_\f &= \L^2  \left[ (n_f-1) (\L L )^{-2} + (\L L )^{(8 - 2n_f)/3}  \right] ,
\nonumber\\ 
 m^2_\s &= \L^2  (\L L )^{(8 - 2n_f)/3} .
\end{align}
The semi-classical domain corresponds to $L\L \lesssim 1$.  In the next subsection we will write a similar formula for the string tension. 

How seriously should we take the $L$-scaling given in formulas like (\ref{masssquare1}) and (\ref{masssquare2})?  For example, if we calculate such observables by using numerical lattice simulations, should we expect to confirm these predictions?  The concern is that, in fact, the perturbative term that we have written as $\cO(g^2)$, when extended to all orders in perturbation theory, is an asymptotic series.  The whole series is divergent, and is non-Borel summable.  The term due to neutral bions, $\cB_{ii} := [\cM_i\bar\cM_i]$, also multiplies an asymptotic series, which is also non-Borel summable.  We are then entitled to ask what this mass formula really means and what does it  really approximate?  Below, we argue that the result (\ref{masssquare2}) is actually physical and meaningful due to the BZJ prescription and its extension, as described in section \ref{etm}.  In section \ref{Ah Istanbul} we give a more extended discussion of the mathematical framework of transseries and Borel-\'Ecalle summation \cite{Ecalle:1981, Sternin:1996, Costin:2009} and how it can serve to make expressions like (\ref{masssquare1}) and (\ref{masssquare2}) the leading terms in a convergent expansion.

An expansion for the mass of the $\f$-fluctuations, which may actually make sense, is
\begin{align}\label{massphi}
L^2 m^2_\f &= \sum_{q=0}^\infty  a_{0, q} g^{2q}
+ e^{-2S_0} g^{-2r_1} 
\sum_{q=0}^\infty a_{1,q} g^{2q}
+ e^{-4S_0} g^{-2r_2} 
\sum_{q=0}^\infty a_{2, q} g^{2q}
+ \ldots,
\end{align}
where in $a_{n,q}$, $n$ labels the topological sector of a given saddle point and $q$ is an index counting the order of perturbation theory for fluctuations around that saddle point.  The $r_n$ are some exponents which are determined from quasi-zero mode integrations of multi-instanton configurations as discussed in section \ref{sec5}.  (We have simplified things by setting the action of all $n$-bion configurations to $2nS_0$ where it should more properly be a sum of $2n$ $S_j(\f)$'s given in section \ref{sec4.1}.)

The first term in (\ref{massphi}) is the contribution of perturbation theory around the perturbative vacuum.  This would be the usual text-book result for the mass for the $\f$ fluctuations, and is the analog of the usual Rayleigh-Schr\"odinger perturbation theory in quantum mechanical systems with degenerate minima.  However, by itself, the first sum is meaningless:  it is not Borel summable, and unless we think of it as part of some larger structure, it is devoid of meaning. 

The second term in (\ref{massphi}) is sourced from the dilute gas of neutral and magnetic bions $\cB_{ij} := [\cM_i\bar\cM_j]$ times perturbative corrections to all orders in perturbation theory around it.   The third term is due to the dilute gas of 2-bion molecular events such as $[\cB_{ij}\cB_{k\ell}]$ (times perturbative fluctuations), and so forth. 

The basic idea for how to give meaning to such a series of asymptotic series with exponentially decreasing factors (known as ``transseries" in the math literature) is basically to iterate the BZJ prescription to all orders in the instanton expansion (known as Borel-\'Ecalle summation of transseries in the math literature).   For example, since the first series in (\ref{massphi}) is not Borel summable, it must lead to an ambiguous imaginary part, which we expect to be of the form $\pm i e^{-4S_0}$ due to the large-order behavior of perturbation theory.   However, the third term in the series is also ambiguous as per our prescription for the $[\cB_{ij}\cB_{ji}]$ amplitude discussed in section \ref{etm}, and produces an imaginary part proportional to $\pm i e^{-4S_0}$.  We expect that these two ambiguities must cancel and we must recover an unambiguous result at order $e^{-4S_0}$, as in (\ref{borelbion2}).   We also expect the sub-leading ambiguities in the Borel sum to be cancelled by neutral topological molecules with higher actions.  

Let $\B_{n,\th=0^\pm}$ denote the Borel resummations of the perturbative series $\sum_{q=0}^\infty a_{n, q} g^{2q}$ for complex $g^2$ with phase $\th=0\pm\e$.  Then our expectation is that the imaginary parts of $\B_{0,\th=0^\pm}$ should cancel with the imaginary parts that we obtain through the refined BZJ-prescription, namely, 
\begin{align}\label{rBZJ}
0 = \Im\Bigl(\B_{0, \th=0^\pm}
+ \B_{1,\th=0^\pm} [\cB_{ii}]
+ \B_{2,\th=0^\pm} [\cB_{ij}\cB_{ji}]_{\th=0^\pm}
+ \B_{3,\th=0^\pm} [\cB_{ij}\cB_{jk}\cB_{ki}]_{\th=0^\pm} 
+ \ldots \Bigr) .
\end{align}
Note that only magnetically neutral multi-bion configurations are included in (\ref{rBZJ}) since only this charge sector can mix with the perturbative vacuum sector to which $\B_0$ belongs.  Also, we have suppressed sums over the repeated $i$, $j$, $k$ monopole indices in (\ref{rBZJ});  note that for each distinct choice of these indices, the associated perturbative series arising from fluctuations around that multi-bion saddle point may be different, and so their $\B_n$ resummations should also properly carry $i$, $j$, $k$ monopole indices.  Finally, note that the $\th=0^\pm$ subscript is left off the $n=1$ neutral bion amplitude since, as discussed in sections \ref{sec5.3} and \ref{etm}, $[\cB_{ii}]$ is unambiguous by itself.  Explicit illustrations of these types of cancellations in the context of matrix models, which are instrumental for an unambiguous non-perturbative definition, are presented in \cite{Marino:2008ya, Aniceto:2011nu}.

Going beyond the refined BZJ prescription, it is clear that for a consistent, unambiguous interpretation of the expansion (\ref{massphi}) to exist there must be (infinitely many) cancellations in addition to (\ref{rBZJ}).   For instance, the second term in (\ref{massphi}) receives contributions not only from neutral bions, but also from magnetic bions $[\cB_{ij}]$, with $i\neq j$.  The ambiguity in the Borel resummation, $\B_{1,\th=0^\pm}$, of the perturbative fluctuations around them, should be cured by the imaginary part coming from the appropriate 2-bion molecules in that charge sector, and so forth, giving
\begin{align}\label{brBZJ}
0 = \Im \Bigl(
[\cB_{ij}]  \B_{1, \th=0^\pm}  
+    [\cB_{ik}  \cB_{kj}]_{\th=0^\pm}   
\B_{2, \th=0^\pm}  
+   [  \cB_{ik}  \cB_{k\ell} \cB_{\ell j}]_{\th=0^\pm} 
\B_{3, \th=0^\pm} 
+      \ldots   \Bigr)   
\end{align}
for given $i$, $j$ (and with the repeated $k, \ell,\ldots$ indices summed over).  In section \ref{Ah Istanbul} we review and discuss the idea of ``resurgence" which systematizes the infinite set of consistency relations generalizing (\ref{rBZJ}) and (\ref{brBZJ}) necessary for Borel-\'Ecalle resummation of a transseries like (\ref{massphi}).

The expression for the mass of $\s$ fluctuations is very similar.  The main difference is that it does not receive any contributions to all orders in perturbation theory nor at the leading order in the semi-classical expansion, and so it is an intrinsically non-perturbative second order effect in semi-classics.  The analog of (\ref{massphi}) for $m_\s$ is then given by
\begin{align}\label{}
L^2m^2_\s 
&=  e^{-2S_0} g^{-2s_1} \sum_{q=0}^\infty b_{1,q} g^{2q}    
+ e^{-4S_0} g^{-2s_2} \sum_{q=0}^\infty b_{2,q} g^{2q}
+ \ldots
\end{align}
for some exponents $s_n$ and coefficients $b_{n,q}$.  Letting $\til\B_{n,\th=0^\pm}$ denote the Borel resummations of the perturbative series $\sum_{q=0}^\infty b_{n, q} g^{2q}$, the condition for the ambiguity in the leading term, $\til\B_1$, to cancel is precisely (\ref{brBZJ}) again, but with $\B_n$ replaced by $\til\B_n$.

Once the cancellation of the ambiguous imaginary parts is assured,  the finite results for the $\f$ mass and for the mass gap for gauge fluctuations given in (\ref{masssquare1}) becomes physical, in that it is an approximation to the physical result 
\begin{align}
L^2m^2_\f &=  
\Re \B_0  (|g^2|)  +  
g^{-2r_1} e^{-2S_0} \Re \B_1  (|g^2|)  
+ \ldots 
\\
L^2m^2_\s &=  
\phantom{\Re \B_0  (|g^2|) +\text{}}
g^{-2s_1}   e^{-2S_0}  \Re\til\B_1(|g^2|)  +  g^{-2s_2} e^{-4S_0} \Re\til\B_2(|g^2|) + \ldots
\nonumber
\end{align}
Thus the scaling for the mass gap for gauge fluctuations given in (\ref{masssquare2}) is the leading structure of the $L$ scaling, and up to our understanding of QCD(adj), is actually physical. 

\subsection{Confinement}

As described in Section \ref{sec1}, the dual photon in QCD(adj) lives in  
\begin{align}\label{sigmaspace-ad2}
\s\in\tf^*/(W\ltimes\G_r) 
\end{align}
and is periodic under translation by electric charges, $\s\to\s+\a, \a \in \G_r$.  Apart from this periodicity, the potential $\sim -\sum_{i, j} \cos[2\pi \s(\a^\v_i-\a^\v_j)]$ in (\ref{Leff2}) also possess an invariance under   
\begin{align}
\s \rightarrow \s + \, \o_i, \qquad \o_i  \in \G_w.
\label{magsim}
\end{align}
since  $\o_i(\a^\v_j)=\d_{ij}$.  The presence of the symmetry (\ref{magsim}) in the dual formulation is associated with the fact that the vacuum of the original (electric) theory can be probed by external electric charges distinguished by their (non-vanishing) charges under the center, 
\begin{align}
Z(\tG)=\G_w/\G_r \; , 
\end{align}
listed in table \ref{tabA1}.  
  
A well-known probe of confinement is the area law for large Wilson loops.  Consider the insertion of a Wilson loop $W[C,\o]$ (\ref{wilsonline}) associated with some charge (weight vector) $\o\in\G_w$.  As was explained in (\ref{dualWL}), the insertion of the Wilson loop in terms of original electric variables, is equivalent, in terms of dual magnetic variables to the requirement that  the dual scalar field acquires a non-trivial monodromy,
\begin{align}
\oint_ {C'} d \s = 2 \pi \o  \in \G_w , \qquad \ell(C, C')=1 \; , 
\label{monodromy}
\end{align}
where $\ell(C, C')$ is the  linking number of the two closed curves. The evaluation of the Wilson loop reduces to the minimization of the dual action in the space of field configurations satisfying the monodromy condition (\ref{monodromy}).  Consider a loop $C= \del\Sigma$ bounding a surface $\Sigma$ lying in the $xy$-plane.  Then, the string tension associated with the non-trivial charge $\o$ can be evaluated as 
\begin{align}
T (\o)  =  \min_{\s(z)}   \lim_{\Sigma \to \R^2} \frac{\D S} {\rm Area (\Sigma)}  \Big|_{\D \s = 2 \pi \o}.
\end{align}
Because of translational invariance in the $xy$-plane, the evaluation  of the string tension reduces to finding the action of kink configurations in the corresponding  one-dimensional problem (obtained after dimensional reduction of the $xy$-directions).  We find the tension, in the semi-classical domain $LN\L \lesssim 1$,  
\begin{align}
T (\o)  =  \L^2 ( \L L N)^{(5-2n_f)/3} f (\o),
\end{align}
where  $f (\o)$ is a function that only depends on the conjugacy class of irrep $\o \in \G_w$.

Physically, in a Euclidean description, confinement is due to the Debye mechanism, as in the Polyakov model \cite{Polyakov:1976fu}, but  with one major difference.  The role of the monopole plasma is now played by the magnetic bion plasma.  The Wilson loop in the $xy$-plane generates a magnetic field in $z$ direction.  The magnetic field has a finite penetration depth into the magnetic conductor, which in turn, implies the area law of confinement.

\subsection{Discrete $\chi$SB by topological disorder operators}

The zero mode structure of the monopole operators in (\ref{Leff1}), also given in (\ref{fierz}), is a singlet under $SU(n_f)$, but transforms under $\Z_{2h^\v n_f}$ by a $\Z_{h^\v}$-valued phase as $\det_{f,f'}(\cdots) \rightarrow e^{2 \pi i k/h^\v} \det_{f,f'}(\cdots)$.  Since $\Z_{2h^\v n_f}$ is an exact symmetry of the quantum theory, the topological operators must respect it. This means, the invariance of  (\ref{Leff1}) demands that the magnetic flux part of $\cM_j$ transforms as
\begin{align}
\Z_{h^\v}\ :\quad
\s  &\to \s -  \frac {k}{h^\v} \r, 
\qquad\qquad 
k=1, \ldots, h^\v,
\end{align}
where $\r := \tfrac12\sum_{\a\in\Phi_+}\! \a$ is the Weyl vector, which satisfies $\r(\a^\v_j)=1$.  In the semi-classical small-$S^1$ domain, this implies that the topological disorder operator $\exp[2\pi i\s(\a^\v_j)]$ is an equally good operator to probe the discrete chiral symmetry $\Z_{h^\v}$ realization.  

The magnetic bion induced potential $\sim  -\sum_{i,j} \cos[2\pi\s(\a^\v_i-\a^\v_j)]$ in (\ref{Leff2}) is invariant under the $\Z_{h^\v}$ chiral symmetry and possess $h^\v$ isolated vacua.  In the small $S^1$ domain, the topological disorder operator acquires a vev and breaks the $\Z_{h^\v}$ chiral symmetry completely.  The theory has $h^\v$ isolated vacua $|\Theta_k \rangle$, for which, in Hamiltonian formulation, we may write   
\begin{align}
\langle \Theta_k  |   \exp [2\pi i \s (\a^\v_j)]   |\Theta_k \rangle = 
e^{2 \pi i k/ h^\v} ,  \qquad k=1, \ldots, h^\v.
\label{SB}
\end{align}
The values of $h^\v$ for all simple gauge groups $G$ are given in table \ref{tabA2}. 

This is to some extent a surprising result.  The discrete $\chi$SB,    which is expected to be dynamical in the strong coupling domain in terms of electric variables, maps to a spontaneous breaking by a tree level scalar potential in the weak coupling domain in the dual magnetic formulation.  This shows that discrete $\chi$SB can also take place at weak coupling, and is sourced by the condensation of topological disorder operators.  

We also note that this is  how chiral symmetry is broken in $\cN{=}1$ superYang-Mills, the $n_f=1$ QCD(adj).  This interpretation disagrees with that of \cite{Davies:2000nw}.  In the one-flavor theory, since a monopole operator has two zero modes, the symmetry breaking as in (\ref{SB}) generates a chirally asymmetric mass term for fermions.  Omitting inessential factors, for example, 
\begin{align}
e^{-S_{0, j}} \langle e^{2\pi i \s (\a^\v_j)} \rangle  
\a_j (\psi)  \a_j (\psi) =
e^{-S_{0, j}} \a_j (\psi)  \a_j (\psi) 
\label{mon2}
\end{align}
in one of the isolated vacua, say, $k=0$.  This induces a mass for fermionic fluctuations $m_\psi \sim \L( \L L N)^2$. 

In supersymmetric gauge theories with supersymmetric boundary conditions, there is compelling reason to believe that  the physics is analytic as a function of the radius.  We have just seen that chiral symmetry breaking in the small $S^1$ phase is due to condensation of the disorder operators.  On the other hand, at large $S^1$, the gauge dynamics cannot be described in terms of abelian photons, due to absence of abelianization, and the chiral symmetry breaking is  expected to be due to condensation of ordinary fermion-bilinear $\langle\tr\psi\psi\rangle \neq 0$.  This does not present a puzzle since the $h^\v$ vacua of the theory in the small $S^1$ domain can smoothly interpolate to the $h^\v$ vacua in the large $S^1$ domain.  The expected phase diagram of the theory is thus
\begin{align} 
\xy 
(-120,0)**\dir{-} ?(.75)+(3,3)*{ \langle e^{ i  \s(\a^\v_j)} \rangle \neq 0 , \; \;\;
};
(0,0)*{\; \; \;  L}; 
(-60,0)**\dir{-} ?(0.01)*\dir{<} ; 
(-35,3)*{ \;\; \;\; \;\;   \langle \tr \l \l \rangle \neq   0
} ; 
(-90,-3)*{   \;\;  \langle \tr \Omega \rangle  =  0, }  ; 
(-30,-3)*{  \langle \tr \Omega \rangle  =  0, }  ; 
(-4,2)*{\scriptstyle {\infty}} ; 
\endxy 
\label{phasediag3} 
\end{align}
with no phase transition.

For multi-flavor theories, $n_f >1$, since monopole-instanton induced operators have $2n_f$ zero modes, the discrete  $\chi$SB does not induce a mass term for fermions.  Instead, at distances larger than the inverse dual photon mass, the theory is described by a Nambu--Jona-Lasinio type model, with a chirally symmetric $2n_f$-fermion interaction, to be described below. 

\subsection{Continuous $\chi$S realization}

QCD(adj) with $n_f>1$ also possesses a continuous chiral symmetry, $SU(n_f)$.  In the small-$S^1$ regime $(r L \L  \lesssim 1)$ and at asymptotically large distances (larger than $m_\s^{-1}$), the fermionic theory is described by the Lagrangian
\begin{align}
\cL_\text{fermionic}
=  i \frac{2L}{g^2} \left( \bar\psi_f , \slashed{\del} \psi_f \right) 
+\sum_{i=1}^{r+1}  \left( \cA_{i}  e^{-S_{0, i}}  \det_{f,f'} [ \a_i( \psi_f) \a_i(\psi_{f'}) ] + \text{h.c.} \right) \,. \qquad 
\label{ald}
\end{align}
Let us first consider $2 \leq n_f \leq n_f^*$ where $n_f^*$ is the lower boundary of the conformal window. 

$\bm{2 \leq n_f \leq n_f^*}$: 
In the small  ${S^1}$ regime, the asymptotically long distance theory is an NJL-type model in the weak coupling regime.  At weak coupling, the $2n_f$-fermion interaction does not break chiral symmetry.  Thus, the theory at small $S^1$ exhibits confinement without continuous $\chi$SB.  At large $S^1$, it is expected to exhibit confinement with continuous $\chi$SB, with a breaking pattern: $SU(n_f) \to SO(n_f)$.  There is strong evidence  that the scale of continuous $\chi$SB is an unconventional one, given by  
\begin{align}
L_{\rm c \chi  SB}= c \L^{-1}/r
\end{align}
moving to zero radius as $r := \text{rank}(\gf)\to\infty$.  In this limit, the region of validity of the dual magnetic  lagrangian (\ref{Leff}) shrinks to zero as well.  In other words, QCD(adj) at $r=\infty$ never becomes weakly coupled regardless of of the size of compactification radius.  This is a consequence of large-$N$ volume independence of center symmetric theories.  We expect that the phase diagram of the finite rank theory to be, according to three types of symmetry realization, as follows: 
\begin{align} 
\xy 
(-70,0)*{ \bullet}; 
(-120,0)**\dir{-} ?(.65)+(3,3)*{ \langle e^{ i  \s(\a^\v_j)} \rangle \neq 0  \; \;\; \langle \tr \l^I \l^J \rangle =  0 
};
(0,0)*{\; \; \;  L}; 
(-70,0)**\dir{-} ?(0.05)*\dir{<} ; 
(-35,3)*{ \;\; \;\; \;\;   \langle \tr \l^I \l^J \rangle \neq   0
} ; 
(-100,-3)*{   \;\;  \langle \tr \Omega \rangle  =  0 }  ; 
(-30,-3)*{  \langle \tr \Omega \rangle  =  0}  ; 
(-4,2)*{\scriptstyle {\infty}} ; 
(-68,4)*{  L_{\rm c \chi  SB}} ; 
\endxy 
\label{phasediag2} 
\end{align}

If we add a small mass for fermions, then,  the continuous chiral symmetry will become an approximate symmetry. Consequently,  the low-energy physics as a function of radius will be a smooth  interpolation  between a small-$S^1$ regime of light fermions and a large-$S^1$ regime of pseudo-Goldstone bosons. 

$\bm{n_f^* \leq n_f \leq 5.5}$: 
The theories in this range are expected to flow to CFTs in the $\R^4$ limit.  If the theory has a weakly-coupled fixed point, then the separation of scales that the dual Lagrangian (\ref{Leff}) relies on is still valid (at distances larger than $m_W^{-1}$) even at large radius.  Thus one can take the arbitrarily large $S^1$ limit while using (\ref{Leff}). Consequently, we expect that continuous $\chi$SB does not occur.  The vacua associated with discrete chiral symmetry breaking, upon proper normalization, are seen to be of  runaway type in the $\R^4$ limit.  Consequently, the theory on $\R^4$  is not expected to break any of its global symmetries. 

\section{Resurgence theory and the transseries framework} 
\label{Ah Istanbul}

In this section, without aiming to be  complete, we would like to point out  the interconnections of some of our ideas in QFT, in particular semi-classically calculable 4-d gauge theory on $\R^3  \times S^1$, to \emph{resurgence theory} and the \emph{transseries framework}  developed by \'Ecalle \cite{Ecalle:1981}.  Resurgence theory provides detailed information on Borel transforms and sums, their inter-connection to Stokes phenomena and a set of general summation rules along the directions in the Borel plane where there are singularities.  For a quantum field theorist, perhaps the most interesting aspect of this framework is Borel-\'Ecalle (BE) summability, which provides tools for dealing with non-Borel summable series \cite{Costin:2009}.  We believe our findings in gauge theory --- in particular, what we called the refined BZJ-prescription --- is the first step of BE resummation applied to QFT. 

An intimately related and important idea is \emph{hyperasymptotics} as developed by Berry and collaborators \cite{bh90, b99, bh91}, building upon earlier ideas of Stokes and Dingle on asymptotics, see \cite{Dingle:1973}.  The usual Poincar\'e  asymptotics corresponds to summing an asymptotic series up to a fixed order, call it $M^*$, in the expansion parameter ($\l := g^2 r$, the 't Hooft coupling in our QFT example where $r=\text{rank}\,\gf$).  This gives an error bounded by $\l^{-M^*-1}$.  \emph{Superasymptotics} is a much more accurate approximation achieved by summing up to the least term in the series.  This optimal truncation reduces the error to $e^{-A/\l}$ where $A$ is positive constant.  This optimal truncation can be repeated for the remainder, where $e^{-A/\l}$ multiplies another asymptotic series.  This leads to a nested structure of superasymptotics, and the sequence of these   defines hyperasymptotics \cite{bh91}.  Although at first sight it looks like this process continues ad infinitum, it turns out not to be so.  Berry and Howls showed that in practice, for a finite $\l$, this process terminates after $\log (1/\l)$ stages.  The error in hyperasymptotics is given by $e^{-(1+ 2 \log 2) A /\l} = e^{- 2.386 A /\l}$.   This is still a significant improvement over superasymptotics.

But this also makes it clear that hyperasymptotics and resurgence differ.  Our approach to gauge theory on $\R^3 \times S^1$ is part of the resurgence framework.  For example, in certain gauge theories, we can show that a mass gap for gauge fluctuations is induced by order $e^{-3 A/\l}$ or $e^{-5 A/\l}$ effects, where $e^{-A/\l}$ is a monopole-instanton factor; see \cite{Poppitz:2009uq} for a list.  This cannot be easily extracted from hyperasymptotics for finite $\l$, but in principle, it can be extracted in the semi-classical resurgence framework.  The fact that one can do considerably better within resurgence formalism compared to hyperasymptotics is pointed out in \cite{Pasquetti:2009jg}.  

Poincar\'e asymptotics or superasymptotics are often used in QFT or quantum mechanics.  However, both hyperasymptotics and BE resummation are much more powerful techniques, and there are cases with ordinary and partial non-linear differential equations, as well as with integral equation examples in which the asymptotic transseries expansions supplemented with BE resummation gives the \emph{exact} result.  We do not know if this is the case in QFTs, but we can be optimistic.

\subsection{Intuitive explanation of resurgence in QFT} 

The semi-classical analysis of a typical bosonic observable, $O(\l)$, in QFT on small $S^1 \times \R^3$ is a double expansion --- a \emph{transseries} in the resurgence framework --- which is a combination of a perturbative expansion in $\l$ and a non-perturbative expansion in $e^{-2A/\l}$:
\begin{align}\label{gen-ob}
O(\l) &= \sum_{n=0}^{\infty} e^{-2nA/\l} 
 \l^{-r_n} [\log(\pm\l)]^{\til r_n}  
 P_n(\l),
\qquad\text{with}\qquad
P_n(\l) = \sum_{q=0}^\infty a_{n,q} \l^q .
\end{align}
We can consider a real observable $O(\l)$ so that all the $a_{n,q}$ perturbative coefficients are real.
In the current application, the exponentials are the  (multi--)monopole-instanton factors from various saddle point contributions and the $P_n(\l)$ come from the perturbative fluctuations around a given saddle point.  So $n$ labels the saddle points and $q$ counts the order of perturbation theory.  $P_0$ is thus the usual perturbation theory series around the perturbative vacuum.  For $SU(r+1)$ QCD(adj), for example, $A = 8\pi^2$ and only multiples of $2A$ appear in (\ref{gen-ob}) since only (multi--)monopole--anti-monopole saddle points can contribute to bosonic observables in this theory.  Then the prefactors of the $P_n(\l)$ series are the multi-bion amplitudes $[\cB\cB\cdots]$ discussed in section \ref{sec5}, heuristically
\begin{align}\label{Btopsect}
[\cB^n] = a_{n,0} e^{-2nA/\l} \l^{-r_n} [\log(\pm\l)]^{\til r_n} .
\end{align}
(More detailed examples of transseries appeared in the expressions for the scalar masses in section \ref{sec7.1} where the dependence on the different magnetic charge sectors was spelled out.  In this section, for the sake of simplicity, we will ignore these complications and pretend the saddle points are organized by a single integer $n$, counting the number of bions. )  The exponents $r_n$, $\til r_n$ are determined from quasi-zero mode integrations of multi-bion configurations as discussed in section \ref{sec5}.  

The main outcome of resurgence, which we wish to explain in more detail in this section, is: 
\begin{quote}
All the divergent series $P_n(\l)$ appearing in the transseries (\ref{gen-ob}) are interrelated.  The parameters $a_{n, q}$ are related, in a calculable way, to $a_{n', q'}$ for topological sectors $n'>n$:  the $a_{n,q}$ for large values of $q$ are determined by the $a_{n',q'}$ for small values of $q'$.\footnote{In theories with fermions, as in QCD(adj), the leading singularity in the Borel plane may cancel, so only topological sectors $n'-n=2,3,\ldots$ have related perturbative expansions.  The general circumstances where this happens can be deduced from our discussion in section \ref{etm}.}  In particular, the large-$q$ asymptotics for an observable in the perturbative vacuum, $a_{0,q}$, is dictated by the exponential (monopole-instanton)  factors.
\end{quote}
In a quantum field theory, we are then led to expect that the perturbative expansions around all non-perturbative sectors are actually related in a systematic way.  The fact that the perturbative expansion around the perturbative vacuum reappears in a slightly modified manner as a perturbative expansion around an instanton sector, and so forth, was called \emph{resurgence} by \'Ecalle.  A transseries expansion is therefore sometimes called a resurgent expansion.  

In the semi-classical transseries expansion of quantum field theory  (\ref{gen-ob}), there are two types of non-perturbative ambiguities:
\begin{list}{$\bullet$}{\itemsep=0pt \parsep=0pt \topsep=2pt}
\item {\bf the ambiguity in the Borel resummation of perturbation theory} around the perturbative vacuum, or around an instanton or multi-instanton saddle point; and
\item {\bf the ambiguity in the definition of the non-perturbative amplitudes} (\ref{Btopsect}) associated with neutral topological molecules, or molecules which include neutral sub-components. 
\end{list}
The main idea of resurgence in QFT is that these ambiguities are related in such a way that the physical observables are ambiguity-free.

\paragraph{The ambiguity in the Borel resummation of perturbation theory.}

Let $BP_n(t)$ denote the Borel transform of an asymptotic  perturbative series $P_n(\l)$,
\begin{align}\label{}
BP_n(t) := \sum_{q=0}^\infty \frac{a_{n,q}}{q!} \l^q .
\end{align}
We assume that the formal power series $P_n(\l)$ all satisfy the ``Gevrey-1" condition \cite{Sternin:1996, Costin:2009}, $|a_{n,q}| \le C_n R_n^q q!$ for some positive constants $C_n$ and $R_n$, so that the $BP_n(t)$ all have finite radius of convergence around the origin.  Thus the $BP_n(t)$ can be analytically continued away from the origin of the complex $t$-plane.  We assume, furthermore, that the the set of Borel transforms $\{BP_n(t)\}$ are ``endlessly continuable", which basically means that as a set they have only discrete singularities on all Riemann sheets of their continuations in $t$.  There are plausible reasons to expect that the Gevrey-1 condition will be satisfied by QFT perturbation expansions \cite{'tHooft:1977am}, but the condition of endless continuability of their Borel transforms, which requires the absence of natural barriers in the Borel plane, seems less easy to justify a priori. 

Assume that a number of the singularities of the set $\{BP_n(t)\}$ are located on the ray $\R^+$ in the Borel plane, i.e., that they are at some points $t = t_m$ indexed by $m\in\Z^+$ with $t_m$ an increasing sequence of positive real numbers.  Then the first ambiguity manifests itself as the ``jumps" in the directional Borel sum,
\begin{equation}\label{directionalBorel}
\B_{n, \th}(\l) = \int_0^{\infty \cdot e^{i\th}} BP_n(t\l) e^{-t} dt ,
\end{equation}
as the angle $\th$ of the contour of integration passes through $\th=0$.  The function $\B_{n\pm}:=\B_{n,\th= 0^\pm}(\l)$, associated with contours just above and just below a ray of singular points, are also called ``lateral Borel sums".  Equivalently, one can think of $\B_n(\l)$ as an analytic function in the complex $\l$-plane with a branch cut along the positive real axis, and $\B_{n\pm}(\l)$ as the values of this function as $\l$ approaches the cut from above or from below.  

The discontinuity of $\B_n$ across $\R^+$, or the jump in the lateral Borel sums, can be written
\begin{align} \label{Stokes}
\text{Disc}\B_n(\l) := \B_{n+}(\l) - \B_{n-}(\l)  
&= 2\pi i\sum_{m=1}^\infty f_{n,m}(\l) e^{-t_m/\l} ,
\end{align}
where the $f_{n,m}(\l)$ are some real analytic functions (for positive real $\l$); so
\begin{align}\label{Stokes2}
\Im \B_{n\pm} = \pm \pi \sum_{m=1}^\infty f_{n,m}(\l) e^{-t_m/\l}.
\end{align}  
This follows from (\ref{directionalBorel}) by a contour deformation argument so that each term picks up the contribution due to a single singularity $t_m$ and from the reality of the $a_{n,q}$.  Since there are (infinitely) many singularities on $\R^+$, there are many different choices for how to do the contour deformation.  No single contour deformation respects the reality of $O(\l)$ (i.e., the symmetry under $\l \to \bar\l$), so this must be restored by taking appropriate averages of different contour deformations.  The different ways of doing this translate into different functions $f_{n,m}(\l)$; they are not uniquely defined by (\ref{Stokes}) since they can differ by pieces which are asymptotically small, $\sim \exp\{-t_{\til m}/\l\}$ with $\til m > m$, as $\l\to0$.

Note that $\B_{n+}(\l)$ and $\B_{n-}(\l)$ are different functions of $\l$ with the same asymptotic behavior since they differ only by exponentially suppressed terms.  The different behavior of $\B_{n,\th}(\l)$ in different $\th$ sectors and the ensuing jumps as one crosses a ray of singularities in the Borel $t$-plane is associated with Stokes lines and Stokes jumps in the complex $\l$-plane.  The jump in (\ref{Stokes}) and the connection of sectorial solutions is encoded in the ``Stokes automorphism" in resurgence terminology.

Finally, it will be useful to note that the discontinuity, $\text{Disc}\B_n(\l)$, in the Borel resummation of $P_n(\l)$ can be related to the coefficients $a_{n,q}$ of $P_n(\l)= \sum_{q=0}^\infty a_{n,q} \l^q$ by a dispersion relation.  Since $\B_n(\l)$ has a cut along the positive real axis, we may use Cauchy's theorem and a contour deformation to write
\begin{align}\label{dispersion1}
\B_n(\l) = \frac{1}{2\pi i} \int_0^\infty d\l' 
\frac{\text{Disc}\B_n(\l')}{\l' - \l}  
- \frac{1}{2\pi i} \oint_{C_\infty} \frac{\B_n(\l')}{\l' - \l}
\end{align}
where $C_\infty$ is a loop at infinity and $\l$ is a point off the positive real axis.  Since the Taylor series of $\B_n$ around the origin gives the asymptotic series $P_n(\l)$, the coefficients of $P_n(\l)$ can be found by taking derivatives with respect to $\l$ and sending $\l\to0$.  This is justified as long as $\B_n(\l)$ grows more slowly than $1/\l$ as $\l\to0$.  Also, the contribution from the contour at infinity does not contribute as long as $\B_n$ descreases faster than $1/\l$ as $\l\to\infty$.  These two conditions can be met by making appropriate subtractions of leading terms of $\B_n$ and dividing by an appropriate power of $\l$; see, e.g., \cite{ZinnJustin:2002ru}.
This then allows us to express the coefficients of $P_n(\l)$ as 
\begin{equation}\label{dispersion3}
a_{n,q} = \frac{1}{2\pi i} \int_0^\infty d\l 
\frac{\text{Disc}\B_n} {\l^{q+1}} 
\qquad \text{for}\qquad q \geq \text{few},
\end{equation}
where the exact value of ``few" depends on the above-mentioned subtractions needed. 

\paragraph{Aside on the behavior of $\B_n$ at infinity.}

If $\B_n$ grows exponentially as $\l\to\infty$, no division by a power of $\l$ will remove the contribution of the integral at infinity.  In many cases in quantum mechanics a scaling argument assures the power-law behavior of $\B_n$ at infinity \cite{Bender:1969si, bw73}, but in QFT the situation is a priori not clear.  Consider the gauge theory on $\R^3 \times S^1$ further compactified down to quantum mechanics on $\R \times T^2 \times S^1$ such that the $T^2$ is much larger than the $S^1$ (so abelianized dynamics is operative at the scale of the $T^2$), but smaller than inter-monopole separations on $\R^3$ such that within the volume of the $T^2$ there will typically be a single monopole-instanton event.  The monopole-instanton in QFT descends to flux-changing events in the associated quantum mechanics, where the flux is defined as $\Phi(t) = \int_{T^2} B$ and flux-changing events are valued in the co-root lattice $\G_r^\v$.  There is ample evidence that this quantum mechanics is continuously connected to the QFT on $\R^3 \times S^1$, and a fair amount of non-perturbative data of the 4-d theory is encoded within this class of  quantum mechanical systems.  (This connection between quantum field theory and quantum mechanics is new and will be explored in a separate work.)  In this reduced quantum mechanics, we were able to show that the integral around infinity does not contribute by using scaling arguments.  By continuity, we expect that the same conclusion is also valid for QFT.  

\paragraph{The ambiguity in the definition of the non-perturbative amplitudes.}

The second ambiguity arises from the choice of path of analytic continuation in $\l$ needed to define the quasi-zero mode integrals appearing in the evaluation of saddle point contributions.  At least for the simplest cases, it is easy to see \cite{Argyres:2012vv} that this ambiguity in choice of path in the complex $\l$-plane can be mapped onto the ambiguity in choice of path --- the directions $\th=0^\pm$ of the ray  in the Borel plane --- in the directional Borel sums (\ref{directionalBorel}).  So the Stokes automorphism also acts on the amplitudes of neutral topological molecules.

For example, as we discussed in section \ref{sec5} for QCD(adj), the one- and two-bion amplitudes have leading forms for small real $\l$
\begin{align}\label{}
[\cB^1] &= a_{1,0} e^{-2A/\l} \l^{-r_1},
\qquad
[\cB^2] = a_{2,0} e^{-4A/\l} \l^{-r_2} \log(-\l),
\end{align}
with $r_1 = 7-4n_f$ and $r_2 = 3$; see (\ref{type-two}), (\ref{sign}), and (\ref{BBbarcont}).  Thus, upon continuing from negative to positive $\l$ either above or below the origin, there is no ambiguity in the $[\cB]$ amplitude, while there is one for the $[\cB^2]$ amplitude,
\begin{align}\label{B2ambig}
[\cB^1]_{+} - [\cB^1]_{-}
&= 0,
\qquad
[\cB^2]_{+} - [\cB^2]_{-}
= 2\pi i a_{2,0} \l^{-r_2} e^{-4A/\l}.
\end{align}
A natural extension of that discussion leads to the expectation that the higher saddle point contributions will have the form given in (\ref{Btopsect}).  In general, the values of the saddle points will vary in a complicated way as a function of complex $\l$.  In particular, there typically occur ``focal points" in the complex $\l$-plane where the values of different saddle points coincide, and emanating from these focal points are ``Stokes lines" where the real parts of different saddle point values coincide.  These are important since on either side of these lines different saddle points dominate the transseries expansion.  More importantly, upon continuing $\l$ around a closed path encircling focal points, and therefore crossing a number of Stokes lines, the saddle points will typically undergo a permutation.  As we will see below, this global information about the behavior of the saddle points under analytic continuation plays a key role in resurgence.

For definiteness (just so we have a simple toy model in which to illustrate resurgence), we will assume that $\til r_n =1$ for $n>1$ and that the $r_n$ are all integers, so that
\begin{align}\label{Bnambig}
[\cB^n]_\pm
= \left[\log(\l) \pm i\pi \right] a_{n,0} \l^{-r_n} e^{-2nA/\l}
\qquad\mbox{for}\qquad
n = 2, 3, \ldots 
\end{align}
But we should note that this simple form for the saddle point values probably does not actually arise from the saddle points of any analytic action functional.

\paragraph{Cancellation of the ambiguities.}

For the field theory to have a sensible non-perturbative definition in the continuum, we must have a cancellation of these two types of  non-perturbative ambiguity.  For an observable $O(\l)$ as in (\ref{gen-ob}) which is real, and for which the ambiguity of the saddle point contributions are always imaginary as in (\ref{Bnambig}), then this cancellation condition is simply the vanishing of the imaginary parts of the Borel sums of the perturbation series against those of the multi-bion amplitudes, 
\begin{align}\label{resurgence1}
0 &= \Im \left(\B_{0\pm}  
+ [\cB]_\pm\, \B_{1\pm}
+ [\cB^2]_\pm \B_{2\pm}
+ [\cB^3]_\pm \B_{3\pm}
+ \ldots\right).
\end{align}
This is just a rewriting of the condition that $\Im O(\l) = 0$.

Since as $\l\to0$ the $n$-bion amplitude is dominated by the $\exp\{-2nA/\l\}$ exponent (\ref{Btopsect}), an asymptotic expansion of this cancellation condition using (\ref{Stokes2}) implies that the singularities of the $BP_n$ in the Borel plane must be at $t_m = 2mA$.  

Now, the positions of the singularities of the $BP_n$ determine the large-order behavior of the $P_n$ series.  This follows from a theorem by Darboux (see chapters 1 and 7 of \cite{Dingle:1973}) which states that two different functions with pole or branch point singularities (but not essential singularities) at the same locations exhibits a \emph{universal} behavior in the late terms of its Taylor series expansion around origin which is independent of the kind of singularity.  As illustration, consider a simple function with Taylor expansion
\begin{align}\label{simple}
BP(t) 
:= \left(1- \frac{t}{A} \right)^\a 
= \sum_{n=0}^\infty \frac{(n-\a-1)!}{n!(-\a-1)!} \left(\frac{t}{A}\right)^n .
\end{align}
$BP(t)$ has a pole or branch point at $t=A$ when $\a$ is not a non-negative integer, but regardless of this value of $\a$, the leading behavior of the Taylor coefficients as $n \to \infty$ are all alike.  They are dictated only by the position of the singularity in the Borel plane, and are independent of the nature of the singularity.  The inverse Borel transform of $BP(t)$ is
\begin{align}\label{simple2}
P (\l) = \sum_{n=0}^\infty \frac{(n-\a-1)!}{(-\a-1)!} 
\left(\frac{\l}{A}\right)^n .
\end{align}
So the late terms of the asymptotic series $P(\l)$, just like the Taylor series for $BP(t)$, are also universal and only dictated by the positions of the singularities in the Borel plane.

Thus the result that the values of the saddle points, $2mA/\l$, are the locations of the singularities in the Borel plane means that consistency of the transseries expansion of $O(\l)$ relates the multi-bion amplitudes to the perturbative expansions.  We will now explain how this relation is made much more precise using resurgence relations.

Upon inserting the multi-bion amplitudes (\ref{Bnambig}), the consistency condition (\ref{resurgence1}) reads
\begin{align}\label{resurg2}
0 &= \Im\B_{0\pm}  
+ \l^{-r_1} e^{-2A/\l}\,\, \Im\B_{1\pm}
+ \sum_{n=2}^\infty \l^{-r_n} e^{-2nA/\l}\left(
\log\l \,\, \Im\B_{n\pm}
\pm \pi \,\, \Re\B_{n\pm} \right).
\end{align}
Now using the transseries expansion of $\Im\B_{n\pm}$ in (\ref{Stokes2}) and the identification $t_m=2mA$, as well as the 
formal identification of $\Re\B_{n\pm}$ with its (defining) asymptotic expansion, $\Re\B_{n\pm} \sim P_n(\l)$, (\ref{resurg2}) becomes
\begin{align}\label{resurg3}
0 &= 
\sum_{m=1}^\infty f_{0,m}(\l) e^{-2mA/\l}
+ \l^{-r_1} e^{-2A/\l}
\sum_{m=1}^\infty f_{1,m}(\l) e^{-2mA/\l}
\nonumber\\
&\qquad\text{}+ \sum_{n=2}^\infty \l^{-r_n} e^{-2nA/\l}
\left(
\log\l  
\sum_{m=1}^\infty f_{n,m}(\l) e^{-2mA/\l}
+
P_n(\l)
\right).
\end{align}
Collecting powers of $e^{2A/\l}$ then gives
\begin{align}\label{f0resurgence}
0 &= f_{0,1}
\nonumber\\
0 &= f_{0,2} + P_2 \l^{-r_2} + f_{1,1} \l^{-r_1} 
\\
0 &=f_{0,m}
+ P_m \l^{-r_m}
+ f_{1,m-1} \l^{-r_1} 
+ \tsum_{n=1}^{m-2}
(\log\l) f_{m-n,n} \l^{-r_{m-n}}
\qquad\text{for}\quad m \ge 3,
\nonumber
\end{align}
expressing $\text{Disc}\B_0$ in terms of $P_{n\ge2}$ and $\text{Disc}\B_{n\ge1}$.

What may be less obvious is that (\ref{resurgence1}) is not the only consistency condition following from demanding an unambiguous $O(\l)$.  As discussed after (\ref{Stokes2}), there is not a unique definition of the real analytic functions $f_{n,m}(\l)$ appearing in $\text{Disc}\B_n$.  So (\ref{f0resurgence}) applies equally to all choices of $f_{n,m}$ arising from different contour choices in the Borel plane.  Furthermore, these different contour choices are related to one another by the condition that the (Borel-\'Ecalle resummed) $O(\l)$ be a single-valued function in the complex $\l$-plane.  For then as $\l$ is continued around focal points, the saddle points contributing to $[\cB^n]$ will be permuted.  Since these saddle point values determine the locations of the singularities in the Borel plane, a monodromy in $\l$ is accompanied by a motion permuting the Borel plane singularities.  This in turn drags the contours used in the definiton of the $f_{n,m}$ into a new set of contours.  The single-valuedness of $O(\l)$ then implies additional relations among the $f_{n,m}$.  These are encoded in the ``resurgence relations" or ``bridge equations" and give a set of equations of the form (\ref{f0resurgence}) expressing $\text{Disc}\B_m$ in terms of $P_{n\ge2}$ and $\text{Disc}\B_{n\ge m+1}$ for all $m$.

For example, for $m=1$, the equations take the form
\begin{align}\label{f1resurgence}
0 &= f_{1,1} \l^{-r_1}
\\
0 &=f_{1,m} \l^{-r_1}
+ P_{m+1} \l^{-r_{m+1}}
+ \tsum_{n=1}^{m-1}
(\log\l) f_{m-n+1,n} \l^{-r_{m-n+1}}
\qquad\text{for}\quad m \ge 2.
\nonumber
\end{align}
Then, combining (\ref{Stokes}) with (\ref{f0resurgence}) and (\ref{f1resurgence}) gives to leading order
\begin{align}\label{leadingDisc}
\text{Disc}\B_0 &= - 2\pi i \l^{-r_2} P_2 e^{-4A/\l} 
+ \cO(e^{-6A/\l}) ,
\nonumber\\
\text{Disc}\B_1 &= -2\pi i \l^{-r_3+r_1} P_3 e^{-4A/\l} 
+ \cO(e^{-6A/\l}) .
\end{align}

We can now use these in the dispersion relation (\ref{dispersion3}) to derive relations between the coefficients of the $P_0(\l)$ and $P_1(\l)$  asymptotic expansions and those of the $P_{n>1}(\l)$. 
Just keeping the leading-order terms shown in (\ref{leadingDisc}), we obtain
\begin{align}\label{dispersion4}
a_{0,q} 
& =  \sum_{q'=0}^\infty a_{2,q'}  
\frac{\G(q+r_2-q')}{(4A)^{q+r_2-q'}},
\qquad
a_{1,q}
 = \sum_{q'=0}^\infty a_{3,q'}  
\frac{\G(q+r_3-r_1-q')}{(4A)^{q+r_3-r_1-q'}},
\end{align}
implying the leading large-order behaviors
\begin{align}\label{dispersion7}
& P_0(\l) \sim \frac{a_{2,0}}{(4A)^{r_2}}
\sum_{q=0}^\infty (q+r_2-1)!
\left( \frac{\l}{4A} \right)^{q}, 
\nonumber\\ 
& P_1(\l) \sim \frac{a_{3,0}}{(4A)^{r_3-r_1}}
\sum_{q=0}^\infty (q+r_3-r_1-1)!
\left( \frac{\l}{4A}\right)^{q} .
\end{align}
Thus the large-order behaviors of $P_0$ and $P_1$ are determined by the early terms of the $P_2$ and $P_3$ series, respectively:  the knowledge of a one-loop fluctuation determinant around the bion--anti-bion background determines the leading order of the asymptotic expansion around the perturbative vacuum.  Keeping additional terms from (\ref{dispersion4}) corresponds to sub-leading asymptotics:  e.g., two-loop fluctuations determine the $1/q$ correction proportional to $a_{2,1}$.  Recall that $P_0$ and $P_1$ are asymptotic expansions around different sectors, respectively the perturbative vacuum and the vacuum populated by neutral bion events.  Despite the drastic difference in the background, the asymptotics of the perturbative expansions around their respective sectors have a \emph{universal} behavior.  This is in accord with Darboux's theorem and Dingle's ideas about asymptotics, described above.  Indeed, (\ref{dispersion7})  can be identified with (\ref{simple2}) by an obvious mapping of the location of the singularities and by matching $\a$ with $r_2$ and $r_3-r_1$, respectively. 

The relations relating $P_0$ to $P_2$ and $P_1$ to $P_3$ in (\ref{dispersion4}) came from only keeping the leading terms in the resurgence relations for $f_{0,m}$ and $f_{1,m}$ in (\ref{f0resurgence}) and (\ref{f1resurgence}).  Such leading-term asymptotics is essentially the content of the BZJ prescription described in section \ref{sec5}.  For example, in the quantum mechanics of the anharmonic quartic oscillator, this argument has been used to connect large orders in perturbation theory to the bounce or instanton--anti-instanton amplitude in the unstable quartic theory, see \cite{ZinnJustin:2002ru}.  (The large order prediction obtained in this manner is identical to that of Bender and Wu \cite{Bender:1969si}, which was obtained by other methods.)  

But this by no means captures the full content of the resurgence relations.  With sufficiently precise knowledge of the global behavior (monodromies) of the $[\cB^n]$ saddle point values in the complex $\l$-plane, one can incorporate their contributions to obtain an infinite sum over all multi-bion sectors of the typical form 
\begin{align}\label{dispersion45}
a_{0,q}
&= \sum_{n=1}^\infty  \sum_{q'=0}^\infty 
a_{2n,q'}  \frac{\G(q+r_{2n}-q')}{(4nA)^{q+r_{2n}-q'}} .
\end{align} 
Writing out a few of the leading terms, 
\begin{align}\label{}
a_{0,q} 
&= (4A)^{-q-r_2} \G(q+r_2) 
\left[ a_{2,0} +  \frac{a_{2,1}(4A)}{q+r_2-1} 
+ \frac{a_{2,2}(4A)^2}{(q+r_2-1)(q+r_2-2)} 
+ \ldots  \right]  \nonumber\\
&\quad\text{}+ (8A)^{-q-r_4} \G(q+r_4) 
\left[ a_{4,0} +  \frac{a_{4,1}(8A)}{q+r_4-1} 
+ \frac{a_{4,2}(8A)^2}{(q+r_4-1)(q+r_4-2)} 
+ \ldots \right] \nonumber\\
&\quad\text{}+ \ldots ,
\end{align}
makes it clear that the one-loop fluctuation determinant around the $[\cB^{2n}]$ saddle point determines leading pieces of sub-series exponentially suppressed by a factor $(2n)^{-q}$. 

We note that similar expressions have appeared in the context of matrix models and topological string theory \cite{Marino:2008ya,  Marino:2007te} and by using the bridge equations in the context of resurgence theory in \cite{Aniceto:2011nu}.  In our current example, the difference stem from the fact that the monopole-instanton is actually a fraction of a 4-d instanton, indeed, 
$4A \sim 4\cdot S_{4d}/N = \frac{4}{N} \cdot \frac{8\pi^2}{g^2}$ for $SU(N)$ gauge group.   On the other hand, the fact that these results are almost the same is not a surprise, and reflects universal aspects of the instanton calculus. 

As emphasized in \cite{Aniceto:2011nu}, the powerful relations (\ref{dispersion45}) come about by the straightforward incorporation of all multi-instanton (multi-bion in our case) sectors in the asymptotic formulas.  In \cite{Aniceto:2011nu} these are derived by using \'Ecalle's ``alien calculus"; in our case, this result came about from our improved knowledge of the topological molecule and neutral bion amplitudes.

\paragraph{Implications of resurgence for extended supersymmetric theories.}

Note that there are also theories whose symmetries or dynamics prevent neutral topological molecules from being generated.  Two examples are 4-d $\cN{=}2$ and $\cN{=}4$ superYang-Mills compactified on $\R^3 \times S^1$.  No superpotential is generated on the Coulomb branch and thus no neutral bion effects are present.  
Since in the Borel-\'Ecalle framework, the possible ambiguities in perturbation theory are cancelled by the ambiguities of the neutral bion amplitudes, the absence of neutral molecules in $\cN{=}2$ and  $\cN{=}4$ superYang-Mills implies a better behaved perturbative expansion.  More precisely, the existence of monopole-instantons indicates that the perturbation theory gives a divergent asymptotic series.  However, it does not tell us whether the series is alternating (Borel summable) or non-alternating (non-Borel summable).  This latter, more delicate issue, is tied to the presence or absence of neutral topological molecules.  The absence of the neutral molecules in $\cN{=}2$ and $\cN{=}4$ superYang-Mills implies that both the expansion around the perturbative vacuum as well as the perturbation series around the instanton sectors are Borel summable.  This argument is complementary to and in agreement with exact results in certain extended supersymmetric theories \cite{Russo:2012kj}.

\subsection{Can we non-perturbatively define QFTs in the continuum?}

Currently the only general non-perturbative definition of QFTs is through a lattice formulation.  Lattice field theory is indeed a remarkable resource for QFTs; however, it has well-known difficulties with theories with chiral fermion content, with general supersymmetric theories, and with the topological $\th$-term.  Furthermore, to the extent that it relies on the notion of an RG universality class, it is an indirect definition.

We would like to know if a general non-perturbative continuum definition of an interacting QFT is possible on $\R^d$, $d \geq 2$.  Establishing that this is so is an outstanding problem of mathematical physics.  So far non-perturbative continuum definitions are only known for a restricted set of minimal conformal or integrable models in two dimensions.  But these definitions take the form of self-consistent solutions for complete S-matrices or operator algebras, and it seems doubtful that the bootstrap techniques that underlie these solutions can be applied to general classes of theories (e.g., with a number of adjustable parameters).

Resurgence theory is a relatively new and powerful mathematical and physical idea.  The combination of generalizing the BZJ prescription to all orders in the instanton expansion together with the technique of Borel-\'Ecalle summation of transseries offers the promise of a finite definition of this class of field theories from their semi-classical expansions.  Furthermore, small-circle compactifications of 4-d asymptotically free gauge theories give a large class of theories with well-defined semi-classical expansions.  Also, large-$N$ volume independence indicates that the small-radius semi-classical behavior may be smoothly continuable to large radii in a large subset of these theories.  Together all these ingredients serve at the very least to give a new perspective on the meaning of continuum field theory.


The BZJ prescription in quantum mechanics was more or less concurrently discovered with \'Ecalle's work in the late 1970's and early 1980's.  Since its discovery, resurgence has had many fruitful applications in diverse parts of physics and mathematics, including linear and non-linear ordinary differential equations, WKB methods, Navier-Stokes equations of fluid dynamics, discrete dynamical systems, separatrix splitting,  Kolmogorov-Arnold-Moser theory, optics, statistical mechanics --- i.e., any field which benefits from a saddle point approximation and its improvements.  (See, for example, \cite{s07, Costin:2009, b99, cnp93}.)

The realization of the utility and importance of resurgent functions in quantum field theory and string theory, however, is quite recent.  A few interesting works have appeared recently, predominantly in the context of matrix models and minimal strings by Mari\~no, Schiappa, and collaborators \cite{Marino:2008ya, Pasquetti:2009jg, Aniceto:2011nu}.  All these works address theories without renormalons. 
 
In the context of asymptotically free confining field theories with renormalons, the current work and its two-dimensional companion \cite{Dunne}, to our knowledge, are the first ones combining ideas about resurgence and semi-classical analysis of gauge field theories.   Admittedly, in the present work, we have not used the full power of the resurgence formalism.  By contrast, the very recent work \cite{Aniceto:2011nu} benefits more from the formalism by extending the theory into the complex coupling constant plane, and by studying singularities in the whole complex Borel plane for complex values of coupling constant.  The study of the ``alien (or singularity) calculus" and the bridge equations provides crucial non-perturbative data needed to give a non-perturbative definition of the theory. 
  
In QCD(adj), the lack of development of the machinery of bridge equations and resurgence relations is partly compensated for by our knowledge of the elementary and molecular topological defects.  At present, we have a fair knowledge of the non-perturbative saddle points in gauge theory on small $\R^3 \times S^1$ due to a program that began in \cite{Unsal:2007jx} where magnetic bions were understood.  In this work we throughly analyzed neutral bions and molecular bion--anti-bion events through the BZJ prescription.  The ambiguity associated with certain neutral topological defects is the extra bit of non-perturbative information that we have, in order to define the theory for the real positive coupling and its infinitesimal imaginary neighborhood.  In this regime, whenever the Borel  sum exhibits a Stokes' jump, a topological molecule amplitude also exhibits a jump in the opposite direction rendering the physical observables, such as mass gap and string tension, real and meaningful.


\acknowledgments

It is a pleasure to thank S. Das,  G. Dunne,  
M. Golterman,  D. Gross, D. Harlow,  
S. Hellerman,  D. Kharzeev, C. Korthals-Altes,  M. Mari\~no, 
S. Peris, E. Poppitz, J. Russo, R. Schiappa, N. Seiberg, G. Semenoff, A. Shapere, Y. Tachikawa, R. Wijewardhana, and L. Yaffe for helpful comments and conversations.  We are especially thankful to G. Dunne for his guidance in the hyperasymptotics literature, and to M. Mari\~no and R. Schiappa for their help with the literature on resurgence theory.  The work of PCA is supported in part by DOE grant FG02-84-ER40153.
 
\appendix

\section{Properties of simple Lie algebras and groups\label{secA}}

We assemble here some basic facts about Lie algebras and their associated compact groups that are useful for the body of the paper.  We also include a number of comments on the relation of some of the mathematical language to terminology and conventions appearing in the physics literature.  Some texts covering this subject that we found useful are \cite{Samelson:1990, Fuchs:1992, Humphreys:1972}.

\subsection{Compact groups with simple Lie algebras and charge lattices\label{secA1}}

For each simple Lie algebra, $\gf$, there is a simply-connected compact Lie group, $\tG$.  There are other compact Lie groups, $G$, with Lie algebra $\gf$, given by quotients of $\tG$ by various subgroups of its center, $Z(\tG)$.  $Z(\tG)$ is always a finite abelian group; the possibilities are listed in table \ref{tabA1} below.  
Any subgroup $\cC\subset Z(\tG)$ is a normal subgroup of $\tG$, so defines another group $G:=\tG/\cC$.  $G$ has the same Lie algebra as $\tG$, but $G$ has smaller center, $Z(G):=Z(\tG)/\cC$, and is not simply connected, but has $\pi_1(G) = \cC$.   Thus, in particular, for any group $G$, $Z(G)\ltimes\pi_1(G)=Z(\tG)$.

Any irreducible representation of $\tG$ (other than the trivial representation) represents all the elements of $\tG$ faithfully except perhaps for a subgroup of the center which is represented by the identity element.  So only those irreps which represent $Z(\tG)/Z(G)$ by the identity are irreps of a given global form $G$.

The set of irreps of a Lie group $G$ is reflected in the set of allowed weights of their Lie algebra generators.  In physical terms, these weights correspond to the set of allowed electric charges of fields and sources that can appear in the theory in a Higgs phase where $G\to U(1)^r$.

A generic Lie algebra element $h\in\gf$ determines a unique Cartan subalgebra (CSA) $\tf\subset\gf$ containing $h$.  A CSA is a maximal commuting subspace of $\gf$ and is always of dimension $r=\text{rank}(\gf)$.   Any two CSAs can be mapped to each other by conjugation by some Lie algebra element.   In a given irrep $R$, the representation matrices of $h\in\tf$ can be simultaneously diagonalized giving vectors $\l\in \tf^*$ of simultaneous eigenvalues so that $\l(h)$ is an eigenvalue of $R(h)$.  The set $\{\l\}$ are called the weights of $R$, and their integral span generates a lattice $\G_R\subset\tf^*$, the weight lattice of $R$.  Here $\tf^*$ is the real linear dual of $\tf$ (i.e., the space of linear maps from $\tf$ to $\R$) and $\G^*$ will denote the lattice integrally dual to $\G$ (i.e., $\G$ is the space of linear maps from $\G^*$ to $\Z$).

The group lattice, $\G_G$, is defined to be the union of the weight lattices for all irreps $R$ of $G$,  $\G_G := \cup_R \G_R$, (though, in fact, the union of only a finite number of irreps suffices).

Exponentiation identifies a given CSA, $\tf\subset\gf$, with a maximal torus $T_G \simeq U(1)^r \subset G$. In particular, the eigenvalues in irrep $R$ of a given element $g\in T_G$ are given by $\exp\{2\pi i \l(h)\}$ for some $h\in\tf$ and for $\l\in\G_R$.  The periodicities of the maximal torus are reflected in the lattice of points in the CSA which are mapped to the identity under exponentiation.  For irrep $R$, these points are those $h\in\tf$ such that $\l(h)\in\Z$ for all $\l\in\G_R$.  These $h$ define the dual lattice $\G^*_R\subset \tf$.  The periodicity common to all irreps of $G$ then defines the dual group lattice $\G^*_G = \cap_R \G^*_R$, and the exponential map identifies $T_G \simeq \tf/\G^*_G$.

The smallest (coarsest) possible group lattice is the root lattice, $\G_r$, which is the weight lattice of the adjoint irrep of $\gf$.  It occurs as the group lattice of the ``adjoint group" which is the compact form of the group which has trivial center
\begin{align}\label{}
\Gad := \tG/Z(\tG) . 
\end{align} 
The largest (finest) possible lattice is called \emph{the} weight lattice of $\gf$, $\G_w$, and is the group lattice of the unique simply-connected covering group $\tG$. 
\begin{table}[ht]
\begin{center}
\begin{tabular}{|c|c|c|c|} \hline
$\gf$   & \phantom{\framebox{$\tG$}}$\tG$\phantom{\framebox{$\tG$}}  
&$\Gad :=\tG/Z(\tG)$ &$\pi_1(\Gad)=Z(\tG)$ \\
\hline\hline
$A_{N-1}$ 	& $\text{\it SU}(N)$    	& $\text{\it PSU}(N)$
& $\Z_N$\\
$B_N$	& $\text{\it Spin}(2N+1)$& $\text{\it SO}(2N+1)$    
& $\Z_2$\\
$C_N$	&  $\text{\it Sp}(N)$ or $\text{\it USp}(2N)$ & $\text{\it PSp}(N)$       
& $\Z_2$\\
$D_{2N}$	& $\text{\it Spin}(4N)$	& $\text{\it PSO}(4N)$
& $\Z_2\times\Z_2$\\
$D_{2N+1}$& $\text{\it Spin}(4N+2)$& $\text{\it PSO}(4N+2)$
& $\Z_4$\\
$E_6$		& $E_6$    	& $E_6^{-78}$ & $\Z_3$   \\ 
$E_7$		& $E_7$	& $E_7^{-133}$ & $\Z_2$  \\ 
$E_8$		& $E_8$    	& $E_8$  		& $1$   \\ 
$F_4$ 	& $F_4$    	& $F_4$ 		& $1$  \\ 
$G_2$ 	& $G_2$   	& $G_2$ 		&  $1$  \\ 
\hline
\end{tabular}
\caption{The simple Lie algebras $\gf$ together with common names for their associated compact simply-connected Lie groups $\tG$ and the compact adjoint Lie groups $\Gad$.\label{tabA1}}
\end{center}
\end{table}

From these definitions it follows that the group lattice, $\G_G$, is intermediate between the root and weight lattices of $\gf$ and determines the center and fundamental groups of $G$, $Z(G)$ and $\pi_1(G)$ respectively, by
\begin{align}\label{lattices}
\begin{matrix}
   &&\G_r & \subset & \G_G & \subset & \G_w & \subset &\tf^* \\
   &&\updownarrow* &&\updownarrow*&&\updownarrow* &\\
\tf&\supset &\G^\v_w &\supset &\G^*_G &\supset &\G^\v_r &\\
\end{matrix}
\text{ with }
\left\{\begin{matrix}
Z(G) \ =& \G_G/\G_r \ =&  \G^\v_w/\G^*_G \\
\\
\pi_1(G) \ =& \G_w/\G_G \ =& \G^*_G/\G^\v_r \\
\end{matrix}\right.\ ,
\end{align}
where the lattices connected by vertical arrows are integrally dual.  
We call the various lattices
\begin{align}\label{}
\G_w &= \text{weight lattice,} &
\G^\v_w &= \text{co-weight lattice,}
\nonumber\\
\G_r &= \text{root lattice,} &
\G^\v_r &= \text{co-root lattice,}
\nonumber\\
\G_G &= \text{group lattice,}  &
\G^*_G &= \text{dual group lattice,}
\nonumber
\end{align}
though other names are often used, e.g., ``magnetic weight lattice" for ``co-weight lattice", and ``weight lattice of $G$" for ``group lattice".  

Sometimes, rather confusingly, the co-lattices are called dual lattices; in these cases ``dual" refers to the more special notion of Goddard-Nuyts-Olive (GNO) duality (also known as electric-magnetic or Langlands duality).  GNO duality has its expression in the lattice isomorphisms
\begin{align}\label{GNOduality}
\G_w(\gf) &\simeq \G^\v_w(\gf^\v)
\quad\text{and}\quad
\G_r(\gf) \simeq \G^\v_r(\gf^\v)
\end{align} 
where
\begin{align}\label{GNOduality2}
\gf^\v &:= \gf, \quad \gf\in\{A_n, D_n, E_n,F_4,G_2\},
\quad\text{but}\quad (B_n)^\v := C_n 
\quad\text{and}\quad (C_n)^\v := B_n.
\end{align}
The isomorphisms in (\ref{GNOduality}) are as lattices with inner product up to overall scaling and rotation.  It can be extended to the group lattices,
\begin{align}\label{}
\G_G^* \simeq \G_{G^\v},
\end{align}
where the GNO-dual group, $G^\v$, is the compact Lie group with Lie algebra $\gf^\v$ such that $Z(G^\v)=\pi_1(G)$ and $\pi_1(G^\v)=Z(G)$.  It follows, in particular, that $(G^\v)^\v = G$ and $(\tG)^\v = (G^\v)_\ad$ and $(\Gad)^\v = \til{G^\v}$.

\subsection{Gauge groups, gauge transformations, and center symmetry\label{secA2}}

In a theory with gauge algebra $\gf$, which compact form of the gauge group, $G$, appears is determined by which representations of the Lie algebra both the dynamical fields in the theory as well as any non-dynamical (or very massive) sources belong to.   We will refer to the dynamical fields just as ``fields" and the non-dynamical sources as ``probes" in what follows.  

$G$ has to be at least large enough to admit all the representations of the fields.  Choosing a larger $G$ allows the inclusion of probes in representations other than those of the fields.  Including such probes (enlarging the gauge group) is a matter of choice, reflecting what questions we are allowed to ask of the theory, but should have no effect on the dynamics of the fields.

In the case of QCD(adj) where all the fields are in the adjoint representation, the smallest allowed gauge group is $G=\Gad$.  The $\tG$ form of the QCD(adj) theory admits probes, such as Wilson line operators, in arbitrary representations, while the $\Gad$ theory only admits probes in the adjoint representation (or in representations with weights in the root lattice).  Note that the action of $\tG$ on the fields of QCD(adj) is not faithful since $Z(\tG)$ acts trivially on all adjoint fields.   On the other hand, the action of $\Gad$ on QCD(adj) is faithful:  for any $h \in \Gad$ (and $h\neq1$), there is some field value $\Psi$ such that $h\cdot\Psi \neq \Psi$.

In addition to the choice of the compact form of the gauge group, $G$, there is a separate choice of the group, $\cG$, of gauge transformations.  $\cG$ consists of maps $g(x)$ from space-time $M$ into $G$ that leaves the theory's action invariant when acting point-wise on the fields.  $\cG$ is a group under point-wise multiplication, $g\cdot g'(x)=g(x) \cdot g'(x)$, and has a point-wise action on the fields, $\Psi$, of the theory, $g\cdot\Psi(x) = g(x)\cdot\Psi(x)$, where the multiplication on the right is the group action of $G$ on the representation space that $\Psi$ is valued in.  Unless $M$ is just a point, $\cG$ is much larger than $G$; e.g.\ $\cG=G^N$ for a lattice theory with $N$ lattice points, and is infinite dimensional in the continuum case.  Furthermore, $\cG$ depends not only on $G$, $M$, and on the theory in question, but can also depend on some discrete choices.  For example, one can choose $\cG = \cG_0$ to consist of only those maps $g(x)$ which are continuously connected to the identity map, or, at the opposite extreme, take $\cG=\tcG$, the union of all the connected components of the group of maps.

Consider first the extreme choices where the gauge group $G$ is taken to be either the largest possible, $\tG$, or the smallest possible, $\Gad$.  These are already distinct for $\tG = SU(2)$; for $\tG =  SU(NM)$ or $Spin(2N)$, there are more possibilities intermediate between these extremes, since then the center of $\tG$ has proper subgroups.

\paragraph*{Adjoint group.}

When $G = \Gad$ and $M = \R^3\times S^1$, then $\tcG$ is the set of all continuous maps from $M$ to $G$.  Discontinuous maps from $M$ to $G$ cannot be included in the set of gauge transformations since if $g$ were discontinuous at $x$ by some $h \in \Gad$ ($h{\neq}1$), i.e., $\lim_{\e\to0} [ g(x{+}\e) = h\cdot g(x{-}\e) ]$, then since $\Gad$ acts faithfully on QCD(adj), such a gauge transformation would map a continuous field configuration $\Psi(x)$ to a discontinuous one where $h\cdot\Psi(x) \neq \Psi(x)$.   

$\tcG$ has disconnected components labelled by the elements of $\pi_1(\Gad)$ since any map $g: M \to \Gad$ can be continuously deformed to a map $g': S^1 \to \Gad$ (since $\R^3$ is contractible) and the homotopy classes of these maps are labelled  by elements of $\pi_1(\Gad)$.  Denote by $g_c(x)$ a map in the homotopy class corresponding to $c\in\pi_1(\Gad)$.  If $c\neq 1$, such maps are called ``large gauge transformations".  $\cG_0$ is the component of $\tcG$ connected to the identity map $g(x) = 1$.  As groups, $\tcG/\cG_0 \simeq \pi_1(\Gad)$. 

Convenient representative gauge maps $g_c(x)$ are the following maps that take values solely in a maximal torus $\exp(i\tf)$:
\begin{align}\label{gmu}
g_\m(\x) := \exp i\x \m, \qquad \m \in \G_w^\v,
\end{align}
where $\x := 2\pi x^4/L$ is the angular variable around the $S^1$.  Since $\G_w^\v = \G_r^*$, $\exp2\pi i\m = 1$ in $\Gad$, so $g_\m(\x)$ simply maps the $S^1$ to a non-trivial cycle of the maximal torus.  The homotopy class $c\in\pi_1(\Gad)$ is given by $c \simeq [\m]\in\G_w^\v/\G_r^\v$, the coset that $\m$ belongs to.

If we choose $\cG_0$ as the group of gauge transformations, then the large gauge maps in the other components of $\tcG$ are not gauged.  In particular $\tcG/\cG_0 \simeq\pi_1(\Gad)$ will act on the theory as a global discrete symmetry, and enlarging the group of gauge transformations to be $\tcG$ is equivalent to gauging this discrete symmetry, i.e., projecting the Hilbert space of the theory onto only the discrete symmetry singlet states.  

\paragraph*{Covering group.}

When $G = \tG$ and $M = \R^3\times S^1$, then $\tcG$ is the set of all potentially discontinuous maps from $M$ to $G$ but which map all continuous fields to continuous fields.  Since only the center, $Z(\tG)$, of $\tG$ acts trivially on the fields of QCD(adj), the only allowed discontinuities in $\tcG$ are by elements of $Z(\tG)$.  

$\cG_0$ is the component of $\tcG$ which is continuously connected to the identity map.  Note that discontinuous maps can still be continuously connected to the identity.  Any\footnote{The following argument assumes that the locus of discontinuity of the maps in $\tcG$ is nowhere dense in $M$.  For if not, then upon doing the deformation to $\R^3\times p$ there can be an accumulation of an infinite number of discontinuities whose product may not converge to a definite net discontinuity.  Issues like this may give a reason to prefer using the smallest gauge group, faithful on the fields, over non-faithful ones like $\tG$ for QCD(adj).} discontinuous map $g$ from $\R^3\times S^1 \to \tG$ with jumps only in $Z(\tG)$ can be deformed to a map $g'$ where the locus of all discontinuities is shrunk to an arbitrarily small neighborhood of $\R^3\times p$ where $p$ is a point on the $S^1$.  At any given point of this locus there will be a net discontinuity which will be an element $c \in Z(\tG)$.  Since $Z(\tG)$ is discrete and since away from the $\R^3\times p$ locus $g'$ is continuous, it follows that $c$ must be the same along the whole of $\R^3\times p$.  If $c=1$ then $g'$ is continuous and is deformable to the identity in $\tcG$ since $\pi_1(\tG) = 1$; while if $c\neq1$ then $g'$ is not deformable to the identity.  In summary, the disconnected components of $\tcG$ are labelled by elements $c$ of $Z(\tG)$, and can all be characterized as maps $g_c(x,\x)$ which are discontinuous by $c$ around the $S^1$: $g_c(x,\x+2\pi) = c\cdot g_c(x,\x)$.  As groups, $\tcG/\cG_0 \simeq Z(\tG)$.  Convenient representative maps can again be taken to be the $g_\m(\x)$ in (\ref{gmu}) which are now discontinuous when $\m$ belongs to non-trivial cosets $0\neq[\m]\in\G_w^\v/\G_r^\v$, and $[\m]\simeq c\in Z(\tG)$.

If we choose $\cG_0$ as our group of gauge transformations, then the large transformations in the other components of $\tcG$ are not gauged.  In particular $\tcG/\cG_0 \simeq Z(\tG)$ will act on the theory as a global discrete symmetry, and enlarging the group of gauge transformations to be $\tcG$ is equivalent to gauging this discrete symmetry.

\paragraph*{General group.} 

The general compact form of the group can have both a non-trivial fundamental group $\pi_1(G)$ and center group $Z(G)$.  In this case there are both large and discontinuous gauge transformations, and there is a global discrete symmetry $\tcG/\cG_0 \simeq Z(G) \ltimes \pi_1(G)$.  Thus in all cases the discrete symmetry is isomorphic to the center of the covering group, $Z(\tG)$, with representative gauge maps $g_c$, $c\in Z(\tG)$ given by (\ref{gmu}) with $c \simeq[\m]\in\G^\v_w/\G^\v_r$.  This discrete symmetry is called the \emph{center symmetry} of QCD(adj).  Enlarging the group of gauge transformations to $\tcG$ is equivalent to gauging the center symmetry.  

As discussed in section \ref{sec2.4}, the center symmetry acts on point electric operators of the effective 3-d theory in the interior of the gauge cell.  (There is also a separate dual center symmetry which acts on magnetic operators of the effective theory.)

\subsection{Roots, Kac labels, Killing form, and co-roots\label{secA3}}

The roots, $\Phi$, are the non-zero weights of the adjoint representation of $\gf$.  They are a set of special non-zero elements of the root lattice, $\{ \a \}$, which are in 1-to-1 correspondence with a basis of generators of $\gf$ not in $\tf$.  $\Phi$ has the property that one can choose (not uniquely) a subset of $r=\text{rank}(\gf)$ simple roots, $\Phi_s :=\{ \a_i, \ i=1,\ldots,r \}$, which are a basis of $\G_r$ and which separate the roots into two disjoint sets: the positive roots, $\Phi_+$, which are those roots which can be written as non-negative integer linear combinations of the simple roots; and the negative roots, $\Phi_-$, which are the negatives of the positive roots.

Given a choice of simple roots, $\Phi_s$, there is a unique lowest root, $\a_0$, such that all other roots are found by adding non-negative integer sums of simple roots to $\a_0$.  This procedure in fact determines the root system $\Phi$ from $\Phi_s$ and $\a_0$.  Thus 
\begin{align}\label{marks}
\sum_{i=0}^r k_i \a_i=0, 
\qquad k_0 := 1,
\end{align}
for some non-negative integers $k_i$ called the Kac labels (or marks, or sometimes Coxeter labels) of the $\a_i$.  The sum of the Kac labels,
\begin{align}\label{defcoxeter}
h := \sum_{i=0}^r k_i,
\end{align}
is called the Coxeter number. 

In addition to its linear structure, $\tf$ comes with a positive definite real inner product inherited from the Killing form on $\gf$: $(e,f):=\tr({\rm ad}(e){\rm ad}(f))$ for $e,f\in\gf$.  Upon restricting to a CSA $\tf$, one finds that $(\m,\n) = \sum_{\a\in\Phi} \a(\m)\a(\n)$ for $\m,\n\in\tf$.  This inner product is defined up to a single overall normalization for simple $\gf$.  

Choosing a normalization, the inner product can be used to select a canonical identification between $\tf$ and its dual $\tf^*$.  In particular, to each $\l\in\tf^*$, define $\l^*\in\tf$ by
\begin{align}
(\l^*,\f) := \l(\f)\quad \forall \f\in\tf.
\end{align}
Likewise, $\tf^*$ inherits an inner product from $\tf$ via the duality map:
\begin{align}
(\l,\m) :=(\l^*,\m^*) = \l(\m^*)\quad\forall\l,\m\in\tf^*.
\end{align}
(Note that we are using the same symbol for the Killing form on $\tf$ as for the inverse Killing form on $\tf^*$.)

Co-roots, $\a^\v$, are elements of $\tf$ associated to each root, and are defined by
\begin{align}\label{corootmap}
\a^\v := \frac{2\a^*}{(\a,\a)} ,
\end{align}
which is independent of the normalization of the Killing form.  When $\a,\b\in\Phi$, then $\b(\a^\v)$ are integers for all simple Lie algebras.
\begin{align}\label{cartanmatrix}
A_{ij} := \a_i(\a_j^\v), \qquad \a_i \in \Phi_s,
\end{align}
are the elements of the Cartan matrix of the algebra.  By including a row and column for the lowest root $\a_0$ in the same way, one defines the extended Cartan matrix.

The charge lattices described in the last subsection can be computed as follows.  The root lattice, $\G_r$, is the integral span of the simple roots $\{\a_i\}$.  The co-root lattice, $\G^\v_r$, is spanned by the simple co-roots $\{\a^\v_i\}$.  The weight lattice, $\G_w$ is spanned by the fundamental weights $\{\o_i\}$ defined by $\o_i(\a^\v_j) = \d_{ij}$.  Finally the co-weight lattice, $\G^\v_w$, is spanned by the fundamental co-weights $\{\o^\v_i\}$ defined by $\a_i(\o_j^\v)=\d_{ij}$, or, equivalently, by $\o_i^\v = 2\o_i^*/(\a_i,\a_i)$.

The extended Dynkin diagram associated to a Lie algebra consists of nodes corresponding to each simple root and to the lowest root, together with $A_{ij}A_{ji}$ lines linking the $i$th and $j$th nodes.   The extended Dynkin diagrams with the Kac labels for all simple Lie algebras are shown in figure \ref{fig-dynkin}.

The dual Kac labels (co-marks), $k^\v_i$, are defined analogously to the Kac labels by
\begin{align}\label{comarks}
\sum_{i=0}^r k^\v_i \a^\v_i =0,
\qquad k^\v_0 :=1,
\end{align}
and the dual Coxeter number is their sum
\begin{align}\label{defdualcox}
h^\v := \sum_{i=1}^r k^\v_i .
\end{align}
For simply-laced algebras, the Kac labels and their duals are the same.  For non-simply-laced algebras, if the ratio of the lengths-squared of the long roots to the short roots is $p$, then the dual Kac labels for the short roots are $1/p$ times their Kac labels, and are the same for the long roots.  The dual Kac labels are integers by virtue of the integrality of the Cartan matrix.  The Coxeter numbers and dual Coxeter numbers for all simple Lie algebras are given in table \ref{tabA2}.  The dual Coexeter numbers also satisfy the identity
\begin{align}\label{}
r h^\v = n_L + (S/L)^2\, n_S
\end{align}
where $r:=\text{rank}(\gf)$, $n_{L,S}:=$ number of long and short roots, and $L,S:=$ lengths of long and short roots.
\begin{table}[ht]
\begin{center}
\begin{tabular}{|c||c c c c c c c c c|} \hline
$\gf$          & 
$A_r$ & $B_{r>1}$ & $C_r$ & $D_{r>2}$ & $E_6$ & $E_7$ & $E_8$ &$F_4$ & $G_2$\\
\hline\hline
$h$      &
$r{+}1$ & $2r$ & $2r$ & $2r{-}2$ & 12 & 18 & 30 & 12 & 6\\
$h^\v$	&
$r{+}1$ & $2r{-}1$ & $r{+}1$ & $2r{-}2$ & 12 & 18 & 30 & 9 & 4\\
\hline
\end{tabular}
\caption{Coxeter and dual Coxeter numbers of the simple Lie algebras.\label{tabA2}}
\end{center}
\end{table}

The Killing form only appears in the gauge theory lagrangian (\ref{LUV}) multiplied by $1/g^2$, so its normalization can always be absorbed in a rescaling of the gauge coupling.  The instanton number can be written in a normalization-independent way as \cite{Bernard:1977nr} 
\begin{align}\label{}
\n := \frac{(\a_0,\a_0)}{4} \frac{1}{(2\pi)^2}
\int (F \overset{\wedge}{,} F)
=
\frac{(\a_0,\a_0)}{32\pi^2} \int d^4x\,(F_{\m\n},\til F_{\m\n}) \in \Z,
\end{align}
where $\a_0$ is a long root.  (This is normalization-independent since $\a_0\in\tf^*$ and $F_{\m\n}\in\tf$, so the two Killing forms are inverses of one another.)  With this definition, the instanton number, $\n$, is an integer for finite action configurations on $\R^4$ and a $\n=1$ configuration exists for all $\gf$.  Then with the theta-angle term in the action, $S_\th := i\th\n$, $\th$ has period $2\pi$.

Nevertheless, when considering instanton configurations it is convenient (and conventional) to fix a particular normalization of the Killing form (or, equivalently, of the coupling constant) such that the one-instanton action is $8\pi^2/g^2$.  This normalization corresponds to the one in which long roots have length $\sqrt2$:
\begin{align}\label{}
(\a_0,\a_0)=2.
\end{align}

The Dynkin index of the representation $R$, denoted $T(R)$, is expressed in terms of the weights, $\l$, of $R$ by
\begin{align}
T(R) = \tfrac{1}{r} \tsum_{\l\in R} (\l,\l) .
\end{align}
In the above normalization, the index of the adjoint representation is given by the dual Coxeter number,
\begin{align}\label{}
T(\ad) = 2 h^\v ,
\end{align}
and, in general, with this normalization $T(R)$ is an integer which counts the number of zero modes of the Dirac equation for spin-1/2 fermions in the representation $R$ in a 1-instanton background.

\subsection{Affine Weyl chambers and gauge cells\label{secA4}}

Further gauge identifications on $\tf$ are provided by the Weyl group, $W(\gf)$.  $W$ is the group of real linear transformations on $\tf$ which preserves $\Phi$ (i.e., permutes the roots).  It includes a reflection $\s_\a$ for each $\a\in\Phi$ which acts on $\tf^*$ as $\s_\a(\m) := \m - \m(\a^\v)\, \a$ for $\m\in\tf^*$.  $W$ is generated by $\s_\a$ with $\a \in \Phi _s$.  $W$ acts on $\tf$ by defining $\s_\a(\m)(\f) = \m(\s_\a(\f))$, which gives
\begin{align}
\s_\a(\f) := \f-2\a(\f)\, \a^*,\  \text{for}\ \f\in \tf .
\end{align}
This reflection fixes the hyperplane $\a(\f)=0$ through the origin in $\tf$ perpendicular to the root $\a$.

$\tf$ is identified under gauge transformations generated by $\G^\v_r$ lattice translations and $W$ transformations, which together generate the residual discrete gauge group,
\begin{align}
\hW := W\ltimes\G^\v_r .
\end{align}
We call a fundamental domain of the action of $\hW$ on $\tf$ a ``gauge cell" and denote it by
\begin{align}
\hT \simeq \tf/\hW.
\end{align} 
A fundamental domain of the action of $\hW$ on $\tf$ is also called an affine Weyl chamber.  A conveneint choice of affine Weyl chamber is
\begin{align}\label{affineweylchamber}
\hT :=\{ \f\in\tf \ |\ 0\le\a(\f)\ \text{for all}\ \a\in\Phi_s,\ \text{and}\ -\a_0(\f)\le 1 \}
 \simeq \tf/\hW ,
\end{align}
where $\a_0$ is the lowest root \cite{Humphreys:1990}.  The $\a_0(\f)=-1$ wall of $\hT$ are those $\f$ fixed by a combination of a $\G^\v_r$ translation and a $\s_{\a_0}$ Weyl reflection.

The gauge cell $\hT$ is the object we are interested in, since it is the CSA modulo gauge equivalences.  At points in the interior of $\hT$ the unbroken gauge group is $U(1)^r$, while at points on its boundary (i.e., points fixed by some element of $\hW$) the unbroken gauge group will be enhanced.

\section{Explicit root systems and gauge cells for the simple Lie algebras\label{secB}}

\begin{figure}
\begin{center}
\includegraphics[width=20em]{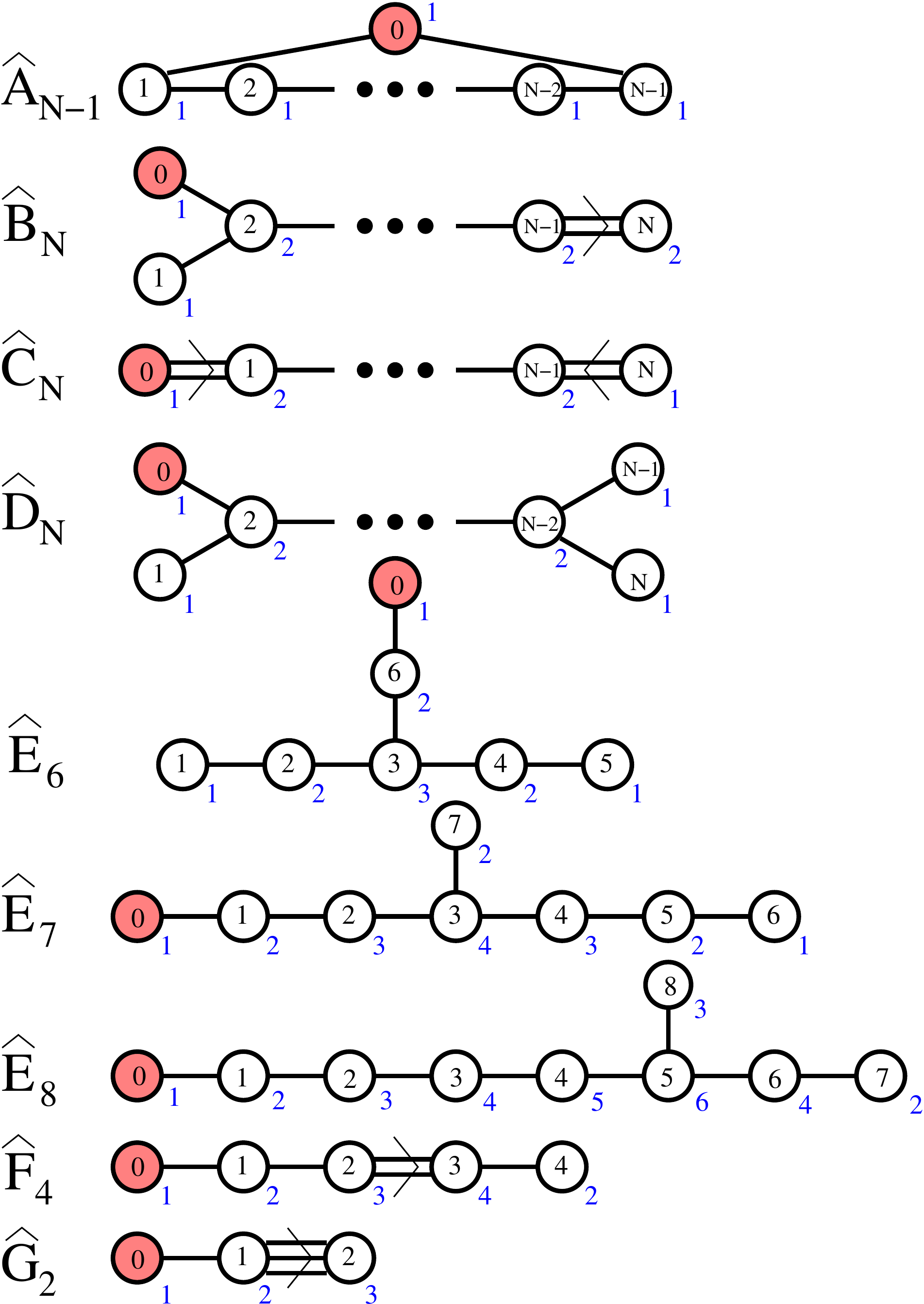}
\caption{Extended Dynkin diagrams for the simple Lie algebras.  Numbers inside the nodes label the simple roots, while the red node with label ``0" is the lowest root.  Numbers besides the nodes are the Kac labels.\label{fig-dynkin}}
\end{center}
\end{figure}

The extended (or untwisted affine) Dynkin diagrams for all simple Lie algebras are shown in figure \ref{fig-dynkin}, with an arbitrary labeling of the simple roots (we follow Dynkin) and the lowest root together with their Kac labels.  The extended Dynkin diagrams have multiple uses.  (Indeed, the whole associated Lie algebra can be reconstructed from them.)  Below we will use them to construct the roots systems of the simple Lie algebras in an explicit basis, and give a coordinate description of their gauge cells.

It is also useful to note that the center of $\tG$ can be read off from the extended Dynkin diagram as the group of diagram symmetries (i.e., disregarding node labels) modulo the symmetry group of the Dynkin diagram with the lowest root node eliminated.  For example, the $\hat E_6$ diagram symmetry is $S_3$, permuting the three ``legs" of the diagram, while the $E_6$ diagram symmetry (without the 0-node) is $\Z_2$, switching the 1-2 nodes with the 5-4 nodes.  Thus $Z(\til E_6) = S_3/\Z_2 = \Z_3$.  Furthermore, the $Z(\tG)$ symmetry action on the diagram nodes can be translated directly to translations of the root lattice by weight vectors using the notion of fundamental weights associated to the Dynkin nodes (which we do not describe here).

Finally, the pattern of gauge symmetry breaking due to a given $\vev\f$ can be easily read off from the extended Dynkin diagram.  If the minimum of the 1-loop potential is at $\f\in \hT$, then eliminate from the extended Dynkin diagram those nodes $i$ such that $\a_i(\f)\notin\Z$.  The remaining nodes form the Dynkin diagram of the unbroken semi-simple subgroup; there are also as many unbroken $U(1)$ factors as needed for the rank of the total unbroken subgroup to be $r$.

In what follows we write the simple roots for the Lie algebras in a convenient basis, and then derive the associated gauge cells.  The bases we use for the classical Lie algebras are standard ones (perhaps up to a relabeling) found, e.g., in \cite{Bourbaki:2002, Bremmer:1985}.  For the $E_6$ and $E_7$ exceptional algebras we use the somewhat more convenient bases used by \cite{Conway:1993}.

In all of what follows $\{e_i\}$ is an orthonormal basis of $\R^N\supset\tf^*$ and $\{e^i\}$ is a basis of $(\R^N)^*\supset\tf$ dual to the $\{e_i\}$ so that $e_i(e^j)=\d^j_i$ and the $e^i$ are also orthonormal.   Thus, with respect to this inner product, $e_i^* = e^i$, and a general element $\f\in\tf$ then has the coordinate expansion
\begin{align}
\f=\f_i e^i .
\end{align}
The normalization of the root systems constructed below is chosen for notational convenience (i.e., keeping coordinates rational) and in particular corresponds to lengths-squared of the long roots being 2 for all algebras except $C_N$ and $G_2$ for which instead the short roots have lengths-squared 2.    

We start with the familiar $A_{N-1}=SU(N)$ algebra for which we give some details to show the method, and then just summarize the results for the other algebras.

\subsection{A$_{\bf N-1}$\label{secB1}}

A convenient choice of coordinates realizing the simple roots and highest root (the negative of the lowest root) invariantly summarized in the affine Dynkin diagram in figure \ref{fig-dynkin} is
\begin{align}
\Phi_s &= \{ \a_i := e_i - e_{i+1}\ ,\  1\le i\le N-1\},
\nonumber\\
-\a_0 &= e_1 - e_N = \sum_{i=1}^{N-1} \a_i.
\end{align}
Subtracting simple roots from the highest root then generates the positive roots
\begin{align}\label{}
\Phi_+ &= \{ e_i - e_j\ ,\ 1\le i<j\le N\}.
\end{align}
Note that in these coordinates the roots span only the $\tf^*\simeq\R^{N-1}$ hyperplane consisting of elements $\f^*:= \f^i e_i\in\R^N$ such that $\sum_i \f^i=0$.  Then 
\begin{align}
\tf = \{ \f=\f_i e^i\ |\ \textstyle{\sum_i} \f_i=0\}.
\end{align}

Given the root lattice, the weight lattice and co-root and co-weight lattices can be deduced from the lattice isomorphisms (\ref{lattices}) and (\ref{GNOduality}).  The basis of co-weights $\{\o^{\v j}\}$ dual to the simple roots, defined by $\a_i(\o^{\v j})=\d^j_i$, is $\o^{\v j}=(\sum_{i\le j} e^i ) - \frac{j}{N}\sum_i e^i$.  They generate the lattice $\G^\v_w \in\tf$ with simpler basis $\G^\v_w = \text{span}\{ e^i - \frac1N\sum_j e^j \}$.  By GNO-duality, a basis $\{\a^{\v i}\}$ of the co-root lattice is given by the co-root map (\ref{corootmap}) $\a^{\v i} = (\a_i)^\v = e^i - e^{i+1}$.   These generate the co-root lattice $\G^\v_r = \text{span}\left\{e^i - e^j\right\}$.  The basis of the weight lattice dual to the simple co-roots is $\o_j=(\sum_{i\le j} e_i ) - \frac{j}{N}\sum_i e_i$.  They generate the weight lattice $\G_w$ with simpler basis $\G_w = \text{span}\{ e_i - \frac1N\sum_j e_j \}$.  (Note that, despite our notation, $\o^{\v i} \neq (\o_i)^\v$.  The co-root map (\ref{corootmap}) only maps roots to co-roots.)

The gauge cell (or affine Weyl chamber) (\ref{affineweylchamber}) is then given by 
\begin{align}
\hT 
&= \{ 1+\f_N \ge \f_1 \ge \f_2 \ge \cdots \ge \f_N \ \text{and}\ 
\textstyle{\sum_i} \f_i =0\},
\end{align}
which implies that $0\ge \f_N \ge \tfrac{1}{N}-1$.  The center symmetry $Z(\tG) \simeq \pi_1(\Gad) \simeq \G^\v_w/\G^\v_r \simeq \Z_N$ acts by translations by elements of $\G^\v_w$ modulo $\hW$, equivalence classes of which are given by $\o^j$,  $j\in\{1,\ldots,N\}$ where the $\hW = W\ltimes \G^\v_r$ action is given by combinations of $\G^\v_r$ translations (i.e., by integral linear combinations of the co-roots) and Weyl group elements which act as permutations on the $1\le i,j\le N$ indices.  Thus the $\Z_N$ center symmetry is generated by $\r=[\o^1]$ (or $\o^{N-1}$) acting as $\r: \f_j \to \f_{\pi(j)} +\d_{\pi(j),1} - \frac1N + n_j$ for arbitrary permutation $\pi$ of the indices and integers $n_i$ such that $\sum n_j =0$.  Choosing $\pi$ to be a generator of the cyclic permutation of all $N$ indices, we have
\begin{align}
\r:\quad\f_j \to \f_j' =
\begin{cases}
\f_N+1-\frac1N & \text{for $j=1$,}\\
\f_{j-1}-\frac1N & \text{for $j>1$,}\\
\end{cases}
\end{align}
which is easily checked to map $\hT_\tG$ to itself, and to obey $\r^N=1$.  A fundamental domain of the action of $\r$ on $\hT$, and can be chosen to be
\begin{align}
\hT/Z(\tG) &= \{ \frac{N-1}{2N} \ge \f_1 \ge \f_2 \ge \cdots \ge \f_N\ge -\frac{N-1}{2N} \ \text{and}\ 
\textstyle{\sum_i} \f_i =0\}.
\end{align}

A center-symmetric vacuum is one which is a fixed point of the $\Z_N$ action, $\r(\f)=\f$, for which there is a unique solution:
\begin{align}
\text{f.p.}(\r) = \bigl\{ \f \ \big|\  \f_j = \tfrac{N+1-2j}{2N} \bigr\}.
\end{align}
(When $N$ is not prime there can exist manifolds of points invariant under non-trivial proper subgroups of $\Z_N$ as well.)

\subsection{B$_{\bf N}$}

The $B_N$ root system and gauge cell are
\begin{align}
\Phi_s &= \{ \a_i = e_i - e_{i+1}\ ,\  1\le i\le N-1,\ \text{and}\ \a_N=e_N\},
\nonumber\\
\Phi_+ &= \{ e_i \pm e_j \ ,\ 1\le i<j\le N, \ \text{and}\ e_i\ ,\ 1\le i\le N \},
\nonumber\\
-\a_0 &= e_1 + e_2 = \a_1 + \tsum_{i=2}^N 2 \a_i .
\nonumber\\
\hT &= \{ 1-\f_2 \ge \f_1 \ge \f_2 \ge \cdots \ge \f_N \ge 0 \},
\nonumber\\
\hT/Z(\tG) & = \{ \tfrac12 \ge \f_1 \ge \f_2 \ge \cdots \ge \f_N \ge 0 \}.
\nonumber
\end{align}
The center symmetry $Z(\tG) \simeq \Z_2$ is generated by $\r$ which 
maps $\hT$ to itself with action and fixed points
\begin{align}\label{BZ2}
\r:\ \f_j &\to 
\begin{cases}
1-\f_1 & \text{for $j=1$,}\\
\f_j & \text{for $j>1$,}\\
\end{cases} &
\text{f.p.}(\r) &= \{ \f \ |\ \f_1 = \tfrac12 \}.
\end{align}
Thus the center-symmetric fixed point set has dimension $N-1$.  Points on the boundaries of $\hT$ have the enhanced gauge symmetries
\begin{align}
\f_j=\f_{j+1}=\cdots=\f_{j+n-1} 
\begin{cases}
=0 & \Rightarrow\ \exists\ \text{unbroken}\  SO(2n+1),\\
=\frac12 & \Rightarrow\ \exists\ \text{unbroken}\  SO(2n),\\
\neq0,\, \frac12 & \Rightarrow\ \exists\ \text{unbroken}\  U(n).
\end{cases}
\end{align}

\subsection{C$_{\bf N}$}

The $C_N$ root system and gauge cell are
\begin{align}
\Phi_s &= \{ \a_1 = e_i - e_{i+1}\ ,\  1\le i\le N-1,\ \text{and}\ \a_N = 2e_N\},
\nonumber\\
\Phi_+ &= \{ e_i \pm e_j \ ,\ 1\le i<j\le N, \ \text{and}\ 2e_i\ ,\ 1\le i\le N \},
\nonumber\\
-\a_0 &= 2e_1 = \tsum_{i=1}^{N-1}2\a_i + \a_N.
\nonumber\\
\hT &= \{ \tfrac12 \ge \f_1 \ge \f_2 \ge \cdots \ge \f_N \ge 0 \}.
\nonumber\\
\hT/Z(\tG) &= \{ \tfrac12-\f_N \ge \f_1 \ge \f_2 \ge \cdots \ge \f_N\}.
\nonumber\
\end{align}
$Z(\tG) \simeq \Z_2$ is generated by $\r$ which 
has action and fixed points
\begin{align}\label{CZ2}
\r:\ \f_j &\to \tfrac12 - \f_{N+1-j}, &
\text{f.p.}(\r) &= \{ \f\ |\ \f_j+\f_{N+1-j} = \tfrac12 \}.
\end{align}
The center-symmetric fixed point set has dimension $\lfloor\tfrac N2\rfloor$.  Points on the boundaries of the gauge cell have the enhanced gauge symmetries
\begin{align}
\f_j=\f_{j+1}=\cdots=\f_{j+n-1}
\begin{cases}
=0 & \Rightarrow\ \exists\ \text{unbroken}\  Sp(2n),\\
=\frac12 & \Rightarrow\ \exists\ \text{unbroken}\  SO(2n),\\
\neq0,\, \frac12 & \Rightarrow\ \exists\ \text{unbroken}\  U(n).
\end{cases}
\end{align}

\subsection{D$_{\bf N}$}

The $D_N$ root system and gauge cell are
\begin{align}
\Phi_s &= \{ \a_i = e_i - e_{i+1}\ ,\  1\le i\le N-1,\ \text{and}\ \a_N = e_{N-1}+e_N\},
\nonumber\\
\Phi_+ &= \{ e_i \pm e_j \ ,\ 1\le i<j\le N  \},
\nonumber\\
-\a_0 &= e_1 + e_2 = \a_1 + \tsum_{i=2}^{N-2} 2 \a_i + \a_{N-1}+\a_N .
\nonumber\\
\hT &= \{ 1-\f_2 \ge \f_1 \ge \f_2 \ge \cdots \ge \f_{N-1} \ge |\f_N| \}, 
\nonumber\\
\hT/Z(\tG) &= \{ \tfrac12 \ge \f_1 \ge \f_2 \ge \cdots \ge \f_N \ge 0 \}.
\nonumber
\end{align}
$Z(\tG) \simeq \Z_4$ or $\Z_2\times\Z_2$, depending on whether $N$ is odd or even, respectively.  If $N$ is odd, then a generator $\s$ of $\Z_4$ 
which maps the gauge cell to itself has action and fixed point sets
\begin{align}\label{DZ4}
\s:\ \f_j &\to
\begin{cases}
\frac12+\f_N & \text{for $j=1$,}\\
\frac12-\f_{N+1-j} & \text{for $j>1$,}\\
\end{cases}
\nonumber\\
\text{f.p.}(\s)
&= \{ \f \ |\ \f_1 =\tfrac12, \f_j + \f_{N+1-j}=\tfrac12, \f_N=0 \},
\nonumber\\
\text{f.p.}(\s^2)
&= \{ \f \ |\ \f_1 =\tfrac12, \f_N=0 \},
\end{align}
which are dimension $(N{-}3)/2$ and $(N{-}2)$ subsets of $\hT$, respectively.  If $N$ is even, then generators $\s_\pm$ of each $\Z_2$ factor 
which map the gauge cell to itself have action and fixed point sets
\begin{align}\label{DZ2Z2}
\s_\pm:\ \f_j &\to 
\begin{cases}
\frac12\pm \f_N & \text{for $j=1$,}\\
\frac12-\f_{N+1-j} & \text{for $1<j<N$,}\\
\pm\frac12\pm \f_1 & \text{for $j=N$,}\\
\end{cases}
\nonumber\\
\text{f.p.}(\s_\pm)
&= \{ \f \ |\  \f_1 \pm \f_N = \f_j + \f_{N+1-j} = \tfrac12 ,\ 1<j<N \},
\nonumber\\
\text{f.p.}(\s_+\s_-)
&= \{ \f \ |\ \f_1 =\tfrac12, \f_N=0 \},
\end{align}
which are dimension $N/2$ and $(N{-}2)$ subsets of $\hT_\tG$, respectively.  Note that points on the boundaries at $\f_1=\frac12$ or $\f_N=0$ do not necessarily preserve all or even part of the center symmetry.  Points on the boundaries of $\hT$ have enhanced gauge symmetry, 
\begin{align}
\f_j=\f_{j+1}=\cdots=\f_{j+n-1} 
\begin{cases}
=0\text{ or }\frac12 & \Rightarrow\ \exists\ \text{unbroken}\  SO(2n),\\
\neq0,\, \frac12 & \Rightarrow\ \exists\ \text{unbroken}\  U(n).
\end{cases}
\end{align}

\subsection{Exceptional algebras}

\paragraph{E$_{\bf 8}$.}

There is no center symmetry and
\begin{align}
\Phi_s &= \{
\a_i=e_{i+1}-e_{i+2}\  \text{\small{($1\le i\le6$)}}\ ,\ 
\a_7=\tfrac12(e_1 -\textstyle{\sum_{i=2}^7}e_i+e_8)\  ,\ \a_8=e_7+e_8\}
\nonumber\\
\Phi_+ &= \{ \tfrac12(e_1+\textstyle{\sum_{i=2}^8}(-)^{n_i}e_i) \ \text{\small{(sum $n_i$ even)}}\ ,\ 
e_i\pm e_j \ \text{\small{($1\le i<j\le8$)}} \}
\nonumber\\
-\a_0 &= e_1 + e_2 = 2\a_1+3\a_2+4\a_3+5\a_4+6\a_5+4\a_6+2\a_7+3\a_8
\nonumber\\
\hT &= \{ 1\ge \f_1+\f_2\ ,\ \f_1+\f_8 \ge \textstyle{\sum_{i=2}^7}\f_i\ ,\ \f_2 \ge \cdots \ge \f_7 \ge |\f_8| \}.
\nonumber
\end{align}

\paragraph{E$_{\bf 7}$.}

The weights are all orthogonal to $\tsum_{i=1}^8 e_i$ in $\R^8$, so $\f_i e^i\in \tf$ have $\tsum_{i=1}^8\f_i=0$, which we use to eliminate $\f_1$ in the description of the gauge cell:
\begin{align}
\Phi_s &= \{ 
\a_i = e_{i+1}-e_{i+2}\  \text{\small{($1\le i\le6$)}}\ ,\ 
\a_7 = \tfrac12(-\tsum_{i=1}^4 e_i + \tsum_{i=5}^8 e_i)\}
\nonumber\\
\Phi_+ &= \{ \tfrac12(-e_1{+}\tsum_{i=2}^8 (-)^{n_i} e_i) \ 
\text{\small{(three $n_i$ odd)}},\ 
e_i{-}e_j\ \text{\small{($2\le i<j\le8$)}},\ 
e_i{-}e_1\ \text{\small{($2\le i\le 8$)}} \}
\nonumber\\
-\a_0 &= e_2-e_1 = 2\a_1+3\a_2+4\a_3+3\a_4+2\a_5+\a_6+2\a_7
\nonumber\\
\hT &= \{ 1-\tsum_{i=2}^8\f_i \ge \f_2 \ge \cdots \ge \f_8 \ ,\ 
\tsum_{i=5}^8\f_i \ge 0 \}
\nonumber\\
\hT/Z(\tG) &= \{ 1-\tsum_{i=2}^8\f_i \ge \f_2 \ge \f_3 \ge \f_4 \ge \tfrac18 \ge \f_5 \ge \cdots \ge \f_8 \ ,\ 
\tsum_{i=5}^8\f_i \ge 0 \}.
\nonumber
\end{align}
$Z(\tG)\simeq\Z_2$ is generated by $\r$ which maps $\hT$ to itself with action and fixed points
\begin{align}\label{E7Z2}
\r:\ \f_j &\to 
\begin{cases}
-\f_{9-j} - \frac34 & \text{for $j=1,8$,}\\
-\f_{9-j} + \frac14 & \text{for $2\le j\le7$,}\\
\end{cases}
\nonumber\\
\text{f.p.}(\r)
&= \{ \f \ |\ \f_j + \f_{9-j} = \tfrac14 ,\ 2\le j\le 7 \}.
\end{align}
The center-symmetric fixed point set is a 4-dimensional subset of $\hT$.

%

\paragraph{E$_{\bf 6}$.}

The weights are all orthogonal to $\tsum_{i=1}^8 e_i$ and to $e_1+e_8$ in $\R^8$, so $\f_i e^i\in \tf$ have $\f_1+\f_8 = \tsum_{i=2}^7\f_i=0$, which we can use to eliminate $\f_7$ and $\f_8$ in the description of the gauge cell:
\begin{align}
\Phi_s &= \{
\a_i = e_{i+2}-e_{i+1}\  \text{\small{($1\ge i\ge5$)}}\ ,\ 
\a_6 = \tfrac12(\tsum_{i=1}^4 e_i - \tsum_{i=5}^8 e_i) \}
\nonumber\\
\Phi_+ &= \{ \tfrac12(e_1+\textstyle{\sum_{i=2}^7}(-)^{n_i}e_i - e_8) \ \text{\small{(three odd $n_i$)}},\ e_i- e_j \ \text{\small{($7\ge i>j\ge2$)}},\ e_1-e_8 \}
\nonumber\\
-\a_0 &= e_1-e_8  =
\a_1 + 2\a_2+3\a_3+2\a_4+\a_5+2\a_6
\nonumber\\
\hT &= \{
\tfrac12 \ge\f_1\ge -\tsum_{i=2}^4 \f_i,\ 
-\tsum_{i=2}^6\f_i \ge \f_6 \ge\cdots \ge \f_3 \ge \f_2
\}
\nonumber\\
\hT/Z(\tG) &= \{ \f\in \hT_\tG \ |\ 1{-}2\f_1\ge \f_3{-}\f_2 \ge {-}2\f_6{-}\tsum_{i=2}^5\f_i\ \text{and}\ \f_4{-}\f_3\ge\f_6{-}\f_5\ge\tsum_{i=1}^4\f_i \}.
\nonumber
\end{align}
$Z(\tG)\simeq\Z_3$ is generated by $\r$ which maps the affine Weyl chamber to itself with action and fixed points 
\begin{align}\label{E6Z3}
\r: &\ 
\begin{pmatrix}
\f_1 \\ \f_2 \\ \f_3 \\ \f_4 \\ \f_5 \\ \f_6 
\end{pmatrix}
\to
\begin{pmatrix}
\phantom{+}\tfrac12 +\tfrac12 \f_2 -\tfrac12\f_3
\phantom{\text{}+\tsum_{i=4}^6\f_i}\\ 
-\tfrac16 +\tfrac12 \f_2 +\tfrac12\f_3 +\tsum_{i=4}^6\f_i\\ 
-\tfrac16 -\tfrac12 \f_2 -\tfrac12\f_3 -\f_6
\phantom{\tsum_{i=4}^6}\\ 
-\tfrac16 -\tfrac12 \f_2 -\tfrac12\f_3 -\f_5
\phantom{\tsum_{i=4}^6}\\ 
-\tfrac16 -\tfrac12 \f_2 -\tfrac12\f_3 -\f_4
\phantom{\tsum_{i=4}^6}\\ 
-\tfrac16 +\tfrac12 \f_2 +\tfrac12\f_3 +\f_1
\phantom{\tsum_{i=4}^6}
\end{pmatrix}
\\
\text{f.p.}(\r)
&= \{ \f \ |\ 
\f_2 = -\tfrac23+\f_1 ,\ 
\f_3 = -\f_6=\tfrac13-\f_1 ,\ 
\f_5 = -\f_4 \
 \}.
 \nonumber
\end{align}
The center-symmetric fixed point set is a 2-dimensional subset of $\hT$.

\paragraph{F$_{\bf 4}$.}
There is no center symmetry and
\begin{align}
\Phi_s &= \{ 
\a_1 = e_2{-}e_3\ ,\ \a_2 = e_3{-}e_4 \ ,\ \a_3 = e_4 \ , \ 
\a_4 = \tfrac12(e_1{-}e_2{-}e_3{-}e_4) \},
\nonumber\\
\Phi_+ &= \{ e_i \ ,\ e_i\pm e_j \ ,\ \tfrac12
(e_1{\pm}e_2{\pm}e_3{\pm}e_4) \}
\nonumber\\
-\a_0 &= e_1{+}e_2 = 2\a_1+3\a_2+4\a_3+2\a_4.
\nonumber\\
\hT &= \{ 1 \ge \f_1+\f_2\ ,\ \ \f_1 \ge \f_2+\f_3+\f_4\ ,\ \f_2\ge \f_3\ge \f_4 \ge 0 \}.
\nonumber
\end{align}

\paragraph{G$_{\bf 2}$.}

There is no center symmetry and the root system and gauge cell in a plane orthogonal to $e_1+e_2+e_3$ in $\R^3$ are given by
\begin{align}
\Phi_s &= \{ \a_1 = 2e_2{-}e_1{-}e_3 \ ,\ \a_2 = e_1{-}e_2\},
\nonumber\\
\Phi_+ &= \{ e_1{-}e_2 \ ,\ e_2{-}e_3\ ,\ e_1{-}e_3\ ,\ 
2e_1{-}e_2{-}e_3\ ,\ 2e_2{-}e_1{-}e_3\ ,\ e_1{+}e_2{-}2e_3 \},
\nonumber\\
-\a_0 &= e_1{+}e_2{-}2e_3 = 2\a_1+3\a_2.
\nonumber\\
\hT &= \{ \tfrac13 \ge \f_1+\f_2\ ,\ \ \f_1 \ge \f_2 \ge 0\ ,\
\text{and}\ \f_1+\f_2+\f_3=0 \}.
\nonumber
\end{align}

\end{document}